\documentclass[11pt,letterpaper]{article}

\usepackage[margin=1in]{geometry} %

\usepackage[style=ieee, url=false]{biblatex}
\usepackage{siunitx} %

\usepackage{authblk}

\usepackage{ifdraft}
\usepackage[markifdraft]{gitinfo2}
\usepackage[status=draft]{fixme}

\usepackage{soul}
\usepackage{ulem}
\usepackage{color}

\definecolor{lightblue}{rgb}{.90,.95,1}

\usepackage[cmex10]{amsmath}   %
\usepackage{mathtools} %
\usepackage{epstopdf}
\usepackage{graphicx}
\usepackage{grffile} %
\setkeys{Gin}{draft=false} %
\usepackage{placeins} %
\usepackage{subcaption} 
\newsavebox{\largestimage} %
\newcommand{\figuredir}{L-G figures, compressed} %

\usepackage{hyperref}
\hypersetup{final=true, %
                    pagebackref=true}

\begin{document}

\title{The decomposition of aberrated or turbulent wavefronts into a spatial mode spectrum using optical cavities}
\author{Merlin L. Mah}
\author{Joseph J. Talghader}
\affil{\footnotesize \textit {merlin@umn.edu, joey@umn.edu}}
\affil{\footnotesize Dept. of Electrical \& Computer Engineering, University of Minnesota \authorcr Minneapolis, MN 55455}

\date{} %
\maketitle

\ifdraft {
    \noindent \uppercase\expandafter{\today} \\
		Git commit: \gitBranch\,@\,\gitAbbrevHash{} (\gitDirty) release: \gitReln{} (\gitAuthorDate) \\
    }

\section{Abstract}

It is shown that an aberrated wavefront incident upon a Fabry-Perot optical cavity excites higher order spatial modes in the cavity, and that the spectral width and distribution of these modes is indicative of the type and magnitude of the aberration. The cavities are purely passive and therefore frequency content is limited to that provided by the original light source, unless time-varying content is introduced. To illustrate this concept, spatial mode decomposition and transmission spectrum calculation are simulated on an example cavity; the effects of various phase delays, in the form of two basic Seidel aberrations and a composite of Zernike polynomial terms, are shown using both Laguerre-Gaussian and plane wave incident beams. The aggregate spectral width of the excited cavity modes is seen to widen as the magnitude of the phase delay increases.

\section{Introduction}

The power levels of diode-pumped solid-state and fiber lasers have been steadily increasing over the past two decades \cite{Jauregui_Highpowerfibrelasers_2013}, which has enhanced the technical viability of high-power applications such as LIDAR, directed energy weaponry, and free space communications \cite{Shi_Fiberlaserstheir_2014}. However, high power beams propagating over long distances tend to accumulate wavefront distortion from turbulence and other atmospheric factors, reducing the efficacy of power delivery. 

There are a number of methods to measure the local aberrations in a wavefront, such as Shack-Hartmann sensors \cite{Platt_HistoryprinciplesShackHartmann_2001}, shearing interferometers \cite{Welsh_Fundamentalperformancecomparison_1995}, and holography \cite{Corbett_Designingholographicmodal_2007a} \cite{Ghebremichael_Holographybasedwavefrontsensing_2008a}, some of which can be used in array/subaperture forms. \cite{Tippie_AberrationCorrectionDigital_2012} The first of these is the most canonical solution for wavefront measurement, using an imaging sensor and a lenslet array to measure localized tilts in the beam and using these to infer the shape of the wavefront. \cite{Platt_HistoryprinciplesShackHartmann_2001} The Shack-Hartmann suffers from a weakness in measuring strong wavefront distortions across small apertures, a situation increasingly relevant for long-distance laser applications. Several alternative sensor designs, including pyramid and curvature wavefront sensors, share this inability to handle the many waves of distortion and apparent null-intensity discontinuities---"branch points"---commonly caused by horizontal atmospheric propagation, as shown in Fig.~\ref{fig:DE_aberration_example}. Methods which can handle these, such as quality-metric estimation, plenoptic sensing, or interferometry, tend to add complex processing algorithms and/or a great deal of large, sensitive hardware. \cite{Watnik_Wavefrontsensingdeep_2018}

\begin{figure}[!h]
\centering
\includegraphics[height=5cm]{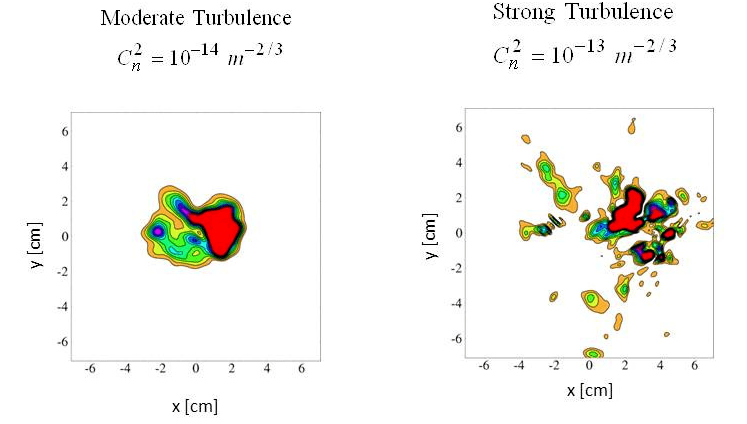}
\vspace{-2mm}
\caption{A high power laser beam wavefront measured after propagating through 520~\si{\meter} of moderate (left; $C_n^2 = 10^{-14} \text{m}^{-2/3}$) and severe (right; $C_n^2 = 10^{-13} \text{m}^{-2/3}$) atmospheric turbulence. Image courtesy Dr. Joseph Pe\~{n}ano.}
\label{fig:DE_aberration_example}
\vspace{-1mm}
\end{figure}

However, there are relatively few technologies to measure natural atmospheric turbulence in real-time. Scintillometry and hot wire measurements are the classic techniques, but their requirements are simply not suited for routine field use. \cite{Fried_Measurementslaserbeamscintillation_1967} \cite{Izquierdo_Cn2determinationdifferential_1987} Some alternatives have been demonstrated, such as using real-time camera images to observe fluctuations in the edges of distant objects \cite{Zamek_Turbulencestrengthestimation_2006}; while these techniques are promising, they require either significant hardware or software processing or both. There is space in this technical area for phenomena that could lead to single point measurements that directly indicate certain features of an aberrated wavefront and the turbulence that caused it.

In this paper, it is shown that an aberrated wavefront coupled to an optical cavity will excite a series of cavity spatial modes. These spatial mode profiles vary from aberration to aberration, and the spectral width of the spatial mode profile tends to increase as the number of waves of aberration increases. Note that in this work, the cavities are perfectly passive and no new frequencies are created that did not exist in the original light beam. Since high-power lasers tend to have large spectral bandwidths---for example, fiber lasers often require intentionally-broadened bandwidths on the order of 10~\si{\giga\hertz\per\kilo\watt} to control stimulated Brillouin scattering---the intrinsic laser spectrum can feed large numbers of spatial modes without the need for additional bandwidth. \cite{Jauregui_Highpowerfibrelasers_2013} \cite{AlOgloza_Sept2018}

\section{Previous works}

Interest in aberration modeling using Gaussian beams has in recent years been boosted by work relating to gravitational wave sensors such as the Laser Interferometer Gravitational-Wave Observatory (LIGO). Bond, Freise, et al. (2011) have simulated the ability of mirror aberrations, expressed using Zernike polynomials, to redirect power from the fundamental Laguerre-Gaussian mode into higher-order helical modes. \cite{Bond_HigherorderLaguerreGauss_2011}. Gatto, Barsuglia, et al. (2014) detailed the intentional use of specific higher-order modes as the primary operational beams in such long-cavity interferometers. \cite{Gatto_FabryPerotMichelsoninterferometerusing_2014} Resonator mode expansions are used in quite a few other measurement systems, such as the laser beam quality and alignment system of Kwee, Danzmann, et al. (2007). \cite{Kwee_Laserbeamquality_2007}

Gaussian beam mode analysis (GBMA) is an established subfield, but its applications are usually focused towards efficient ways of performing diffraction calculations \cite{Martin_Longwaveoptics_1993}   \cite{Tsigaridas_ZscananalysisnearGaussian_2003}. This is possibly exemplified by Trappe, Murphy, and Withington (2003) \cite{Trappe_Gaussianbeammode_2003}, who use GBMA to calculate the propagation of a beam with Seidel aberrations and compare the results to traditional diffraction equations. 
In shorter resonators, Talghader and Liu (2006) examined the effects of a cavity on the spatial mode makeup of a Gaussian beam, with the goal of detecting imperfections in tunable micromirror cavities; however, a relatively limited selection of aberrations could be easily treated. \cite{Liu_Spatialmodeanalysismicromachined_2006}
Takeno, Shirai, et al. (2011) also adopted the Fabry-Perot cavity's mode discrimination for aberration sensing, but this technique used the aggregate amplitudes of the reflectivities of the modes to distinguish how much of the incident light represented the fundamental mode. \cite{Takeno_Determinationwavefrontaberrations_2011}

\section{Cavity Mode Decomposition and Transmission Spectra}

The eigenmodes of a stable optical resonator are a set of Gaussian beams: the Laguerre-Gaussians in cylindrically-symmetric systems, and the Hermite-Gaussians in Cartesian coordinates \cite{Padgett_PoyntingvectorLaguerreGaussian_1995}. For the former, the field for transverse mode index integers $n$ ($\ge 0$) and $\alpha$ is given by 
\begin{align}
\begin{split}
E_{n,\alpha} (r, \phi, z) = & E_0 \frac{w_0}{w(z)} \sqrt{\frac{2 n!}{\pi ( \lvert \alpha \rvert + n )!}} \left( \frac{\sqrt{2} r }{w(z)} \right)^{\lvert \alpha \rvert}
L_n^{\lvert \alpha \rvert} \left( \frac{2 r^2 }{w^2(z)} \right)\\ 
& \exp \left( - \frac{r^2}{w^2 (z)} - i k \frac{r^2}{2 R(z)} - i k z - i \alpha \phi - i ( 2 n + \lvert \alpha \rvert + 1 ) \, tan^{-1} \frac{z}{z_0} \right) 
\end{split}
\label{eqn:L-G_def}
\end{align}

\noindent where $L_n^\alpha$ is the generalized Laguerre polynomial, $z_0 = \pi w_0^2 n / \lambda$ is the Rayleigh range, $k = 2 \pi n /\lambda$ is the wavenumber, and $w(z)$, $w_0$, and $R(z)$ are the familiar Gaussian parameters of spot size, radius at beam waist, and wavefront radius of curvature respectively \cite{Yariv_PhotonicsOpticalElectronics_2006} \cite{Goldsmith_QuasiopticalSystems_1998}.
The Laguerre-Gaussians supported within the resonator cavity will have varying frequencies (some degenerate), each of which will host a peak in the resonator's spectral transmission curve. For a resonator with spherical mirrors of radii $r_1$ and $r_2$ separated by distance $d$, these resonant frequencies are given by 
\begin{align}
\nu_{q, n, \alpha} = \frac{c}{2 n_{\text{medium}}d} \left( q + \frac{l + m + 1}{\pi} \, cos^{-1} \bigl( \sqrt{(1 - \frac{d}{r_1}) \, (1 - \frac{d}{r_2})} \bigr) \right)
\label{eqn:cavity_resonant_freqs}
\end{align}

\noindent where $q$ is the longitudinal mode index. \cite{Yariv_PhotonicsOpticalElectronics_2006} 

Each of the Hermite- and Laguerre-Gaussian families form a mutually orthogonal basis set which may be used to express any arbitrary optical beam meeting the paraxial conditions \cite{Yariv_PhotonicsOpticalElectronics_2006} \cite{Siegman_Lasers_1986}.
The electric field amplitude of each Gaussian mode $E_{n,\alpha}(r, \phi, z)$ can be computed by the overlap integral \cite{Liu_ElectricallyTunableMicromirrors_2004}
\begin{align}
c_{n,\alpha}=\int_{0}^{r=a} \int_{0}^{\phi=2 \pi} E_{\text{in}}(r,\phi,z) E_{n,\alpha}(r,\phi,z)^{*} r \, dr \, d\phi
\label{eqn:L-G_overlap_integral_cavity}
\end{align}

\noindent where $a$ is the radius of the cavity or limiting pupil. The resulting power being coupled into the mode is then $\lvert c_{n, \alpha} \rvert ^2$.

If a laser beam of finite spectral bandwidth, with an imposed wavefront aberration, is incident upon a lossless resonator, it will couple some of its energy into the eigenmodes of the cavity. Aberrations or other phasefront deformations to the incident wave will change the spatial distribution of energy, and thus the amount of coupling---where supported by the source spectrum---into the higher-order transverse modes of the cavity. The excitation of these modes can be easily seen in the transmission of light at the resonant frequencies. 

The process suggests a reversible correlation. Since each cavity mode corresponds (usually degenerately) to a transmitted resonant frequency, the spectral content of the transmitted output of the cavity will correlate to the strength and nature of any aberrations imposed on the input beam. Given spectral transmittance information from such a cavity, it will be possible to deduce the presence of wavefront aberrations and limited information about their number and magnitude.

\section{Concept Demonstration - Single Aberrations}

We can illustrate this concept by numerical application of the above equations. The choice of the optical cavity that we might use to analyze a wavefront is certainly not unique; one must consider cavity issues of the longitudinal mode spacing, spatial mode spacing, spectral width of each resonance, and other practical factors such as mirror shape and cavity materials. The laser wavelength and lineshape are also important to the design and analysis.

Let us consider an aberrated wavefront originally emitted from a high power laser at wavelength 1.064~\si{\micro\meter} with a spectral linewidth of about 100~\si{\giga\hertz}. Since we will be analyzing the wavefront using the transverse modes of an optical cavity, the cavity length (and material) are chosen such that the longitudinal mode spacing is somewhat greater than the spectral linewidth of the laser and excitation is thus limited to a single longitudinal mode number. 
A fused silica blank of thickness 1~\si{\milli\meter} would lead to a longitudinal mode spacing of about 103~\si{\giga\hertz}. We wish to have a large number of spatial modes in the system, so erring on the side of too many, we can choose one spatial mode every 1~\si{\giga\hertz}. We also wish to allot a large enough spectral width to each cavity mode so that we will have reasonable light throughput for each mode but not so much that they overlap and are difficult to distinguish. With this in mind, a spectral width of approximately 300~\si{\mega\hertz} would be useful. Using $\Delta \nu = \frac{c}{2 \pi n z_0}$ for the separation between transverse modes for near-planar mirrors and $\Delta \nu_{1/2} = \frac{c}{2 \pi n l } \frac{1 - R}{\sqrt{R}}$ for the full-width at half-maximum (FWHM) \cite{Yariv_PhotonicsOpticalElectronics_2006}, these parameters lead to a cavity with mirror reflectivity of $R=$~99.1\%, fed by a beam with waist size $w_0 \approx$ 100~\si{\micro\meter}. 

For calculation of cavity transmission spectra, we presume that the laser supplies, with equal intensity, every wavelength necessary to observe the transmission spectrum from the range of modes that we sample. A more realistic light source would typically have variations in intensity with frequency, but these deviations from uniformity will vary from source to source and can be easily handled mathematically after a spectral power measurement of the laser beam. To justify knowledge of the total beam power entering the cavity, we will assume that the beam may be sampled for a power reading. Fig.~\ref{fig:Conceptual_diagram} shows a conceptual diagram of the measurement setup which might enable our simulated scenarios. 

\begin{figure}[!h]
\centering
\includegraphics[height=6cm]{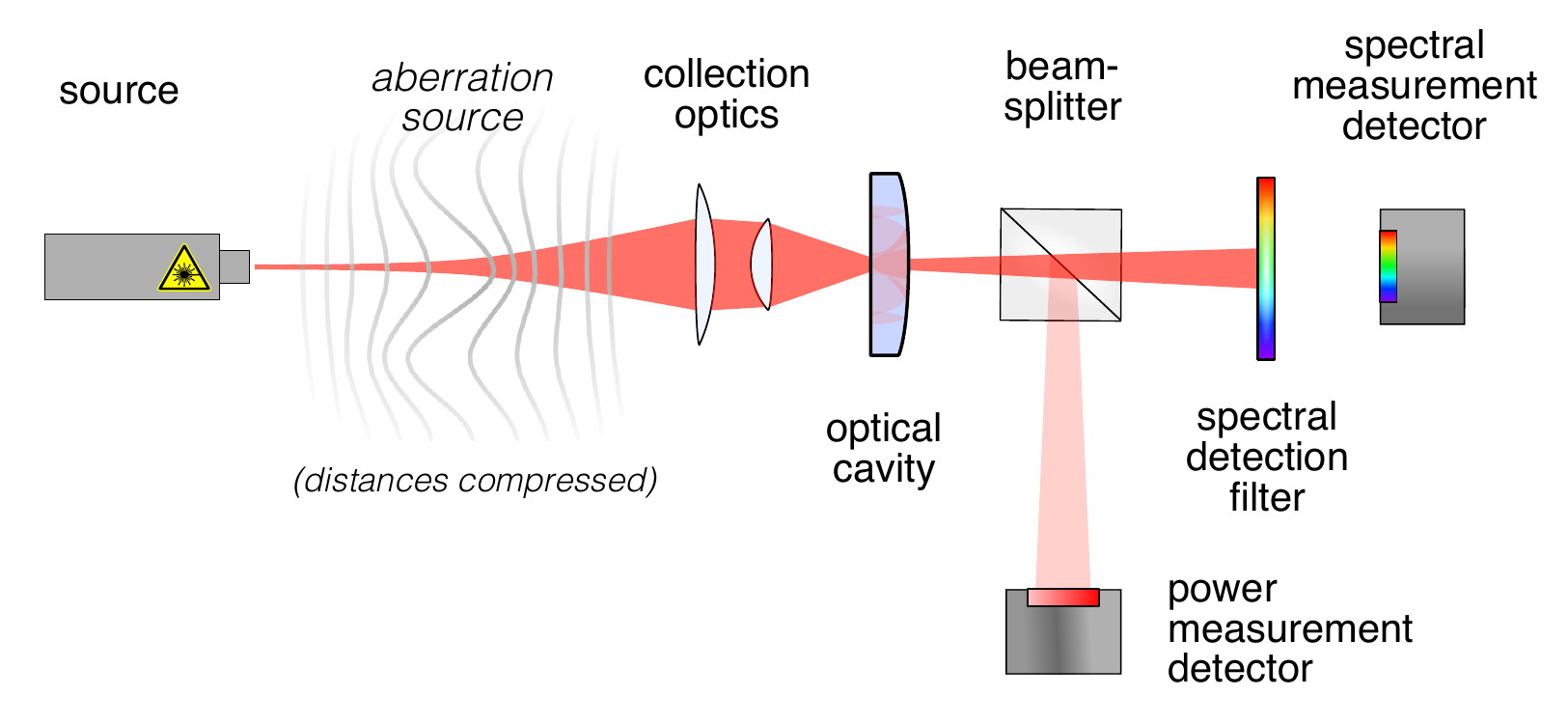}
\vspace{-2mm}
\caption{A conceptual diagram of the measurement setup emulated by the simulations below. A light source provides a source beam which has some phase aberrations imposed upon it. The light is fed---optionally with the help of collection optics---into an optical cavity and the transmitted output is spectrally analyzed using some kind of selectable spectral filter and a detector. A small portion of the light may be sampled to measure the total beam power.}
\label{fig:Conceptual_diagram}
\vspace{-1mm}
\end{figure}

To simplify our initial case, we will assume that the source beam has exactly the correct $w_0$ and $R$ to transmit all of its power into the cavity's fundamental mode. Without aberration or perturbation, we confirm from evaluation of the overlap integral eq.~\ref{eqn:L-G_overlap_integral_cavity} that only the fundamental mode $(n, \alpha) = (0, 0)$ is excited, as we would expect. 

We may now add a basic wavefront modification and observe its effects. The Seidel aberrations, the five simple "third-order" transverse ray aberrations almost omnipresent to some degree in every optical imaging system, seem a natural choice. The Seidels are listed, in terms of the exit pupil spatial coordinates $(x, y)$ and image plane coordinate $x_0$ appropriate for an imaging system, in Table~\ref{tab:Seidels} \cite{Wyant_BasicWavefrontAberration_1992}.

\begin{table}[h]
\centering
\setlength\tabcolsep{12pt}
    \begin{tabular}{ll}
        \hline 
            Seidel aberration name & Wavefront functional form \\ 
        \hline
            Spherical aberration & $r^4$ \\ 
            Coma & $x_0 \cdot r^3 \cos \phi$ \\ 
            Astigmatism & $x_0^2 \cdot r^2 \cos^2 \phi $ \\ 
            Field curvature & $x_0^2 \cdot r^2 $ \\ 
            Distortion & $x_0^3 \cdot r \cos \phi $ \\ 
        \hline
    \end{tabular}
\caption{The five Seidel aberrations and their functional forms, in the context of a lens or imaging system. $r$ and $\phi$ are the polar spatial coordinates in the system's exit pupil and $x_0$ is one of the spatial coordinates of the subsequent image plane (the image plane coordinate system having been assigned such that $y_0$ reduces to 0, leaving only $x_0$ and $z_0$.) After \cite[p.17]{Wyant_BasicWavefrontAberration_1992}. }
\label{tab:Seidels}
\end{table}

The Seidels are most often seen in the context of the lenses or imaging systems they plague, but free propagation through does not possess the same refocusing or imaging properties. Consequently, we use the Seidels here only to describe the surface of a phasefront modification, multiplicatively applied to the propagated field of our light source. Re-examining the Seidel functional forms while disregarding the image plane spatial coordinate $x_0$, we notice that there are two basic components combined in various powers: the directionally-sensitive term $r \cos \phi$ and the circularly symmetric $r^2$ term. These are seen isolated in distortion and field curvature, respectively, with the other three Seidel phasefronts able to be synthesized by variously multiplying them together; therefore, we will choose distortion and field curvature for our first examples. Each aberration shape is scaled by a coefficient $W$ and multiplied as a phase delay onto our cavity's incident field, producing 
\begin{align}
E_\text{aberrated} (r, \phi) = & E_\text{source} (r, \phi, z = z_\text{cavity})  \cdot \exp \left(i W r \cos \phi \right)
\label{eqn:Distortion_Seidel_def}
\end{align}

\noindent for distortion, and 
\begin{align}
E_\text{aberrated} (r, \phi) = & E_\text{source} (r, \phi, z = z_\text{cavity})  \cdot \exp \left(i W r^2 \right)
\label{eqn:Fieldcurvature_Seidel_def}
\end{align}

\noindent for field curvature. In both cases, $W$ is set to yield a desired number of waves of phase delay (at maximum) over the area where the source beam's intensity is greater than $1/e^2$ of its maximum. The phasefronts, and the results of applying these phase delays to the incident beam, are seen in Figures~\ref{fig:Distortion_LGbasecase_fields}~and~\ref{fig:Fieldcurvature_LGbasecase_fields}. 

\begin{figure}[htb]
	\begin{subfigure}[c]{0.5\textwidth}
		\centering
		\includegraphics[width=6.5cm]{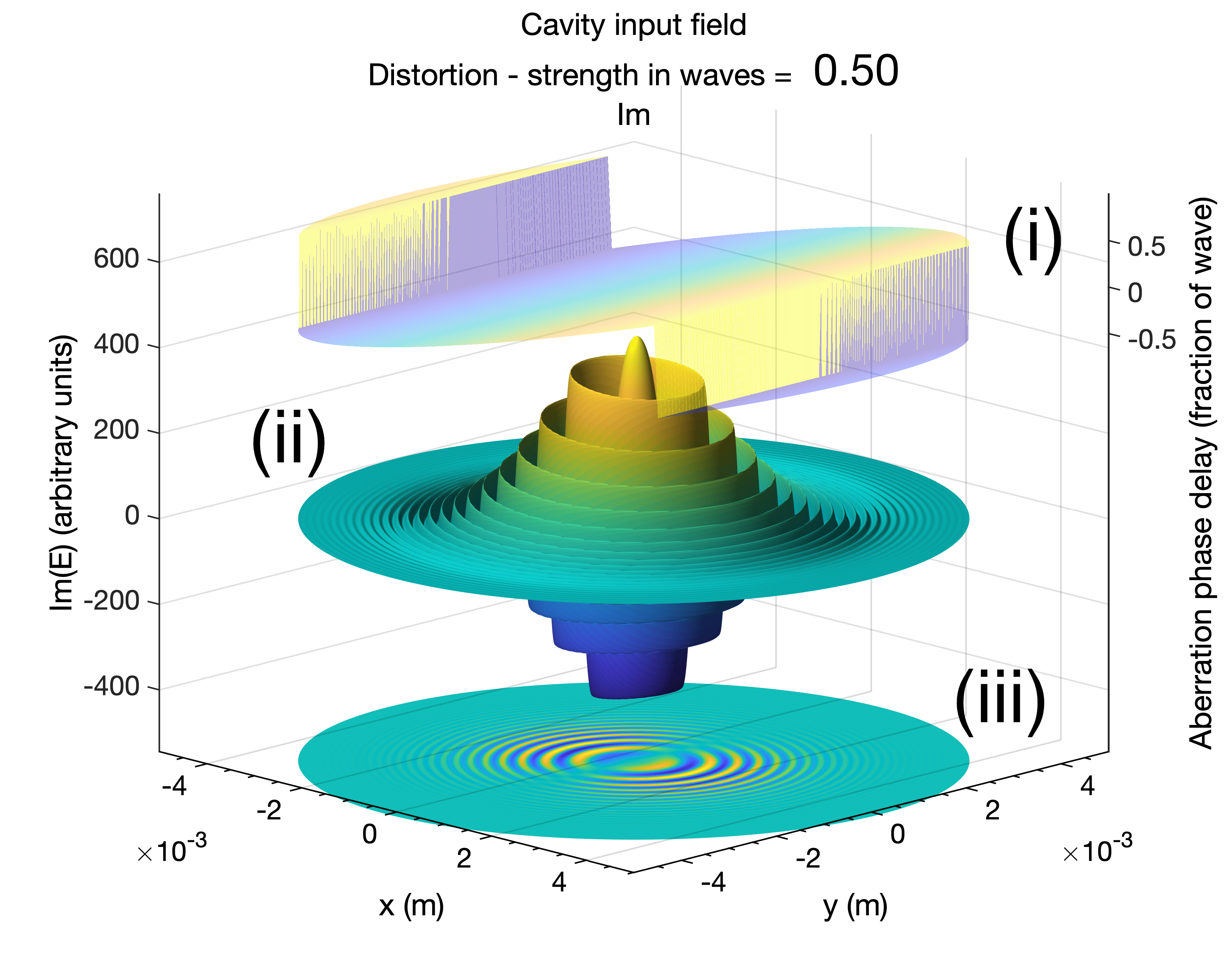}
		\caption{0.5 waves}
		\label{fig:Distortion_LGbasecase_fields_05}
	\end{subfigure}%
	\begin{subfigure}[c]{0.5\textwidth}
		\centering
		\includegraphics[width=6.5cm]{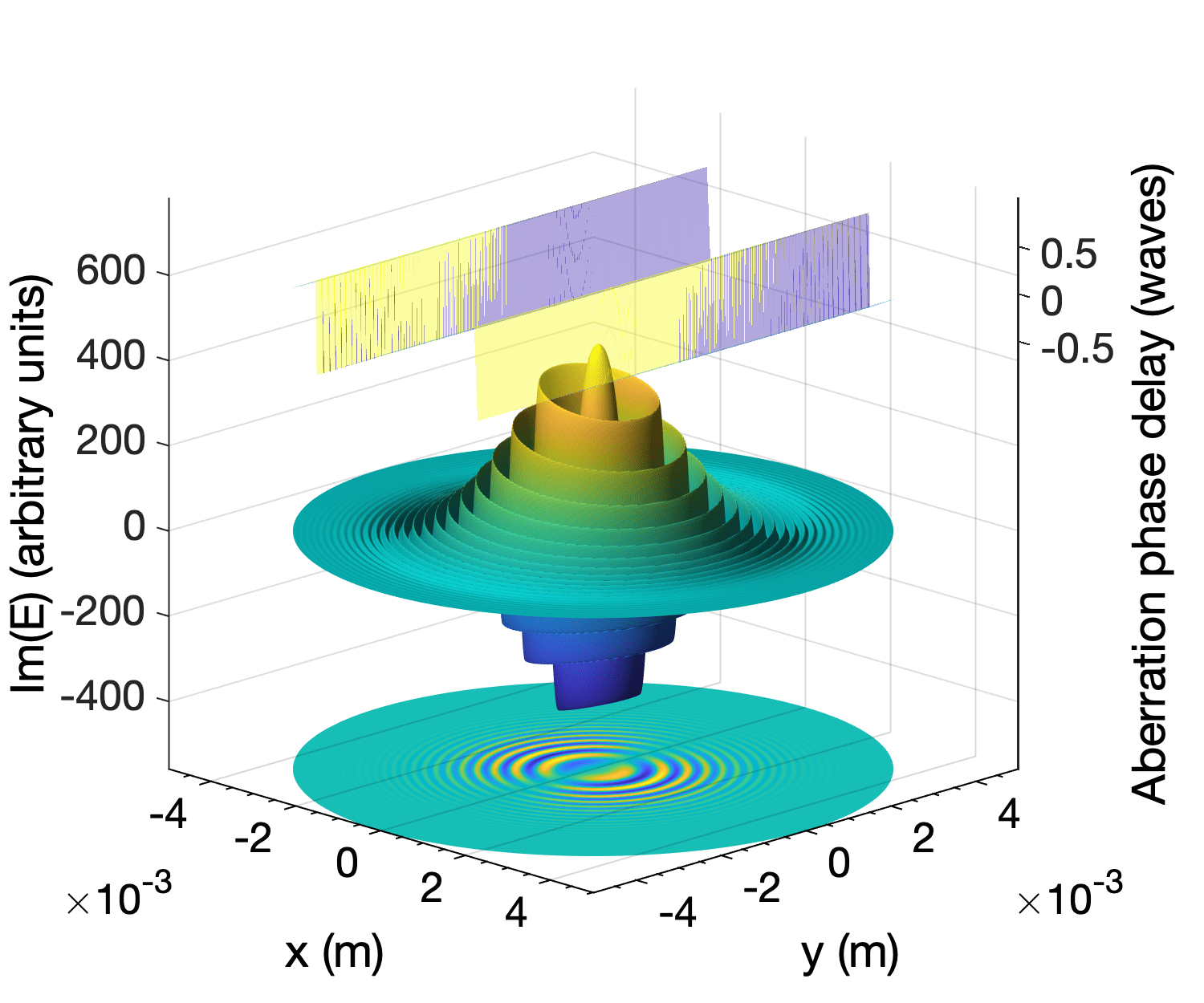}
		\caption{1 wave}
		\label{fig:Distortion_LGbasecase_fields_1}
	\end{subfigure} \\
	\par\bigskip %
	\begin{subfigure}[c]{0.5\textwidth}
		\centering
		\includegraphics[width=6.5cm]{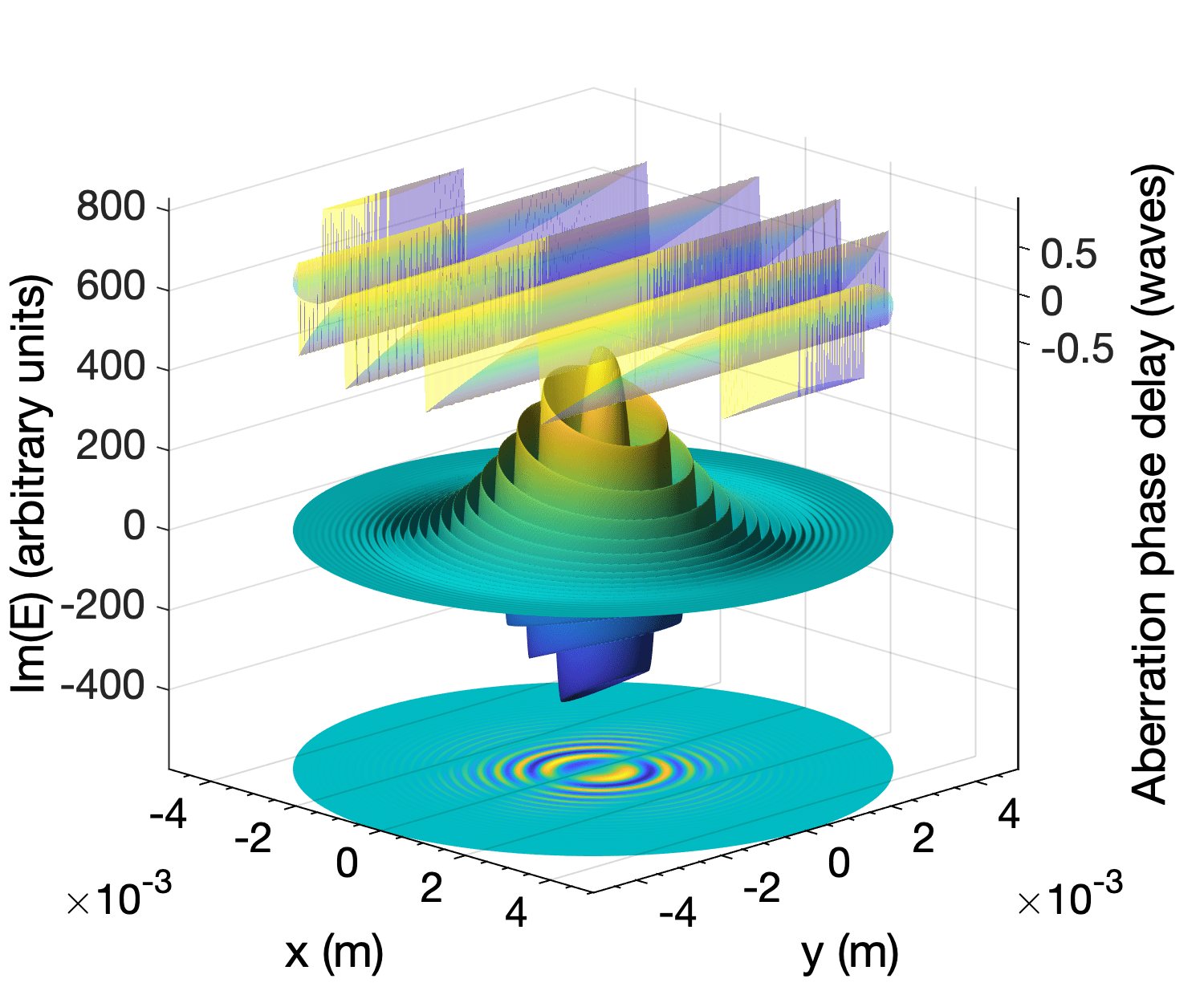}
		\caption{2 waves}
		\label{fig:Distortion_LGbasecase_fields_2}
	\end{subfigure}
	\begin{subfigure}[c]{0.5\textwidth}
		\centering
		\includegraphics[width=6.5cm]{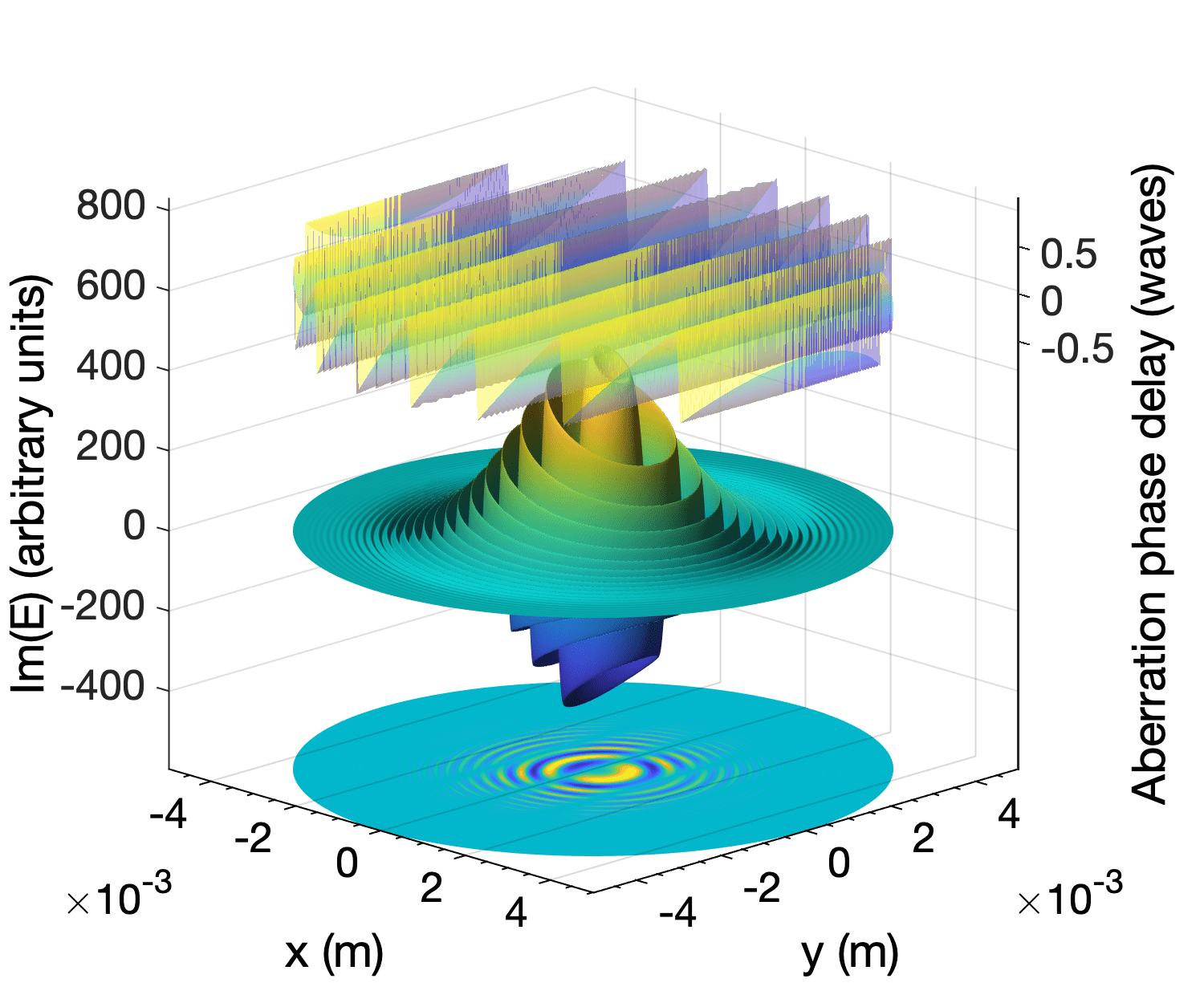}
		\caption{3 waves}
		\label{fig:Distortion_LGbasecase_fields_3}
	\end{subfigure} \\
\caption{The effects of varying strengths of the Seidel aberration distortion. Each plot combines, from top to bottom, and as labeled in the first (0.5~wave) graph: (i) the surface of the phase delay imparted by the distortion aberration, in units of waves per the vertical axis at right, with periodic "folding"; (ii) the imaginary component of the aberrated field, $\Im (E_\text{aberrated} (x, y) )$; and (iii) a flat colormap representation of the change caused in the source beam's imaginary field component by the application of this phase shift, i.e., $\Im (E_\text{aberrated} - E_\text{source} )$ .}
\label{fig:Distortion_LGbasecase_fields}
\end{figure}

\begin{figure}[h!]
	\begin{subfigure}[c]{0.5\textwidth}
		\centering
		\includegraphics[width=6.5cm]{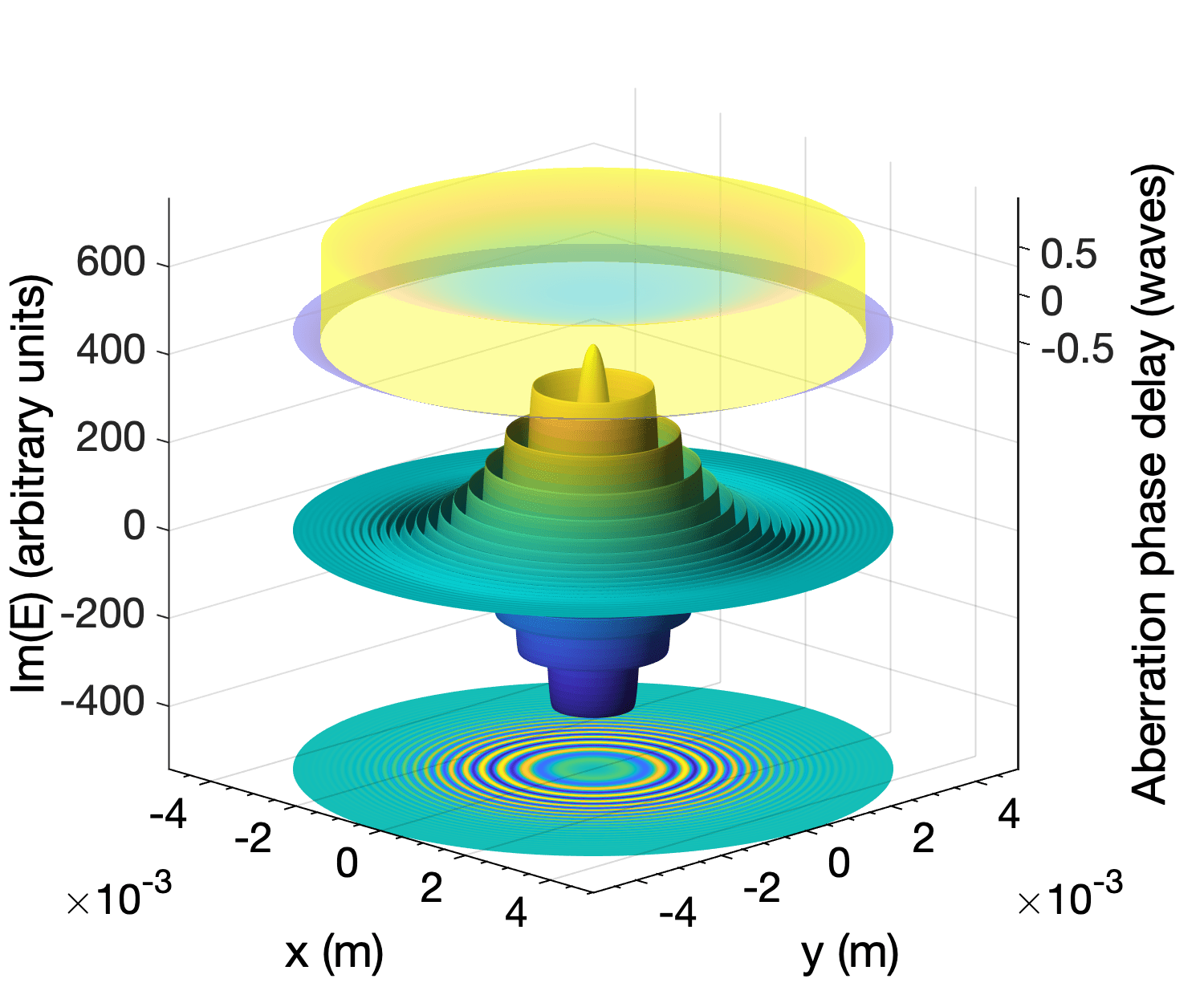}
		\caption{0.5 waves}
		\label{fig:Fieldcurvature_LGbasecase_fields_05}
	\end{subfigure} %
	\begin{subfigure}[c]{0.5\textwidth}
		\centering
		\includegraphics[width=6.5cm]{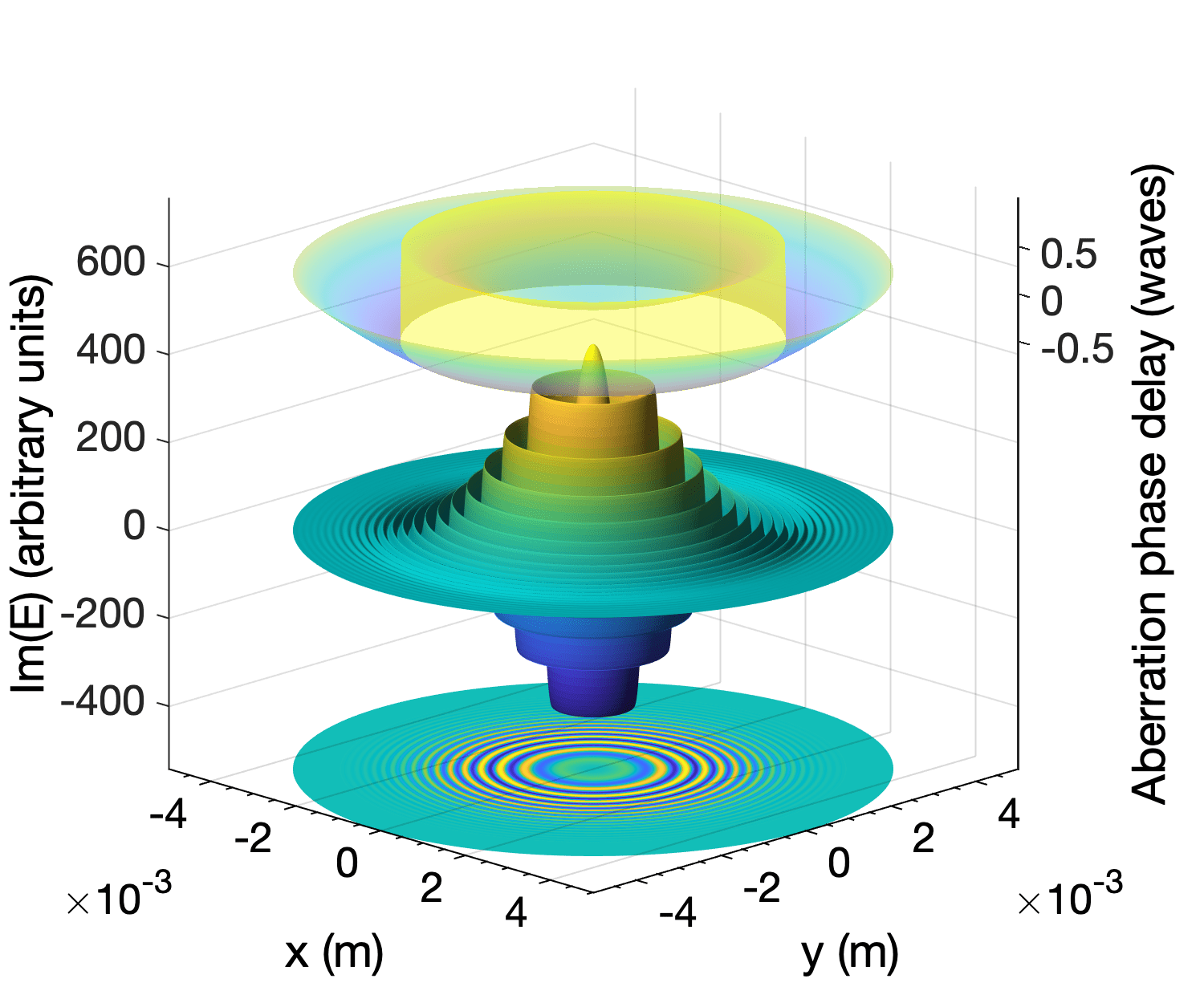}
		\caption{1 wave}
		\label{fig:Fieldcurvature_LGbasecase_fields_1}
	\end{subfigure} \\
	\par\bigskip %
	\begin{subfigure}[c]{0.5\textwidth}
		\centering
		\includegraphics[width=6.5cm]{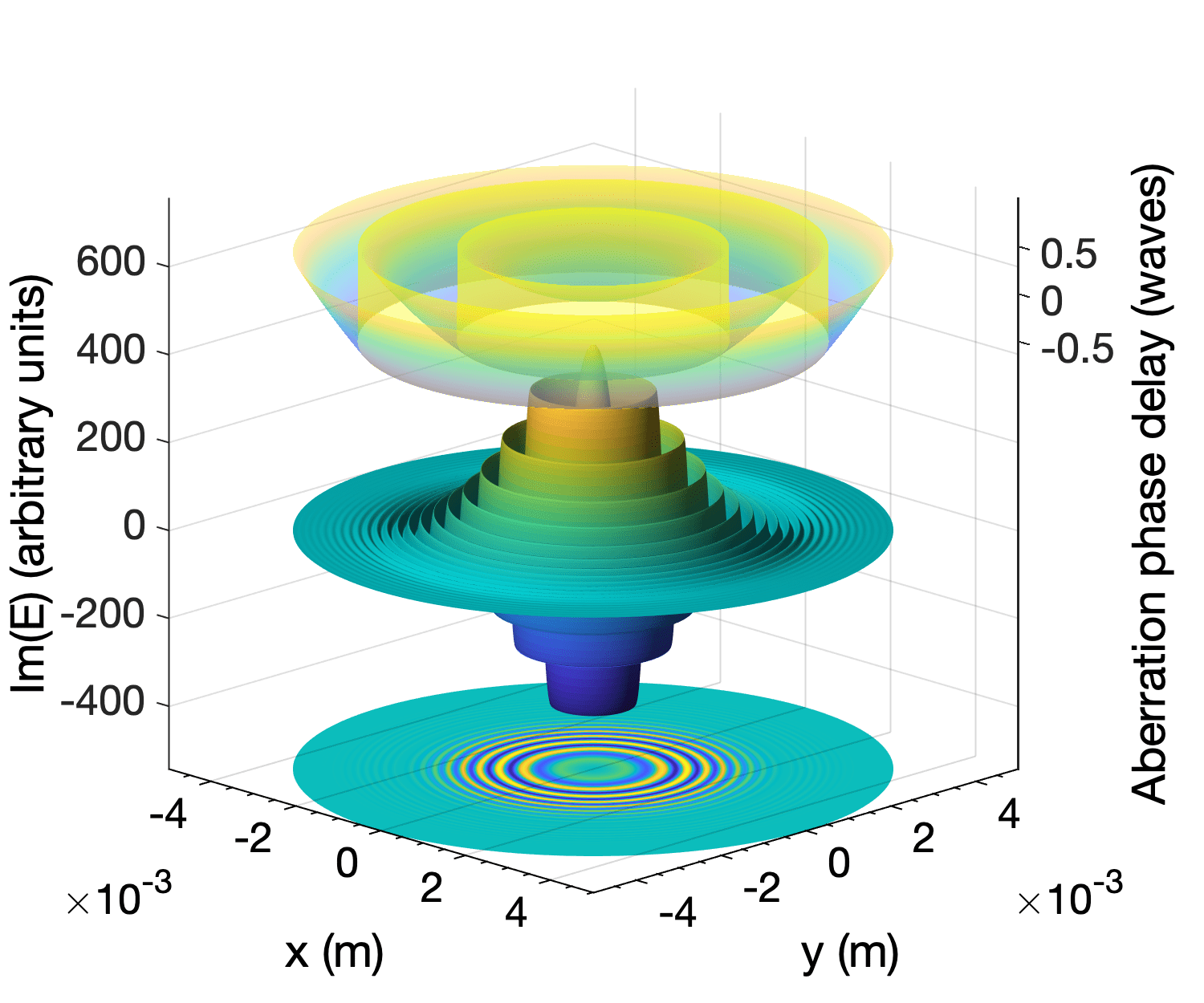}
		\caption{2 waves}
		\label{fig:Fieldcurvature_LGbasecase_fields_2}
	\end{subfigure} %
	\begin{subfigure}[c]{0.5\textwidth}
		\centering
		\includegraphics[width=6.5cm]{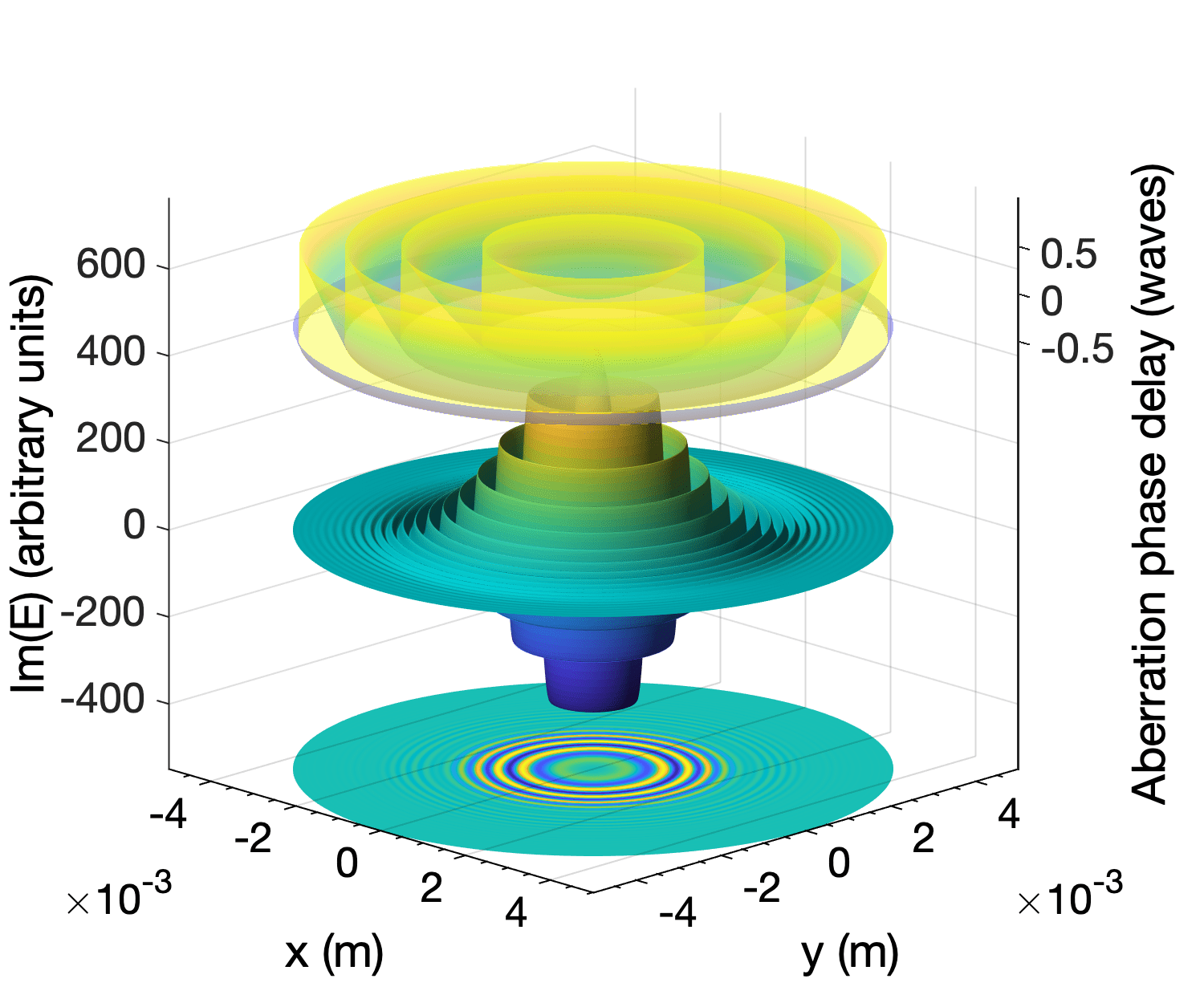}
		\caption{3 waves}
		\label{fig:Fieldcurvature_LGbasecase_fields_3}
	\end{subfigure} %
\caption{The counterparts of the plots in Fig.~\ref{fig:Distortion_LGbasecase_fields} for the Seidel aberration of field curvature.}
\label{fig:Fieldcurvature_LGbasecase_fields}
\end{figure}

\FloatBarrier %

Thus aberrated, these fields now encounter the cavity. We may view the excitation amplitudes of the cavity modes for a variety of different aberration strengths by making use of eq.~\ref{eqn:L-G_overlap_integral_cavity}. The results are shown in Figures~\ref{fig:Distortion_LGbasecase_decomps}~and~\ref{fig:Fieldcurvature_LGbasecase_decomps} for distortion and field curvature, respectively. 

\begin{figure}[!ht]
	\begin{subfigure}[c]{0.5\textwidth}
		\centering
		\includegraphics[width=6.5cm]{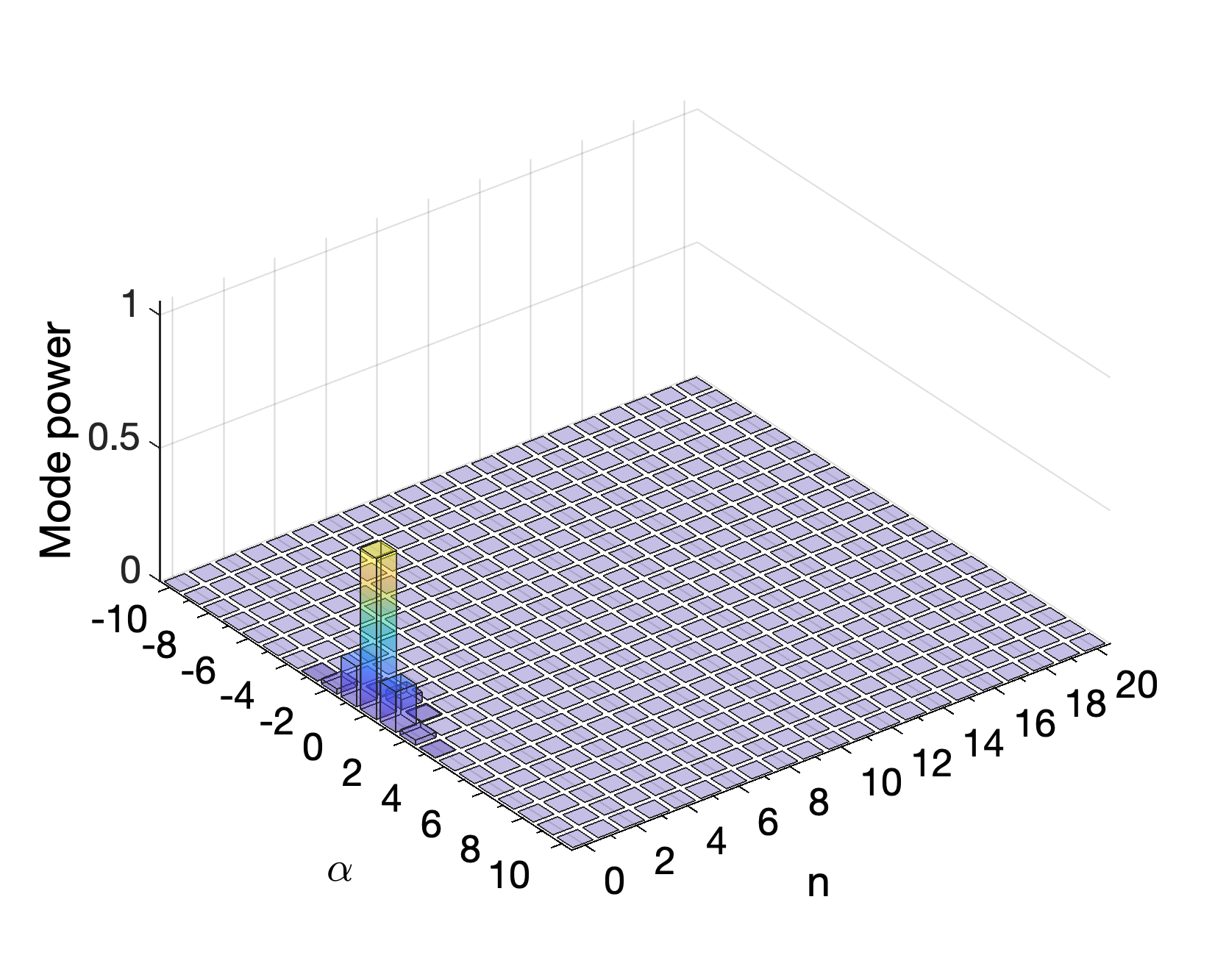}
		\caption{0.5 waves}
		\label{fig:Distortion_LGbasecase_decomp_05}
	\end{subfigure} %
	\begin{subfigure}[c]{0.5\textwidth}
		\centering
		\includegraphics[width=6.5cm]{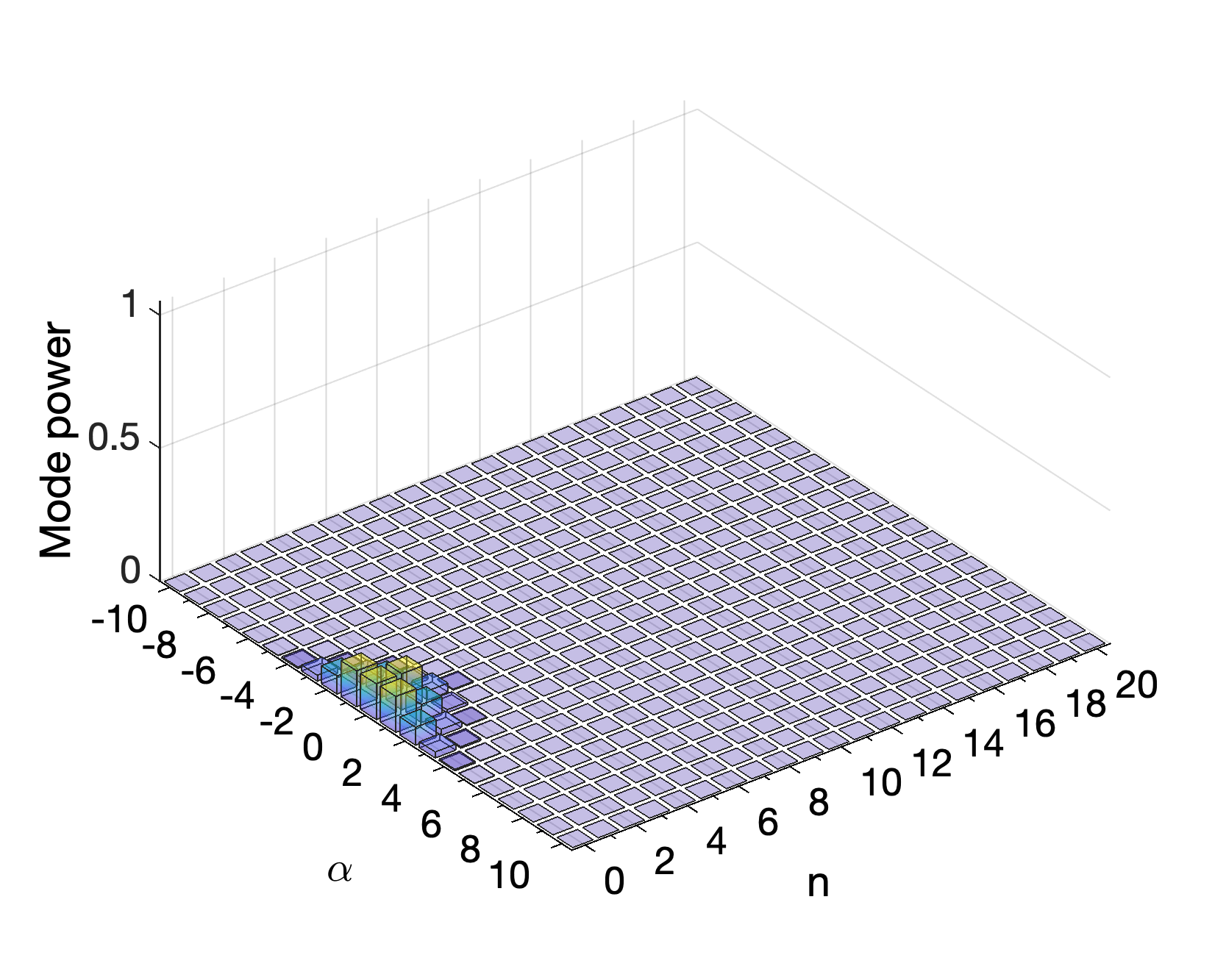}
		\caption{1 wave}
		\label{fig:Distortion_LGbasecase_decomp_1}
	\end{subfigure} \\
	\par\bigskip %
	\begin{subfigure}[c]{0.5\textwidth}
		\centering
		\includegraphics[width=6.5cm]{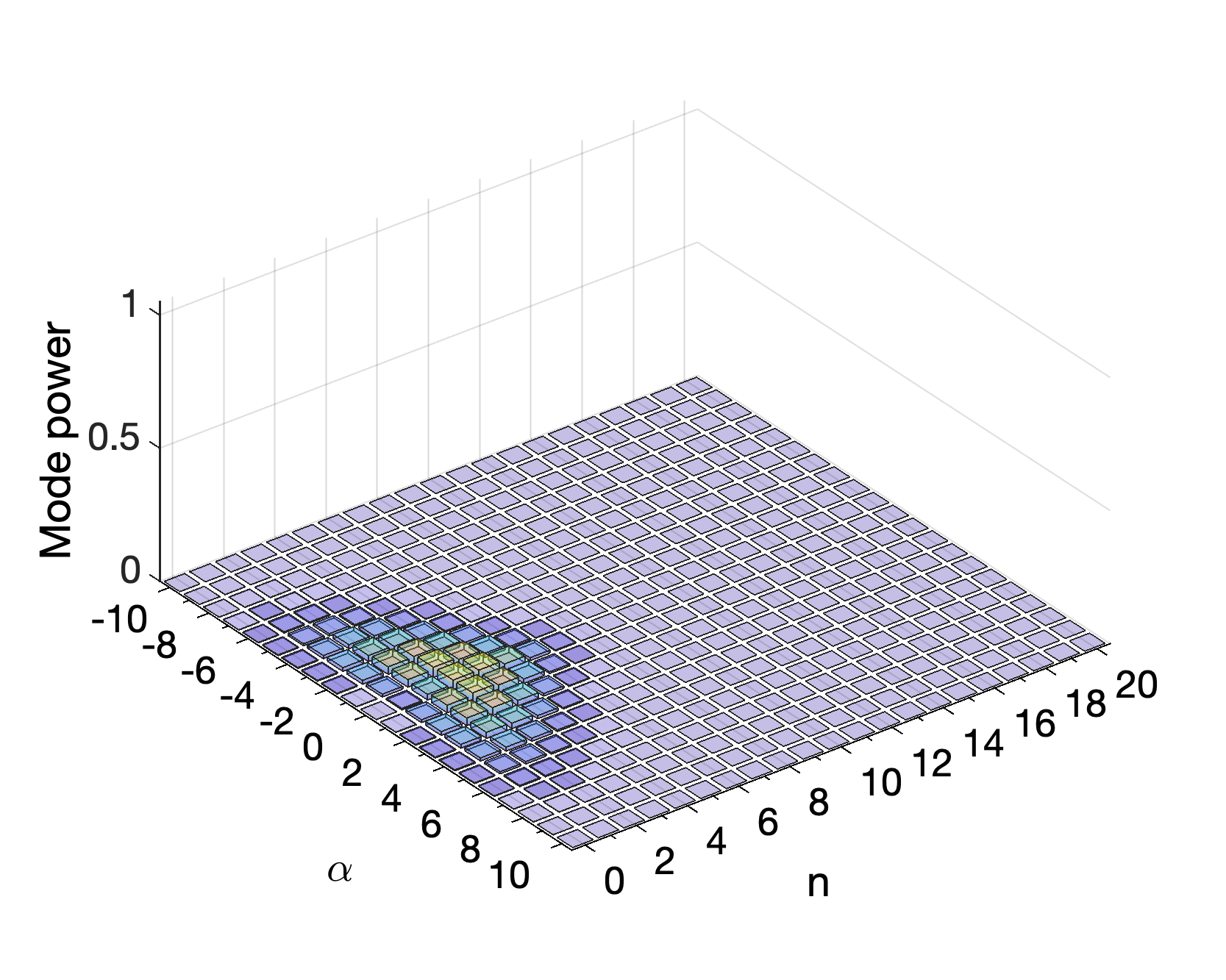}
		\caption{2 waves}
		\label{fig:Distortion_LGbasecase_decomp_2}
	\end{subfigure} %
	\begin{subfigure}[c]{0.5\textwidth}
		\centering
		\includegraphics[width=6.5cm]{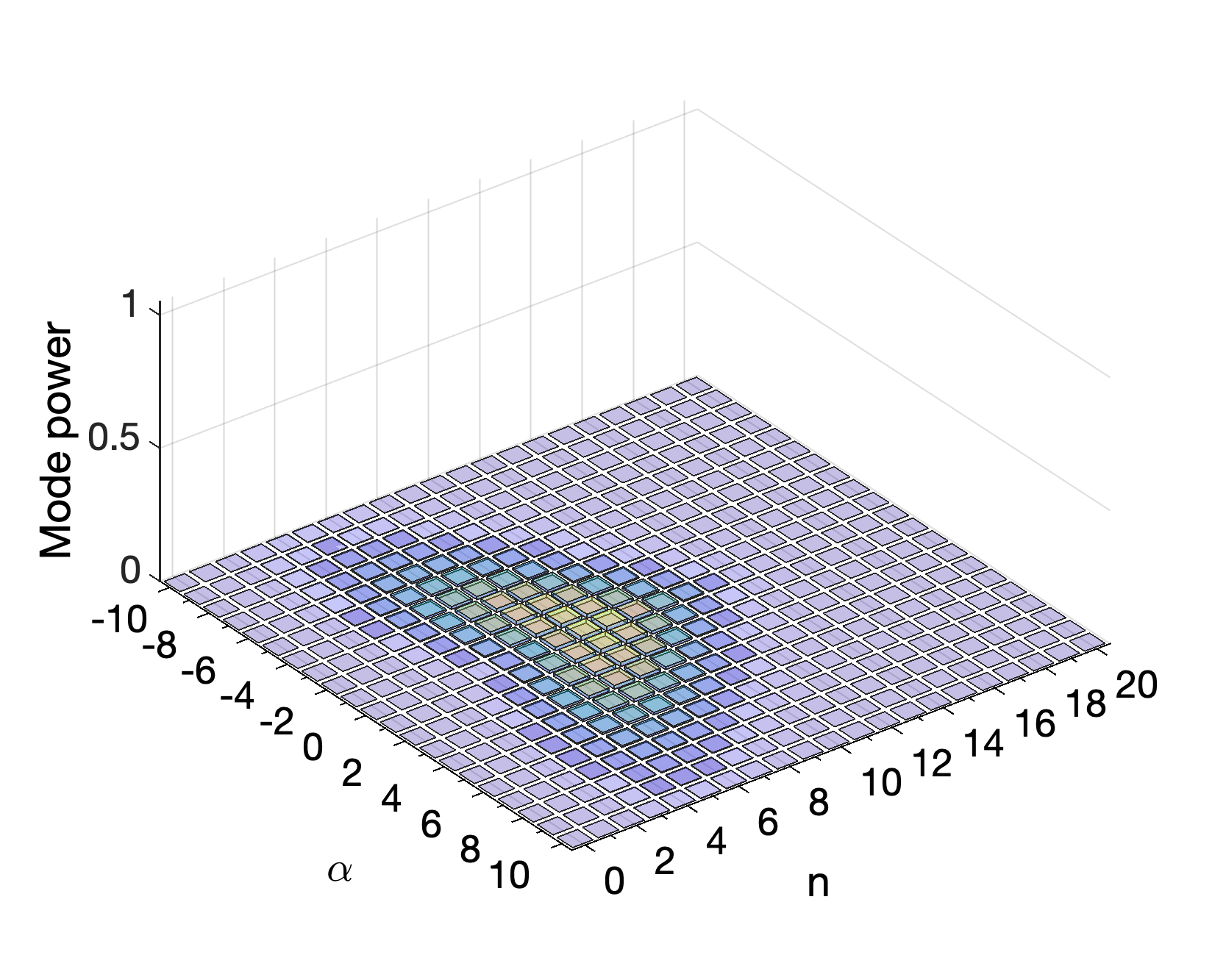}
		\caption{3 waves}
		\label{fig:Distortion_LGbasecase_decomp_3}
	\end{subfigure} %
\caption{The power distribution amongst the transverse cavity modes as the strength of distortion increases from one to five waves of maximum phase delay. }
\label{fig:Distortion_LGbasecase_decomps}
\end{figure}

\begin{figure}[!ht]
	\begin{subfigure}[c]{0.5\textwidth}
		\centering
		\includegraphics[width=6.5cm]{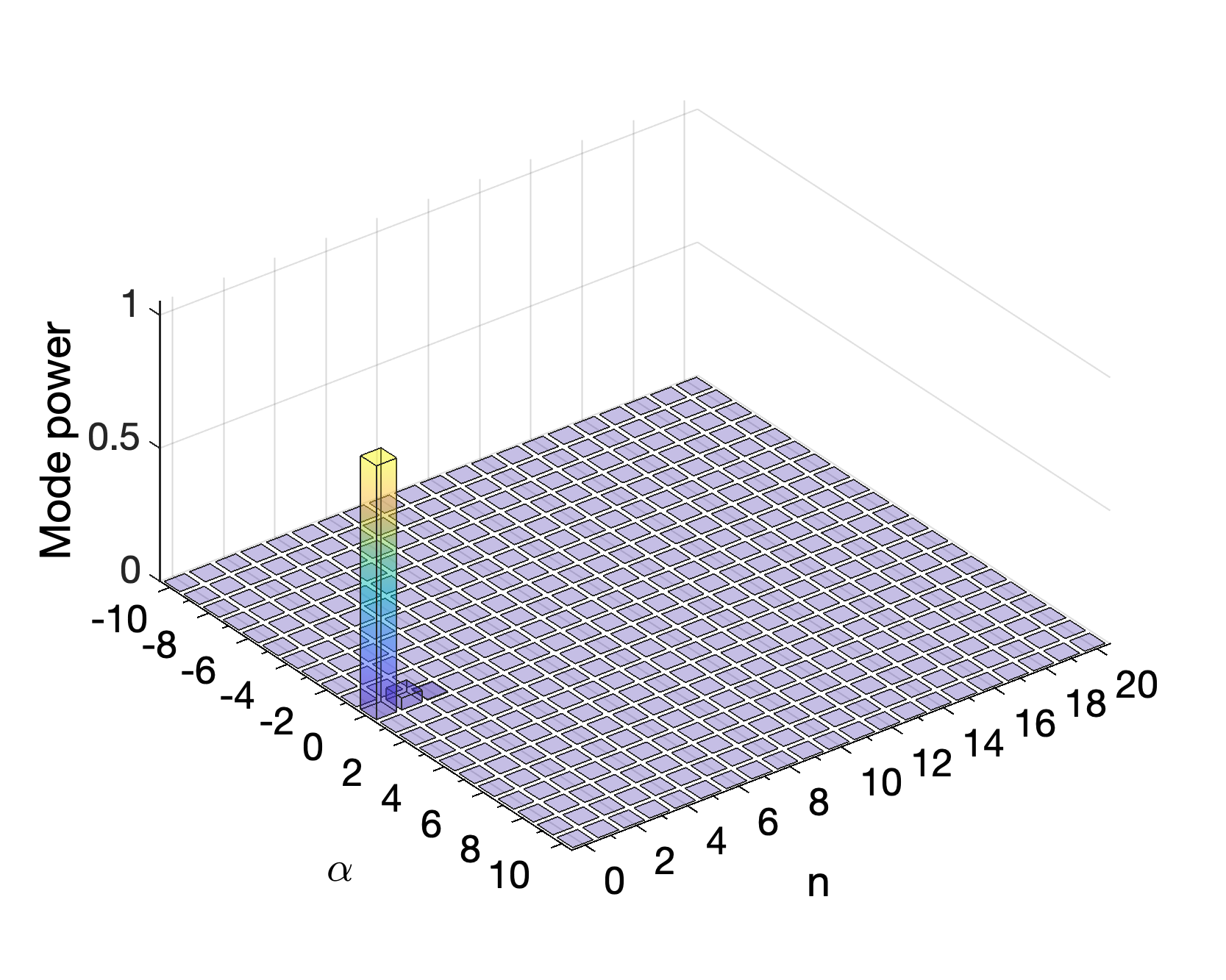}
		\caption{0.5 waves}
		\label{fig:Fieldcurvature_LGbasecase_decomp_05}
	\end{subfigure} %
	\begin{subfigure}[c]{0.5\textwidth}
		\centering 
		\includegraphics[width=6.5cm]{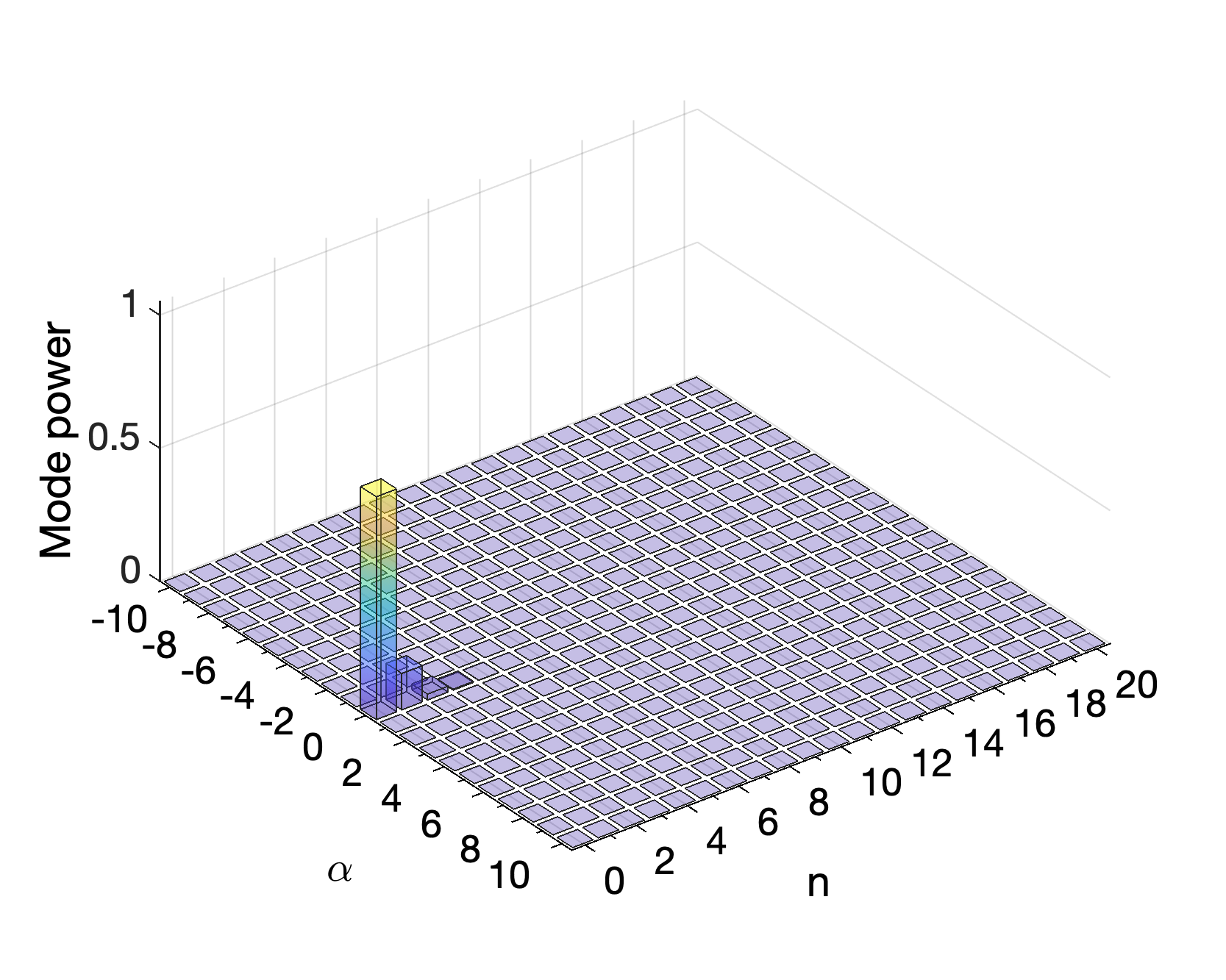}
		\caption{2 waves}
		\label{fig:Fieldcurvature_LGbasecase_decomp_1}
	\end{subfigure} \\
	\par\bigskip %
	\begin{subfigure}[c]{0.5\textwidth}
		\centering
		\includegraphics[width=6.5cm]{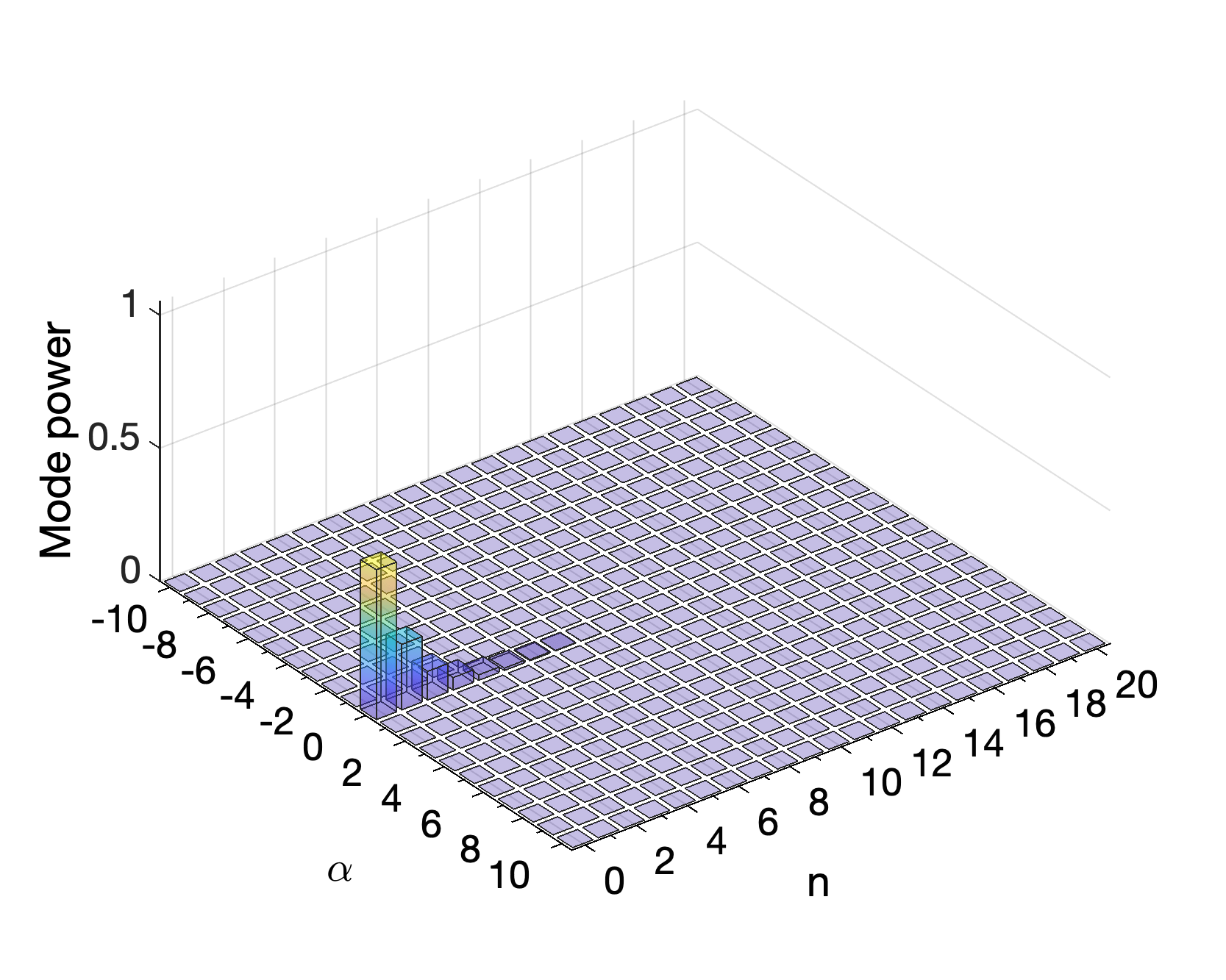}
		\caption{2 waves}
		\label{fig:Fieldcurvature_LGbasecase_decomp_2}
	\end{subfigure} %
	\begin{subfigure}[c]{0.5\textwidth}
		\centering
		\includegraphics[width=6.5cm]{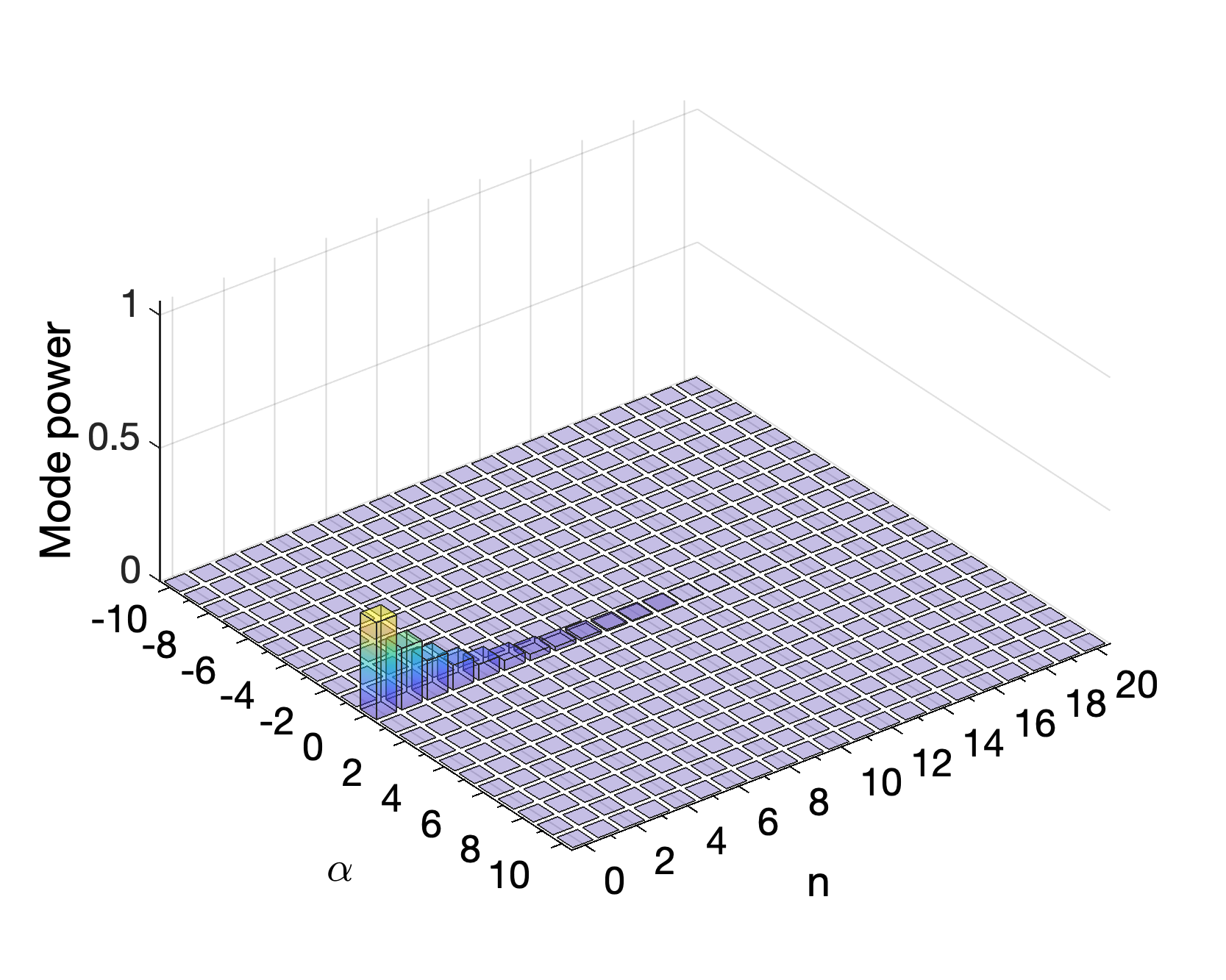}
		\caption{3 waves}
		\label{fig:Fieldcurvature_LGbasecase_decomp_3}
	\end{subfigure} %
\caption{The transverse cavity mode power distribution caused by one to five waves of maximum phase delay, shaped as field curvature.}
\label{fig:Fieldcurvature_LGbasecase_decomps}
\end{figure}

\FloatBarrier %

Since field curvature is circularly symmetric, with a phase delay dependent only upon $r$, we might expect its effects would eschew any modes which are similarly circularly asymmetric. Fig.~\ref{fig:Fieldcurvature_LGbasecase_decomps} confirms this suspicion. As increasing strengths of field curvature are applied, power diffuses upward along $n$ but remains only in modes where $\alpha=0$; the fundamental mode remaining the single most-energetic mode even as it loses power to an increasing number of others. Distortion, on the other hand, involves both $r$ and $phi$ in its phase delay expression, and thus the expansion takes place in both of the respective corresponding mode indices $n$ and $\alpha$. As the aberration strength increases, power is transferred from the source beam's fundamental mode into a "pulse" of additional modes which widens as its center moves upward in mode index $n$. 

Finally, we translate the cavity mode activity information to the experimentally observable transmission spectrum. Eq.~\ref{eqn:cavity_resonant_freqs} gives the resonant frequency of each transverse mode (degenerately shared with others of the same $l+m$); thus, the transmission at each resonant peak is equal to the sum of the intensities of all the cavity modes which share that frequency. We will display here only the peak intensity at each resonance and not concern ourselves with the full transmission curve, which---since we have expressly designed the cavity to separate the resonant peaks---should clearly display the results without materially affect our conclusions. Figures~\ref{fig:Distortion_LGbasecase_spectra}~and~\ref{fig:Fieldcurvature_LGbasecase_spectra} display the resulting transmission spectra for distortion and field curvature, respectively. 

\begin{figure}[h!]
	\begin{subfigure}[c]{0.5\textwidth}
		\centering
		\includegraphics[width=7cm]{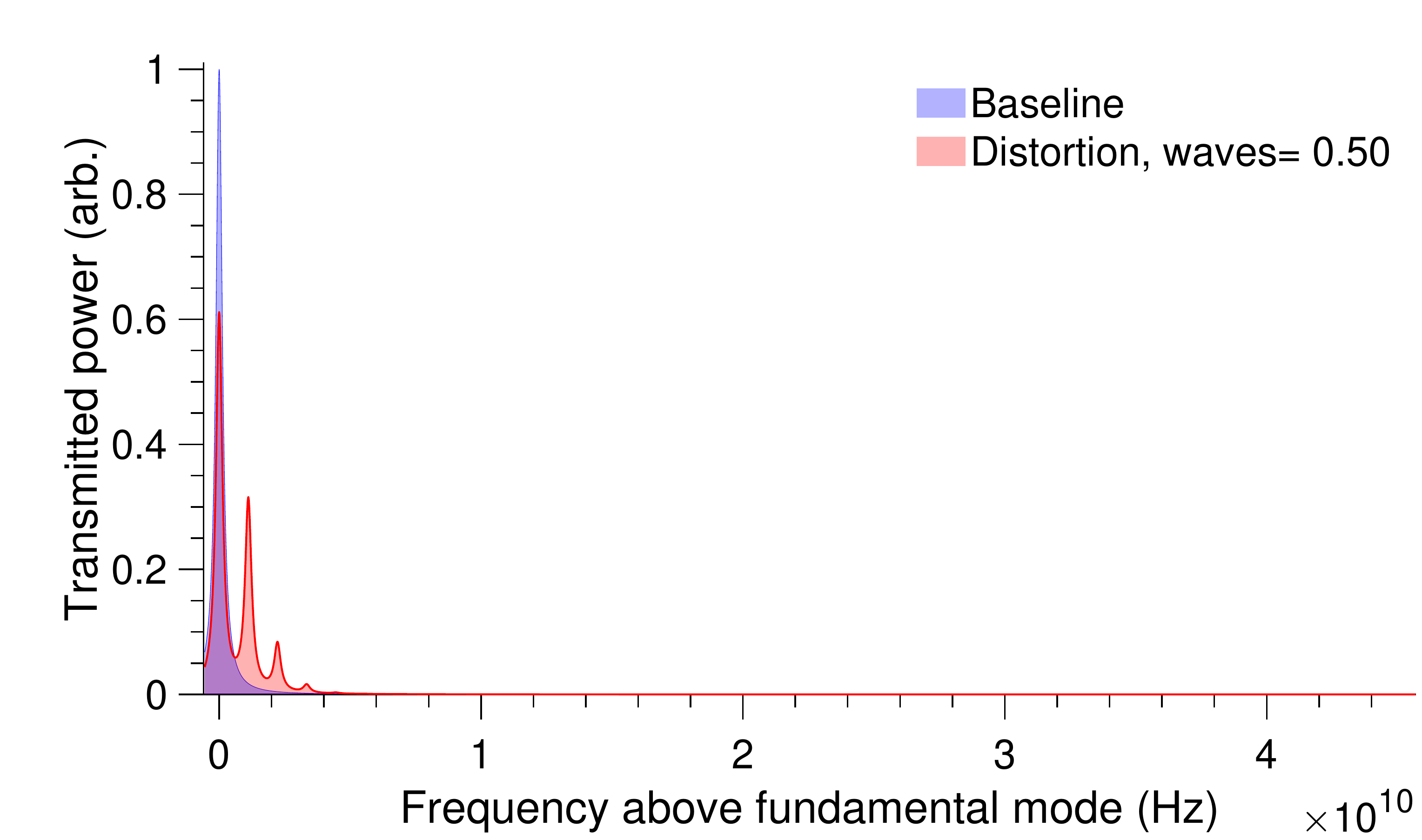}
		\caption{0.5 waves}
		\label{fig:Distortion_LGbasecase_spectrum_05}
	\end{subfigure} %
	\begin{subfigure}[c]{0.5\textwidth}
		\centering
		\includegraphics[width=7cm]{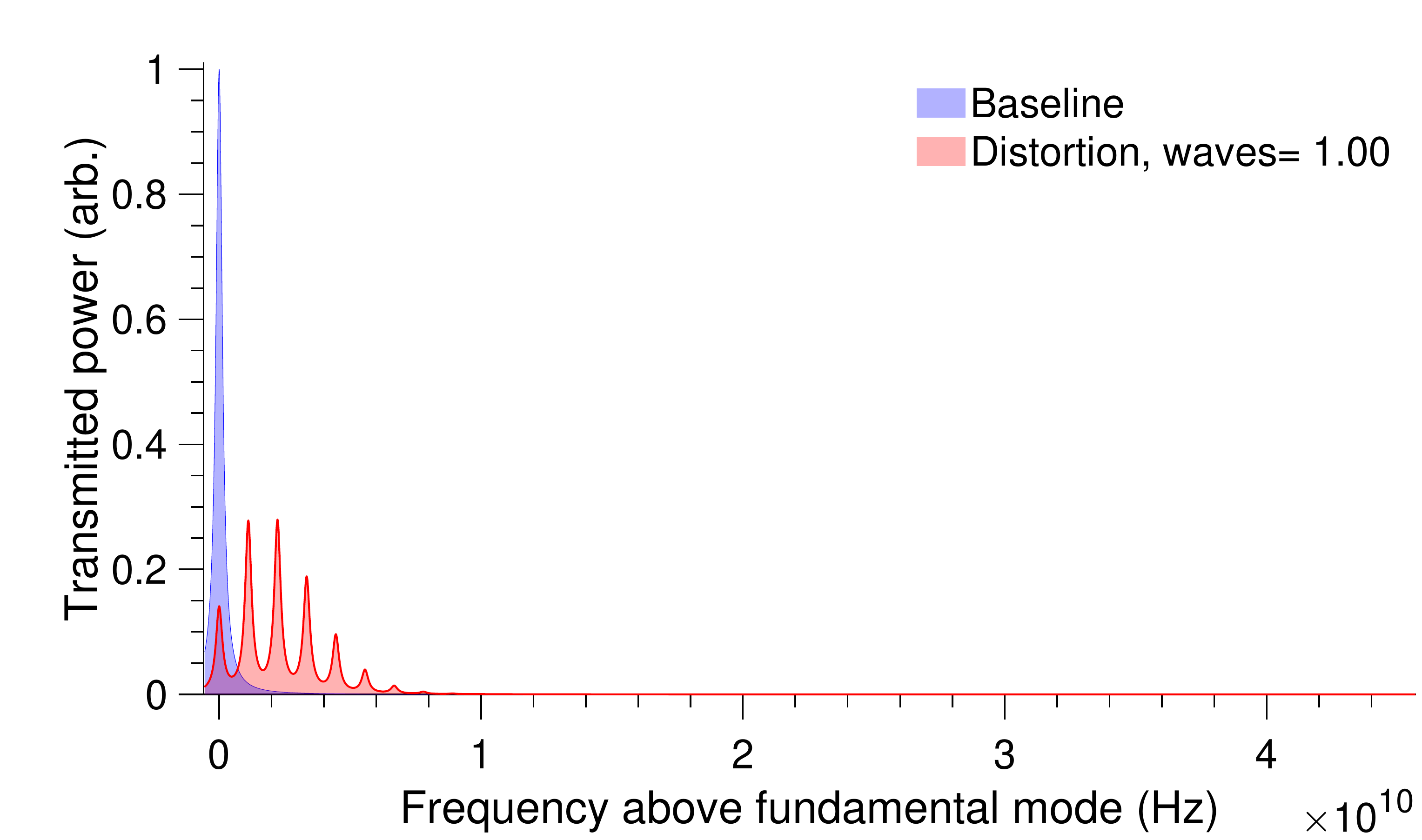}
		\caption{1 wave}
		\label{fig:Distortion_LGbasecase_spectrum_1}
	\end{subfigure} \\
	\par\bigskip %
	\begin{subfigure}[c]{0.5\textwidth}
		\centering
		\includegraphics[width=7cm]{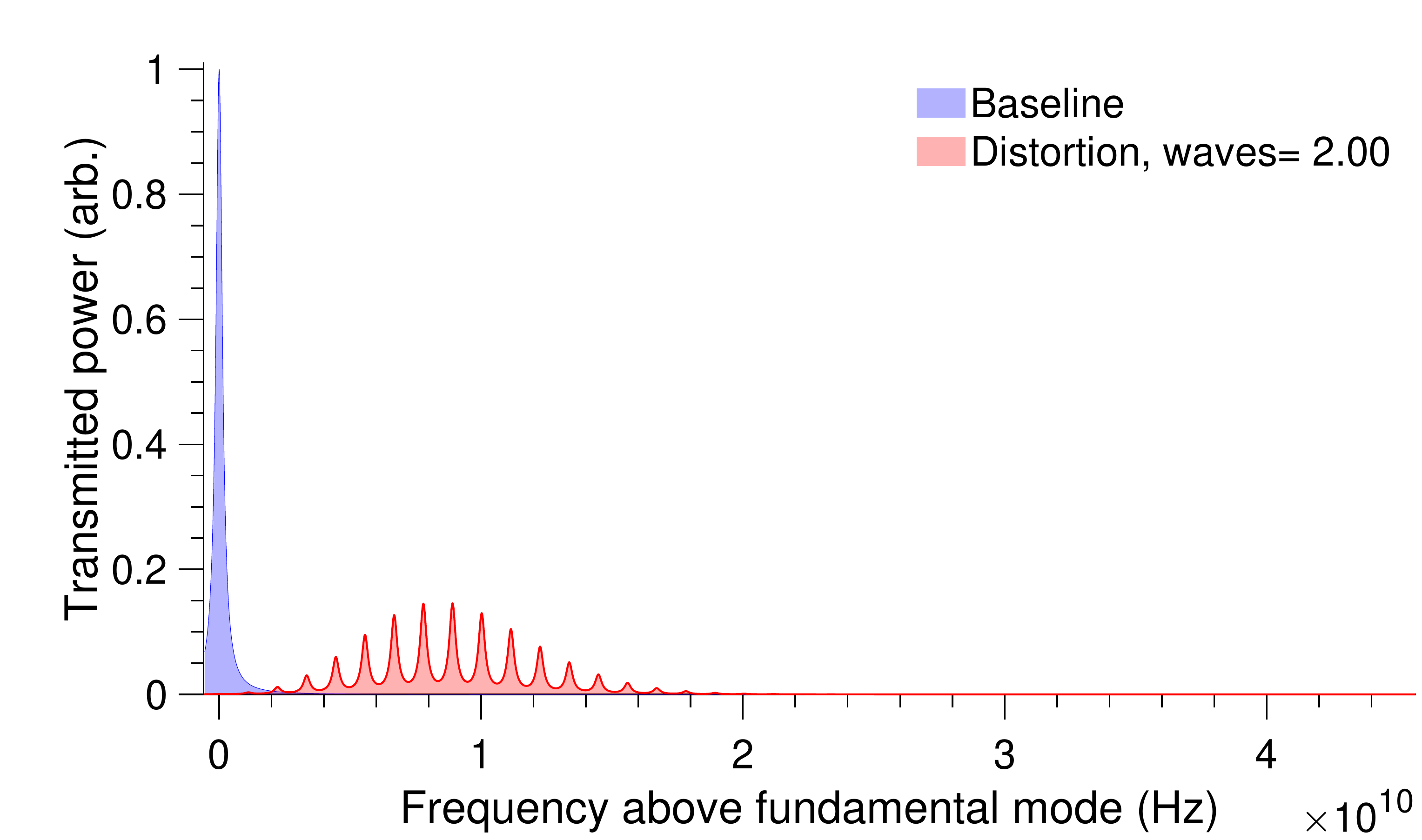}
		\caption{2 waves}
		\label{fig:Distortion_LGbasecase_spectrum_2}
	\end{subfigure} %
	\begin{subfigure}[c]{0.5\textwidth}
		\centering
		\includegraphics[width=7cm]{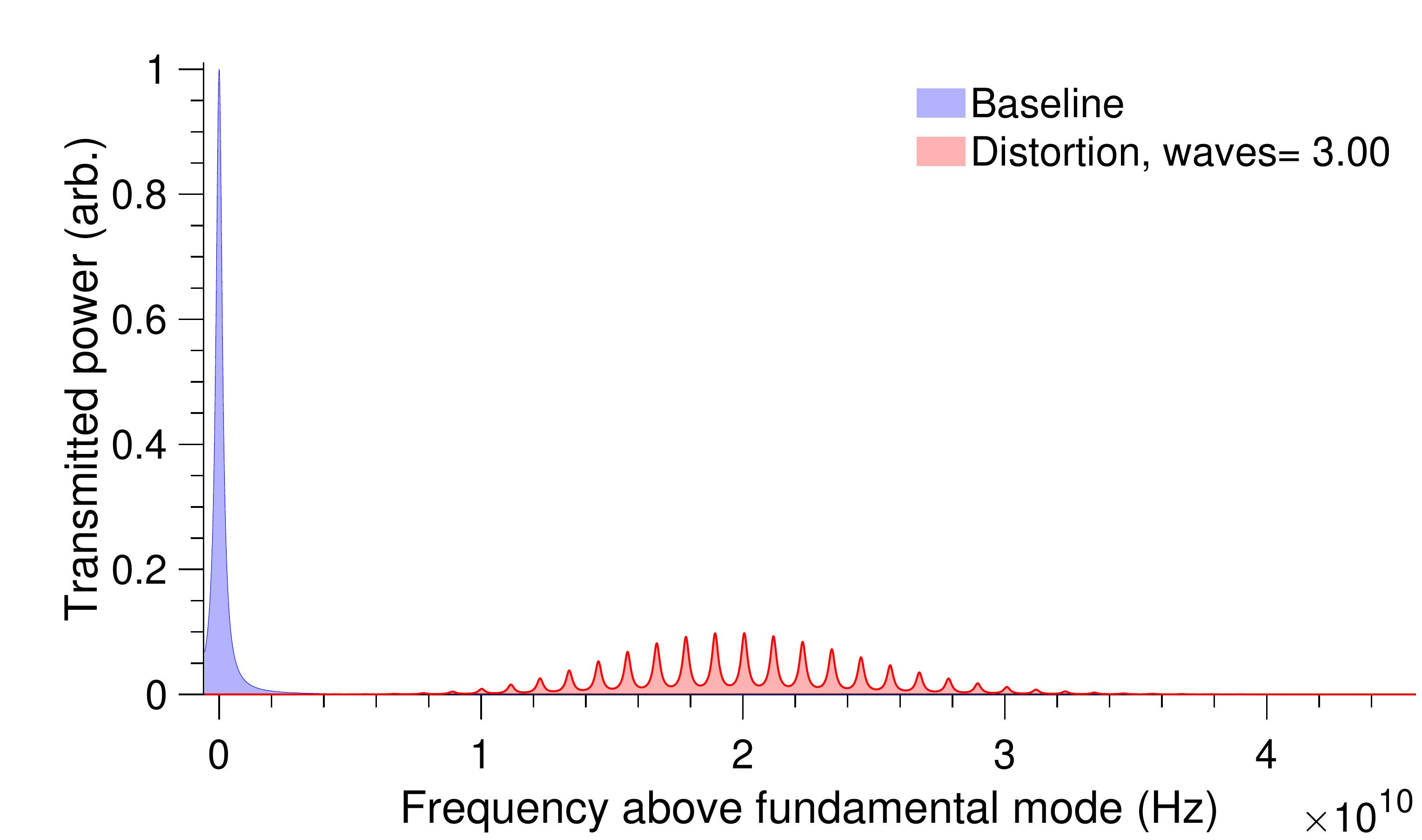}
		\caption{3 waves}
		\label{fig:Distortion_LGbasecase_spectrum_3}
	\end{subfigure} %
\caption{The intensity spectrum transmitted by the cavity as a result of transverse mode excitation by an input Laguerre-Gaussian, with and without varying amounts of distortion. Each spatial mode's transmission peak contribution is taken to have Lorentzian lineshape, approximating the typical response of dielectric mirrors.}
\label{fig:Distortion_LGbasecase_spectra}
\end{figure}

\begin{figure}[h!]
	\begin{subfigure}[c]{0.5\textwidth}
		\centering
		\includegraphics[width=7cm]{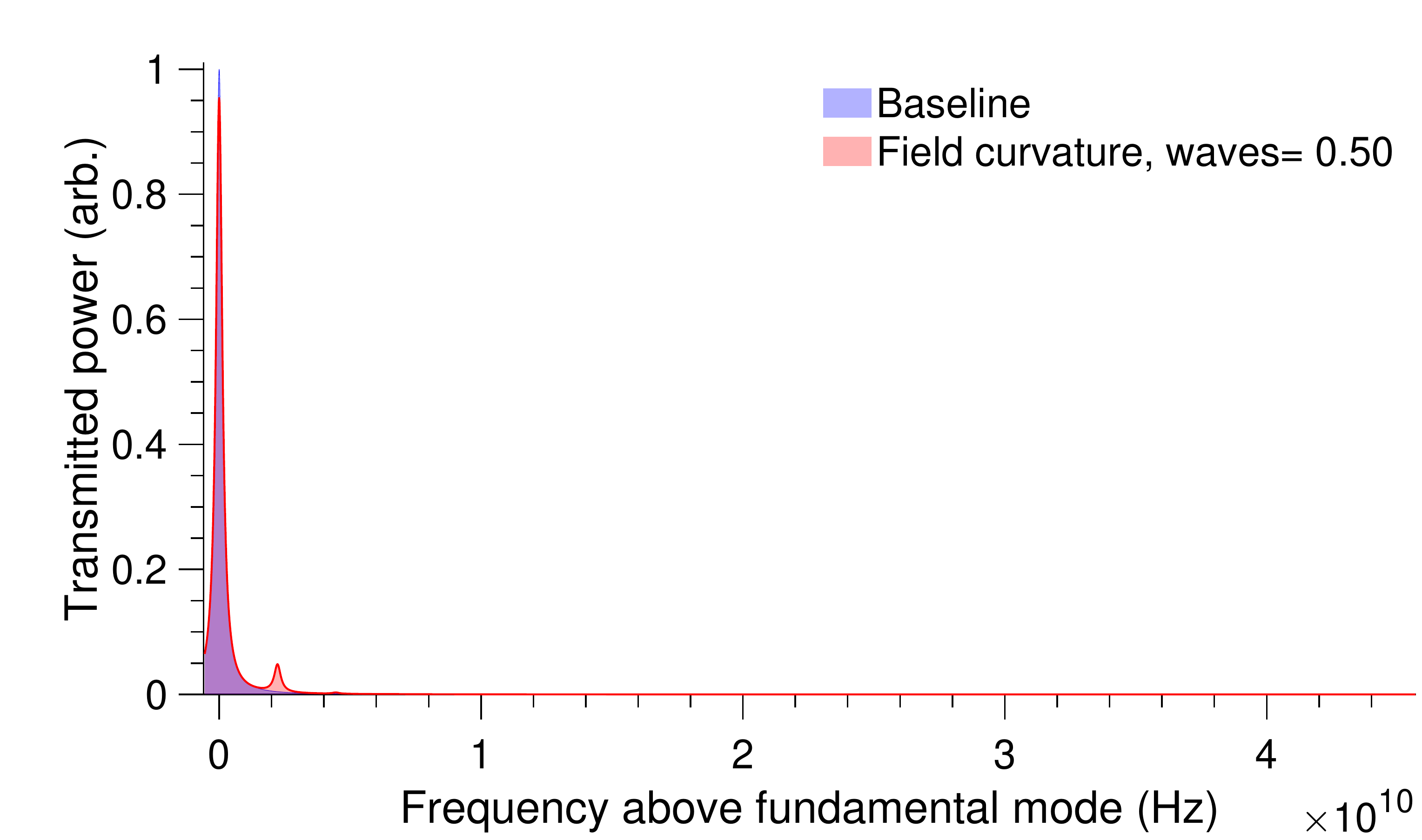}
		\caption{0.5 waves}
		\label{fig:Fieldcurvature_LGbasecase_spectrum_05}
	\end{subfigure} %
	\begin{subfigure}[c]{0.5\textwidth}
		\centering
		\includegraphics[width=7cm]{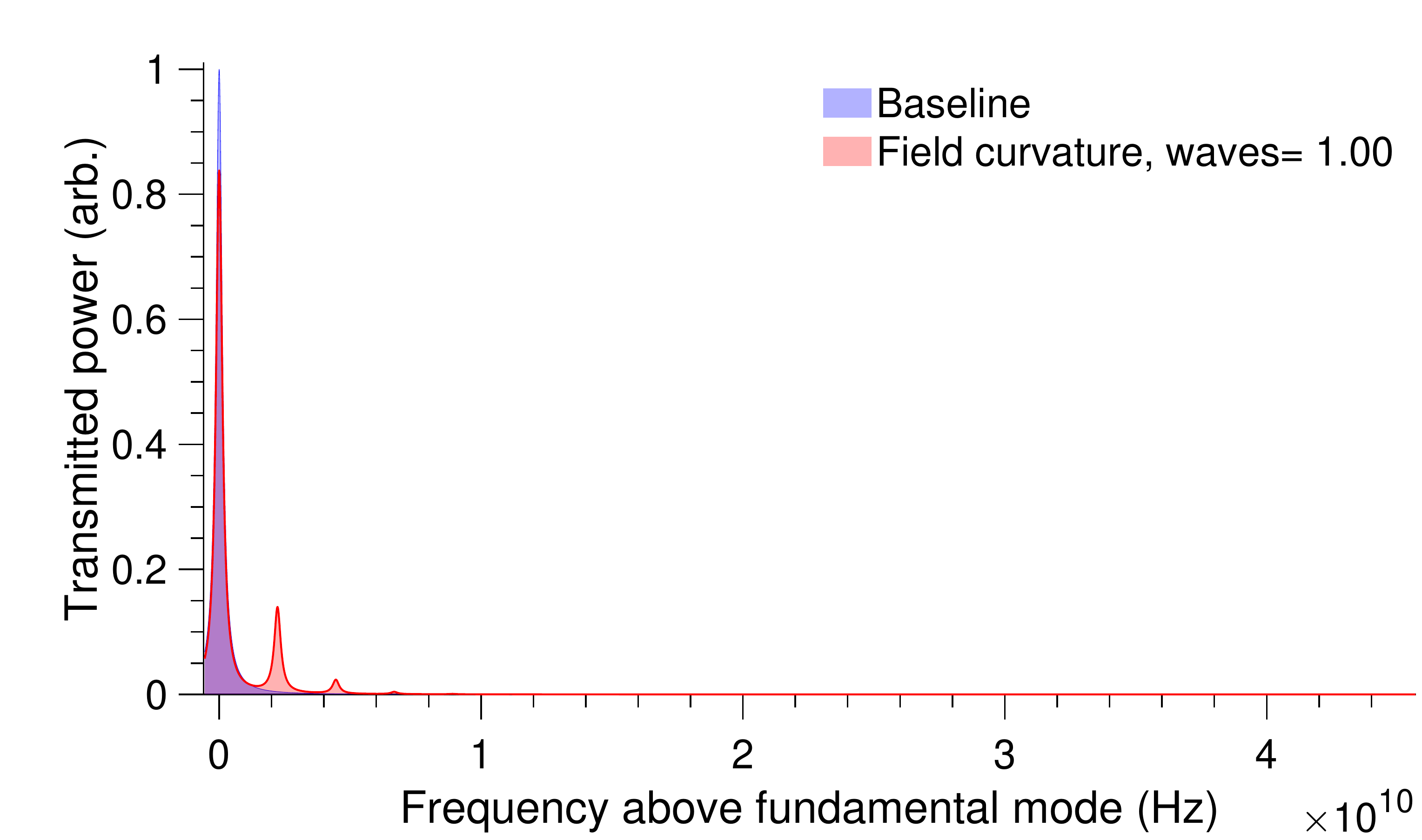}
		\caption{1 wave}
		\label{fig:Fieldcurvature_LGbasecase_spectrum_1}
	\end{subfigure} \\
	\par\bigskip %
	\begin{subfigure}[c]{0.5\textwidth}
		\centering
		\includegraphics[width=7cm]{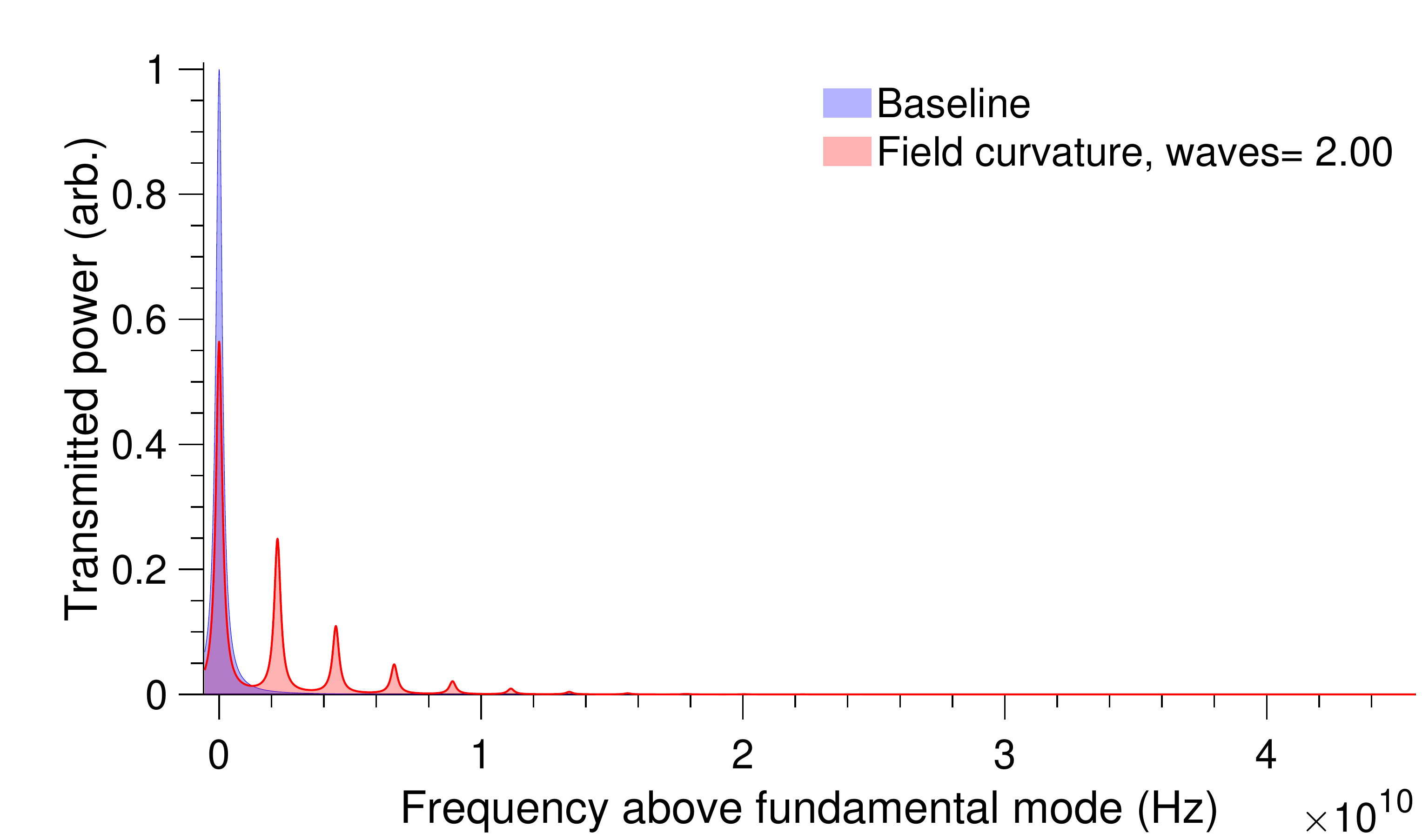}
		\caption{2 waves}
		\label{fig:Fieldcurvature_LGbasecase_spectrum_2}
	\end{subfigure} %
	\begin{subfigure}[c]{0.5\textwidth}
		\centering
		\includegraphics[width=7cm]{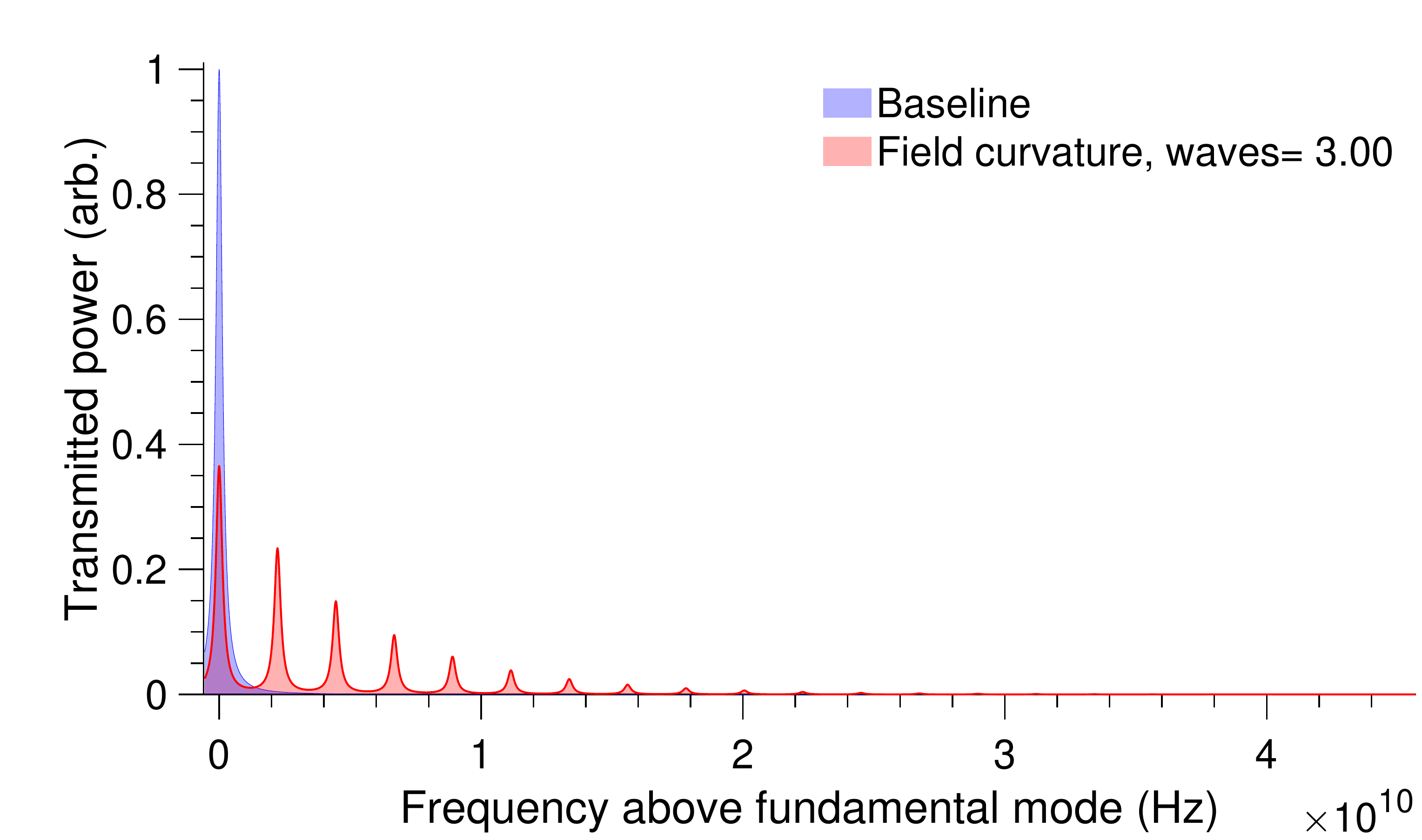}
		\caption{3 waves}
		\label{fig:Fieldcurvature_LGbasecase_spectrum_3}
	\end{subfigure} %
	\caption{The intensity spectrum transmitted by the cavity as a result of transverse mode excitation by an input Laguerre-Gaussian, with and without varying amounts of field curvature.}
	\label{fig:Fieldcurvature_LGbasecase_spectra}
\end{figure}

The transmission spectrum from the distortion aberration is simple to predict: since there is no activity except where mode index $m=0$, each resonance frequency directly corresponds to a value of $l$, and the spectrum reads exactly as the mode activity plot viewed from one side. The "pulse" of power is again clearly visible traveling upwards and spreading in frequency. Field curvature produces a markedly different spectrum, with power diffusing upwards in frequency from the fundamental mode and skipping every other cavity resonance.

Clearly, the cavity response to these two aberrations differs quite starkly, at least between these two Seidel aberrations. Figures~\ref{fig:Distortion_LGbasecase_spectrum_history}~and~\ref{fig:Fieldcurvature_LGbasecase_spectrum_history} display the evolution of the transmission spectra as the aberration maximum phase delay increases. 

\begin{figure}[h!]
	\begin{subfigure}[c]{0.5\textwidth}
		\centering
		\includegraphics[width=6.5cm]{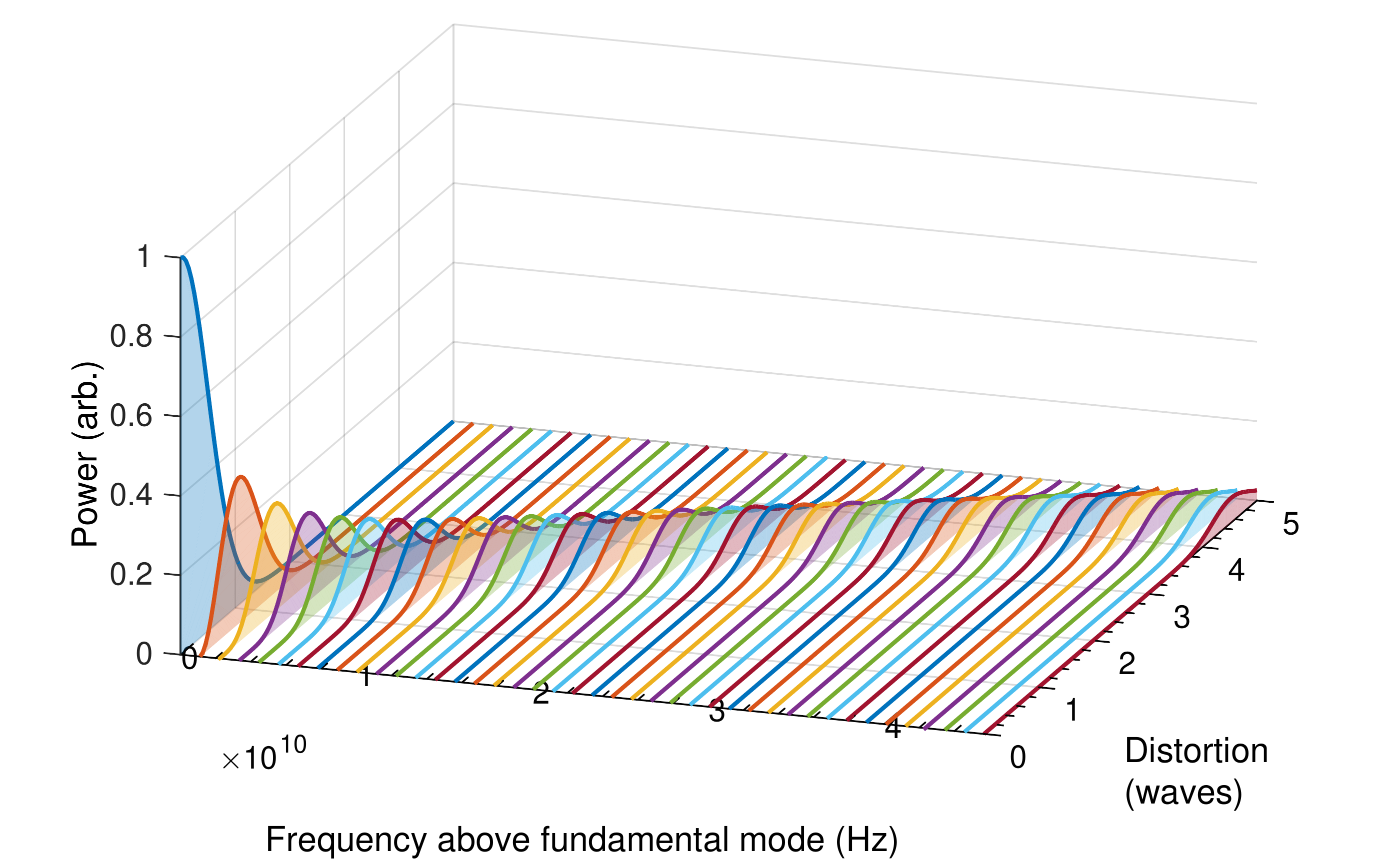}
		\caption{Distortion}
		\label{fig:Distortion_LGbasecase_spectrum_history}
	\end{subfigure} %
	\begin{subfigure}[c]{0.5\textwidth}
		\centering
		\includegraphics[width=6.5cm]{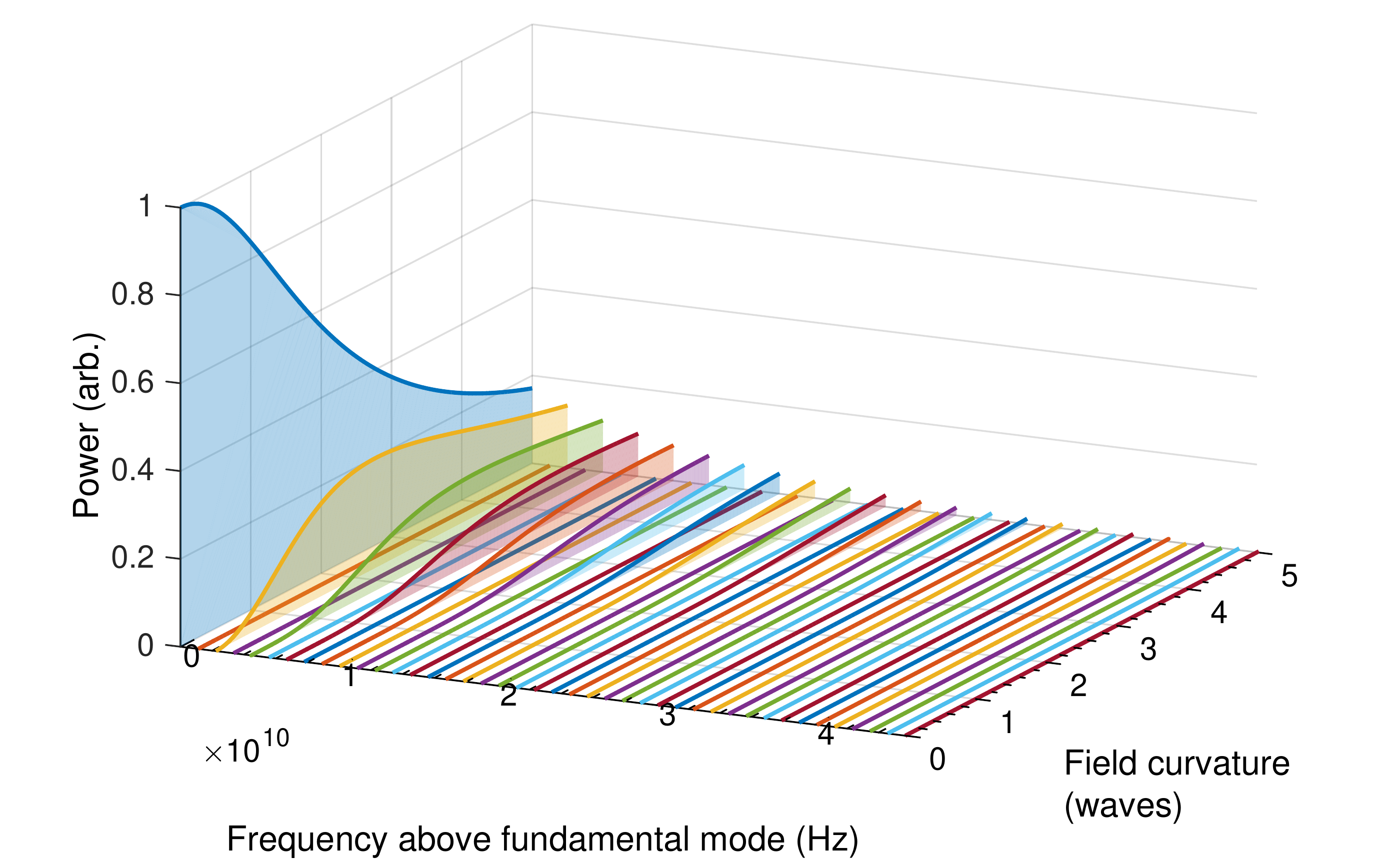}
		\caption{Field curvature}
		\label{fig:Fieldcurvature_LGbasecase_spectrum_history}
	\end{subfigure} %
\caption{The spectral intensities of optical cavity transmission versus the varying amounts of distortion (left) and field curvature (right) applied to a cavity-matched Laguerre-Gaussian beam.}
\label{fig:LGbasecase_spectrum_histories}
\end{figure}

The above data show full transmission spectrum, which appears promising for the identification of a specific single aberration or classes of aberrations. However, in many general turbulence applications, many aberrations are present all at once and the goal is to estimate a statistical characteristic of the turbulence, such as $C_n^2$. For these common scenarios, it may be more useful to measure a single metric of the overall cavity spectrum, such as the spectral power distribution shown in Fig.~\ref{fig:LGbasecase_spectral_singlemetrics}.

\begin{figure}[h!]
	\begin{subfigure}[c]{0.5\textwidth}
		\centering
		\includegraphics[width=7cm]{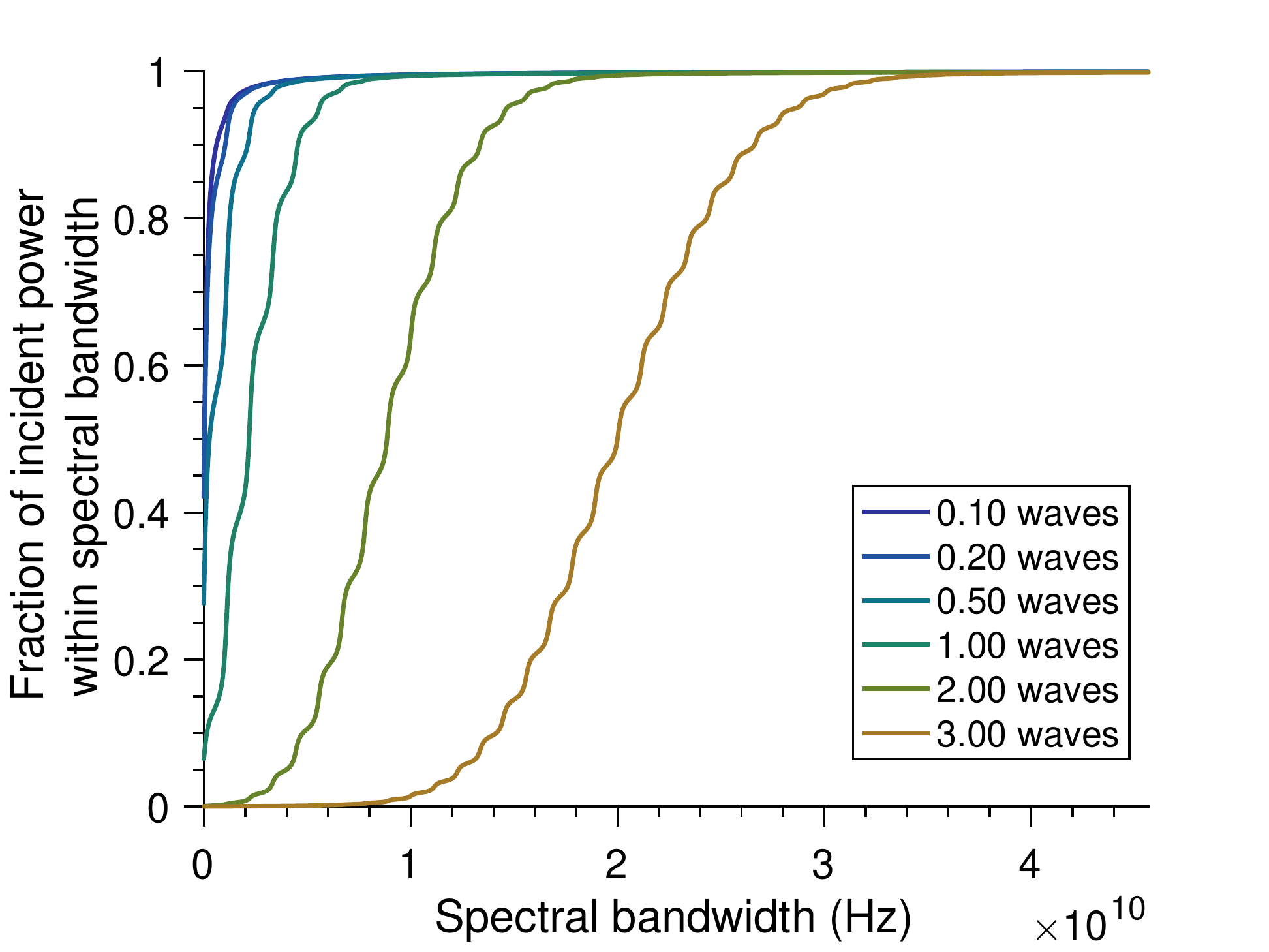}
		\caption{Distortion}
		\label{fig:Distortion_LGbasecase_spectral_power_distr}
	\end{subfigure} %
	\begin{subfigure}[c]{0.5\textwidth}
		\centering
		\includegraphics[width=7cm]{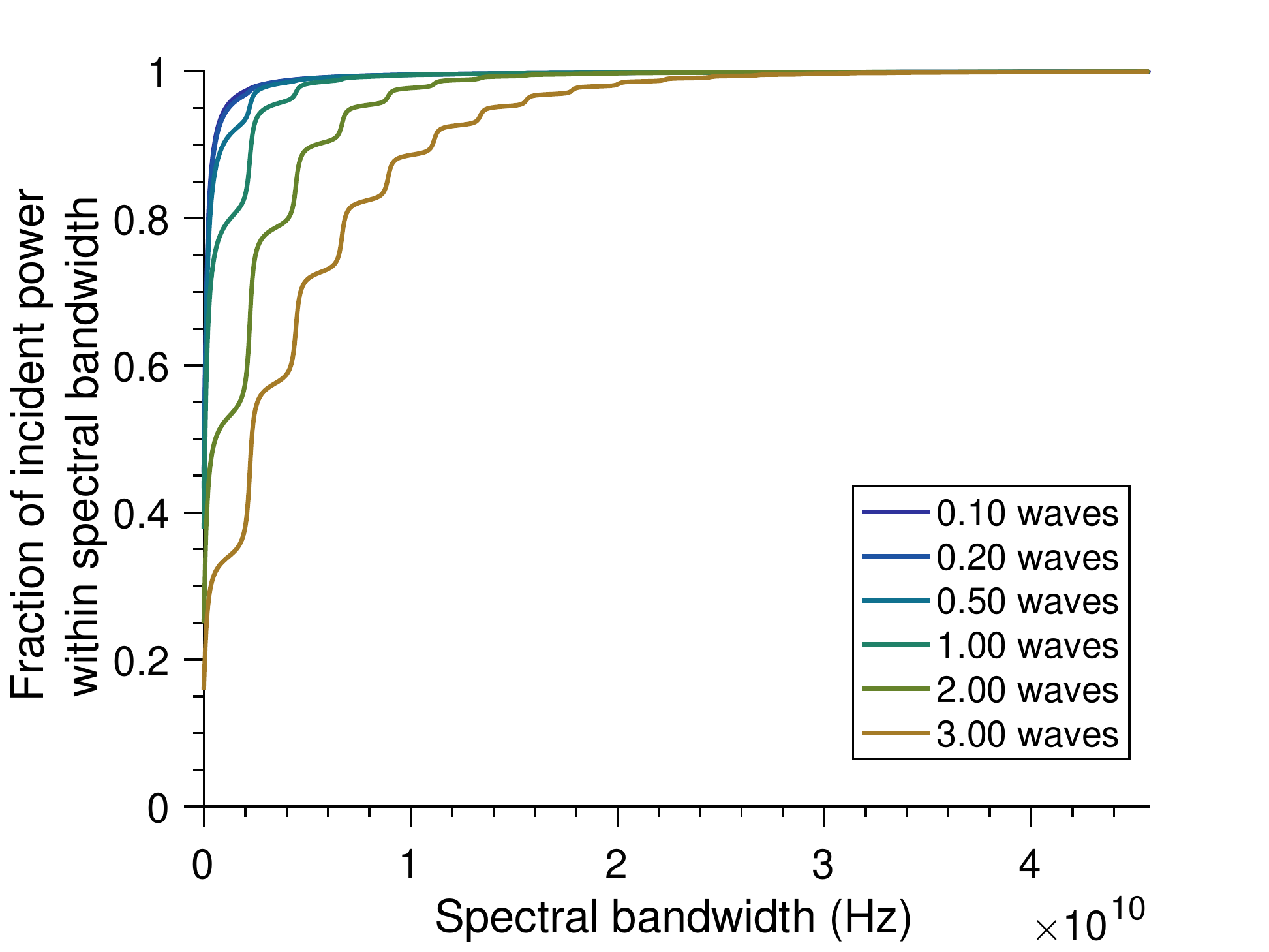}
		\caption{Field curvature}
		\label{fig:Fieldcurvature_LGbasecase_spectral_power_distr}
	\end{subfigure} %
\caption{The power distribution within the cavity transmission spectrum resulting from various strengths of distortion (left) and field curvature (right). The intensity of light which passes within a certain spectral bandwidth above the fundamental, as a portion of the beam power entering the cavity, is plotted for a few selected aberration strengths. In the case of the largest strengths of distortion, power begins to shift beyond our sampled mode range, and thus the lower lines may not reach 1 on the vertical axis within the bandwidth range plotted.} 
\label{fig:LGbasecase_spectral_singlemetrics}
\end{figure}

As we observed in Figs.~\ref{fig:Distortion_LGbasecase_spectra}~and~\ref{fig:Fieldcurvature_LGbasecase_spectra}, the concentration of power breaks quickly away from the fundamental mode under distortion, whereas with field curvature the power gradient becomes flatter but remains anchored at the fundamental mode. We can see this signature in the spectral power distribution plots: distortion's Fig.~\ref{fig:Distortion_LGbasecase_spectral_power_distr} quickly forms a gap containing no power next to the vertical axis, but Fig.~\ref{fig:Fieldcurvature_LGbasecase_spectral_power_distr}'s lines remain more or less adhered to the fundamental mode. For any aberration strength above 1 wave, we may confidently state that we can distinguish distortion from field curvature---and estimate its strength---using this single metric. 

\FloatBarrier %

\section{Concept Demonstration - Distant Source}

The single-mode Laguerre-Gaussian source beam assumed above has the advantage of clearly isolating the effects of aberrations, but it is not likely to be replicated outside of a laboratory setting. For example, a sensor for atmospheric turbulence may deal with light that has propagated over a long distance. The small patch of light sampled would resemble a plane wave regardless of the original source beam. Collection optics would likely be used to gather a slightly larger amount of light and funnel it to the cavity for analysis. 

To better fit this and similar scenarios, let us replace our original Laguerre-Gaussian beam with a plane wave. We further assume that a diffraction-limited optical system is in use to enhance the light-collection area of the cavity, and it has a sufficiently large exit pupil that frequency apodization can be neglected. The input to the cavity can then be taken as the aberrated plane wave, bounded by the exit pupil function and arbitrarily magnified. We set, for convenience, a circular pupil function and a magnification which scales the now-circular field outline to approximately the size of the Gaussian cavity mode. Since our source beam is no longer a perfect match for a cavity mode, we should expect a more complicated mode structure and spectrum even without applied aberrations; the cavity mode decomposition and transmission spectrum plotted in Fig.~\ref{fig:Planewave_baselines} confirm these predictions.

\begin{figure}[!ht]
	\savebox{\largestimage}{\includegraphics[width=6.5cm]{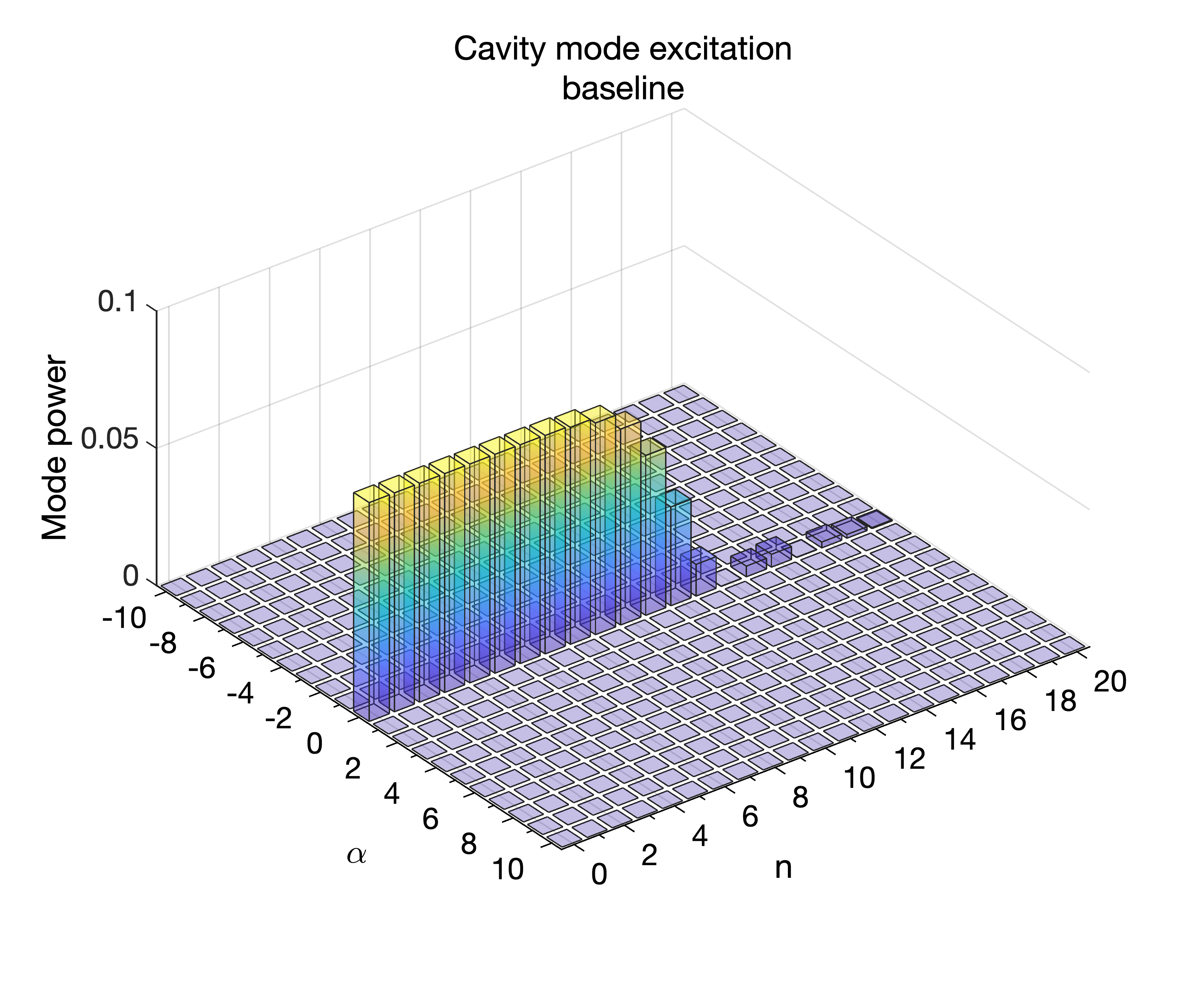}} %
	\begin{subfigure}[b]{0.5\textwidth}
		\centering
		\usebox{\largestimage}
		\caption{Mode decomposition} 
		\label{fig:Planewave_baseline_decomp}
	\end{subfigure} %
	\begin{subfigure}[b]{0.5\textwidth}
		\centering
		 \raisebox{\dimexpr.5\ht\largestimage-.5\height}{\includegraphics[width=7.5cm]{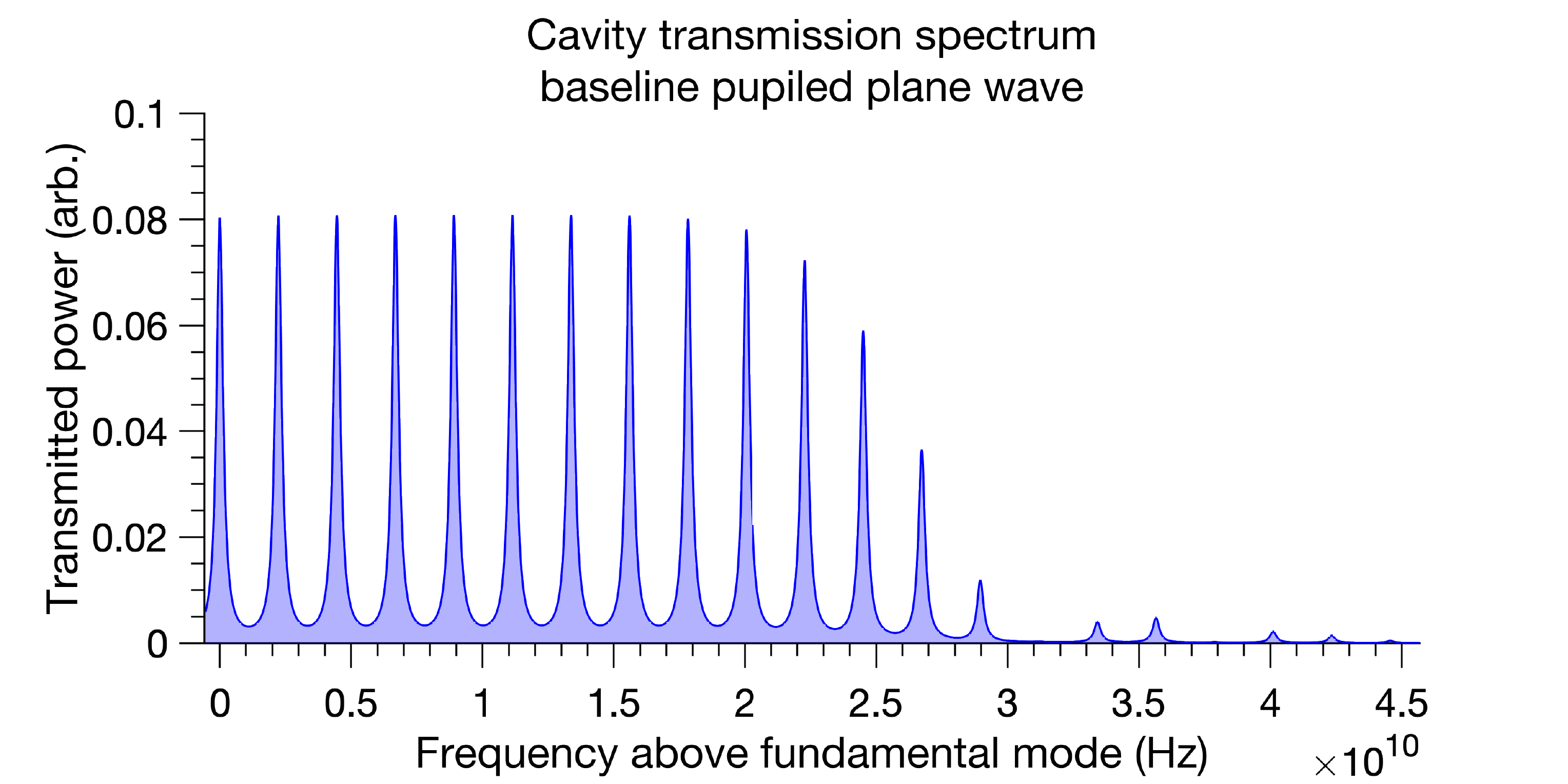}}
		\caption{Transmission spectrum} 
		\label{fig:Planewave_baseline_spectrum}
	\end{subfigure} %
	\caption{The cavity mode decomposition (left) and resulting transmission spectrum (right) for a plane wave with a flat phasefront, bounded by a circular pupil with negligible diffraction effects.}
	\label{fig:Planewave_baselines}
\end{figure}

\begin{figure}[h!]
	\begin{subfigure}[c]{0.5\textwidth}
		\centering
		\includegraphics[width=6.5cm]{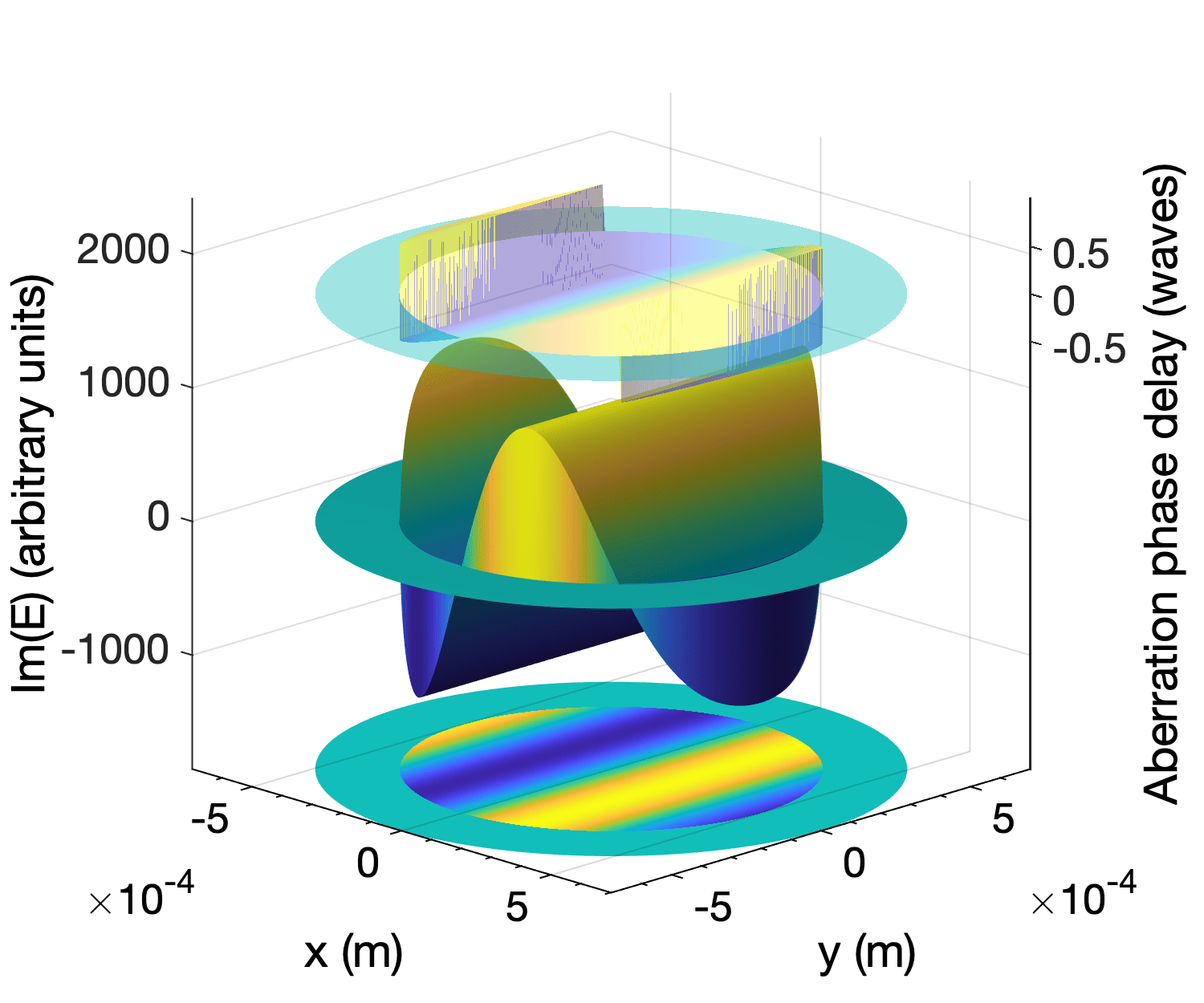}
		\caption{0.5 waves}
		\label{fig:Distortion_planewave_fields_05}
	\end{subfigure}%
	\begin{subfigure}[c]{0.5\textwidth}
		\centering
		\includegraphics[width=6.5cm]{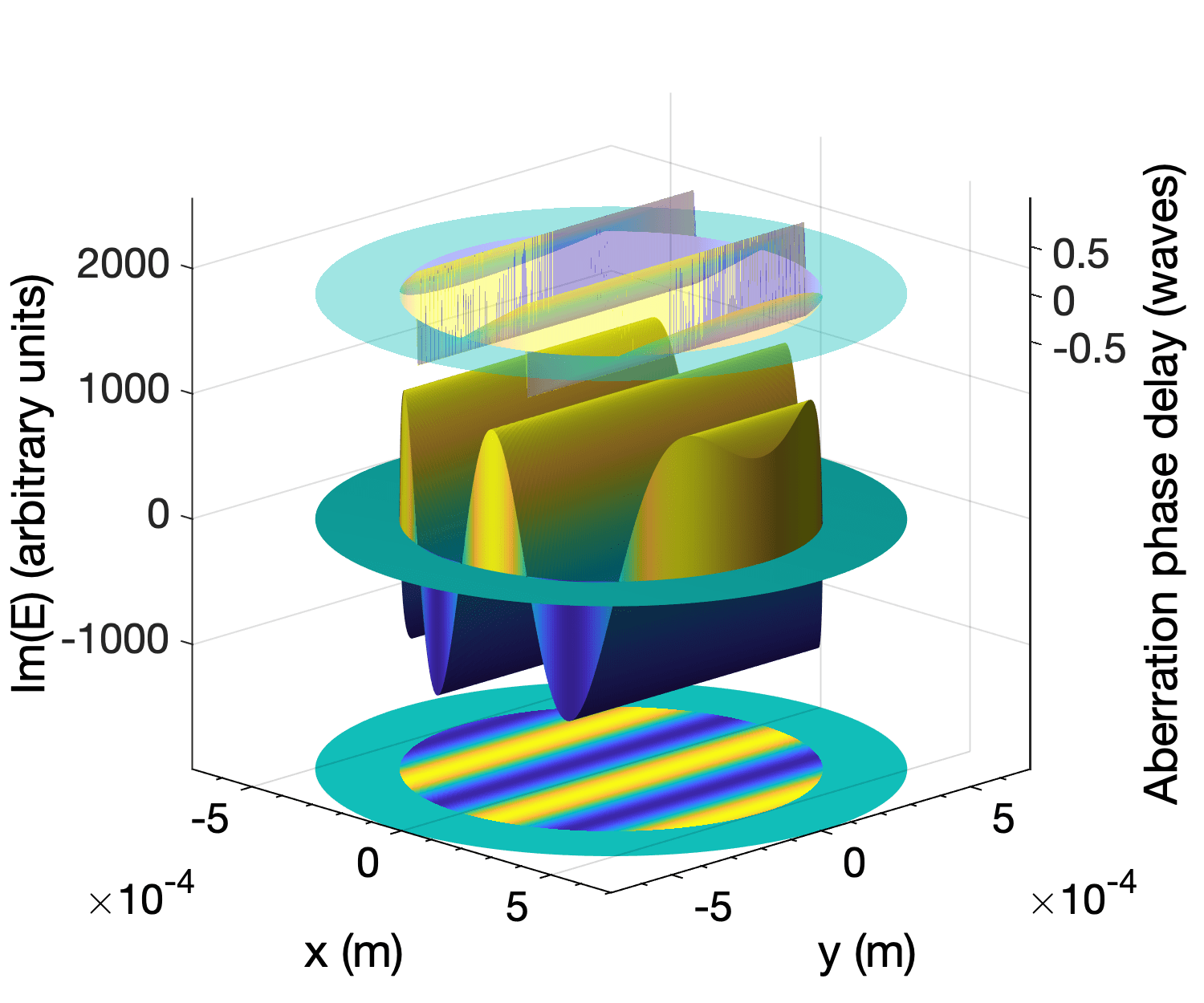}
		\caption{1 wave}
		\label{fig:Distortion_planewave_fields_1}
	\end{subfigure} \\
	\par\bigskip %
	\begin{subfigure}[c]{0.5\textwidth}
		\centering
		\includegraphics[width=6.5cm]{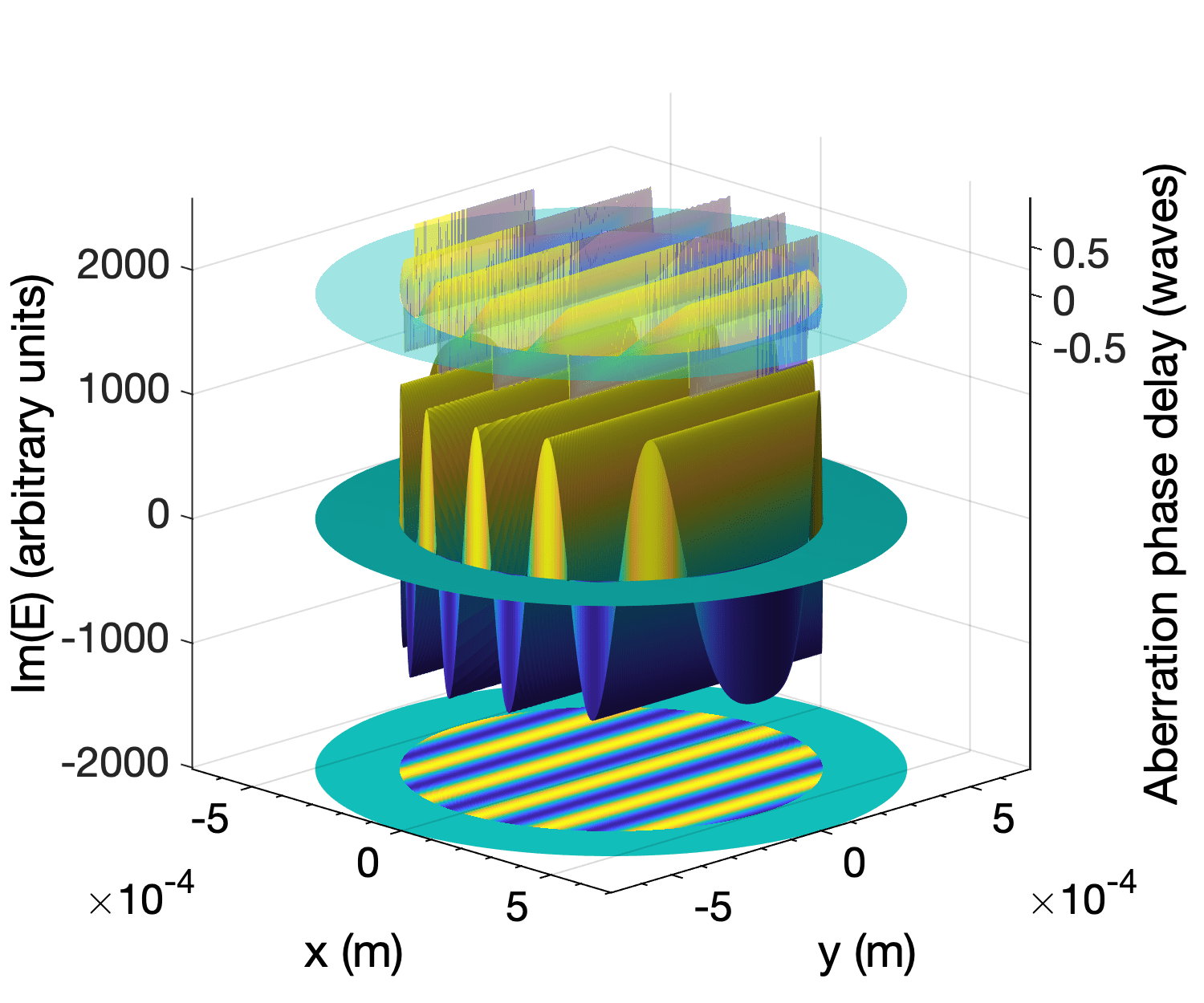}
		\caption{2 waves}
		\label{fig:Distortion_planewave_fields_2}
	\end{subfigure}%
	\begin{subfigure}[c]{0.5\textwidth}
		\centering
		\includegraphics[width=6.5cm]{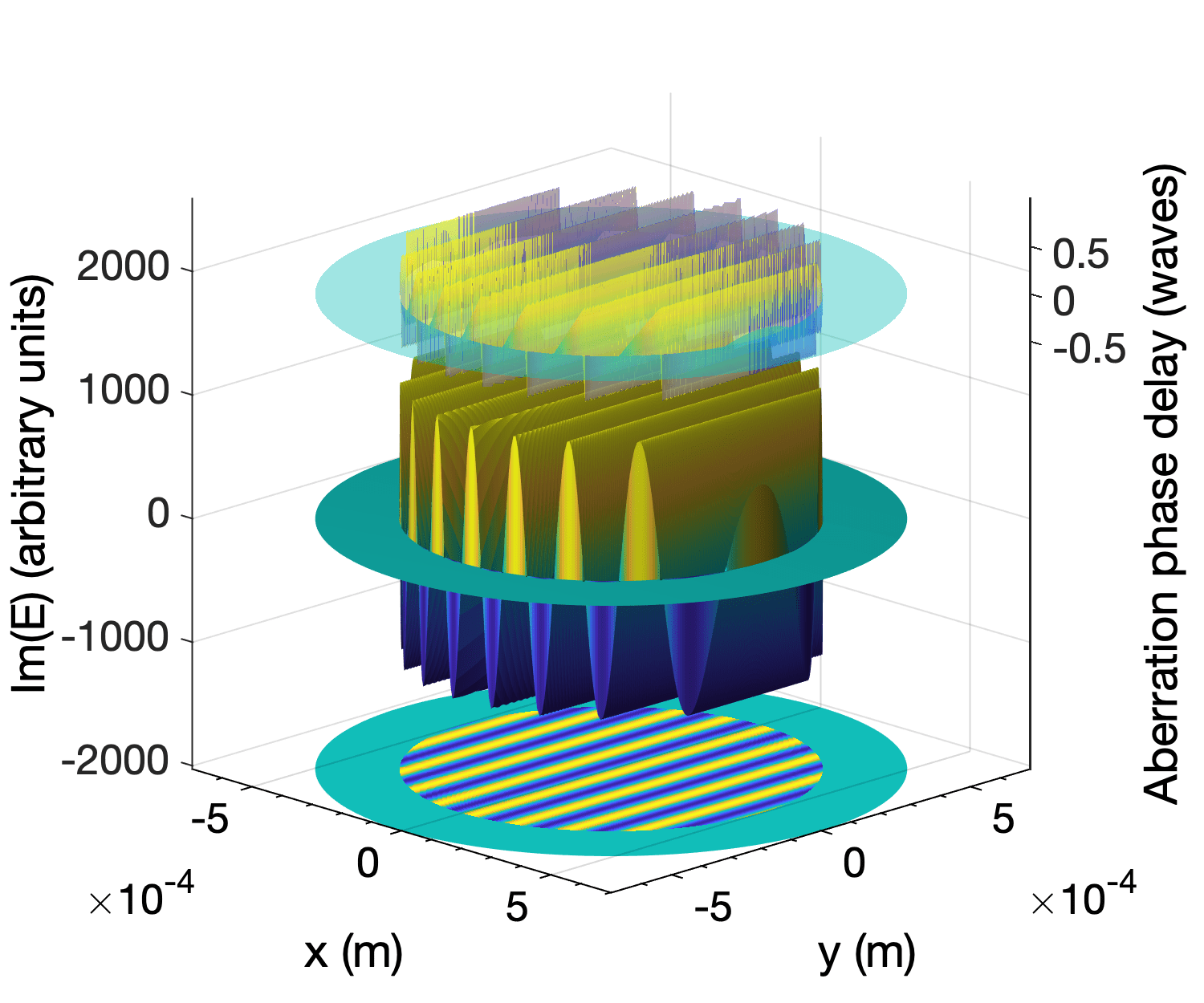}
		\caption{3 waves}
		\label{fig:Distortion_planewave_fields_3}
	\end{subfigure} 
\caption{The effects of varying strengths of distortion on a plane wave, shown with the same three components in each plot as in the earlier Fig.~\ref{fig:Distortion_LGbasecase_fields}~and~\ref{fig:Fieldcurvature_LGbasecase_fields}. With the Laguerre-Gaussian source beam swapped for a plane wave, the field incident upon the cavity now conforms quite clearly to the applied phase delay.}
\label{fig:Distortion_planewave_fields}
\vspace{-2mm}
\end{figure}

\begin{figure}[h!]
	\begin{subfigure}[c]{0.5\textwidth}
		\centering
		\includegraphics[width=6.5cm]{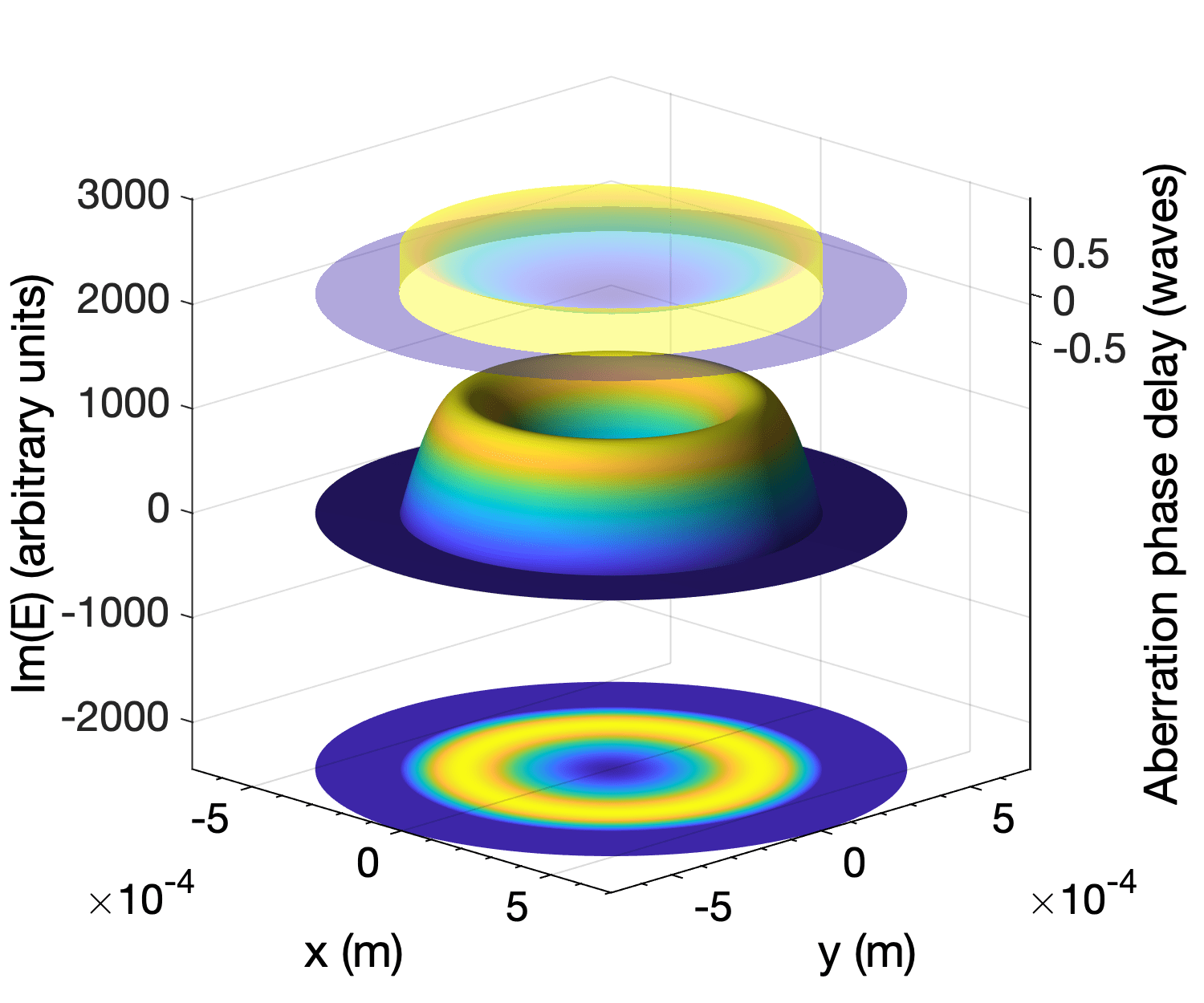}
		\caption{0.5 waves}
		\label{fig:Fieldcurvature_planewave_fields_05}
	\end{subfigure} %
	\begin{subfigure}[c]{0.5\textwidth}
		\centering
		\includegraphics[width=6.5cm]{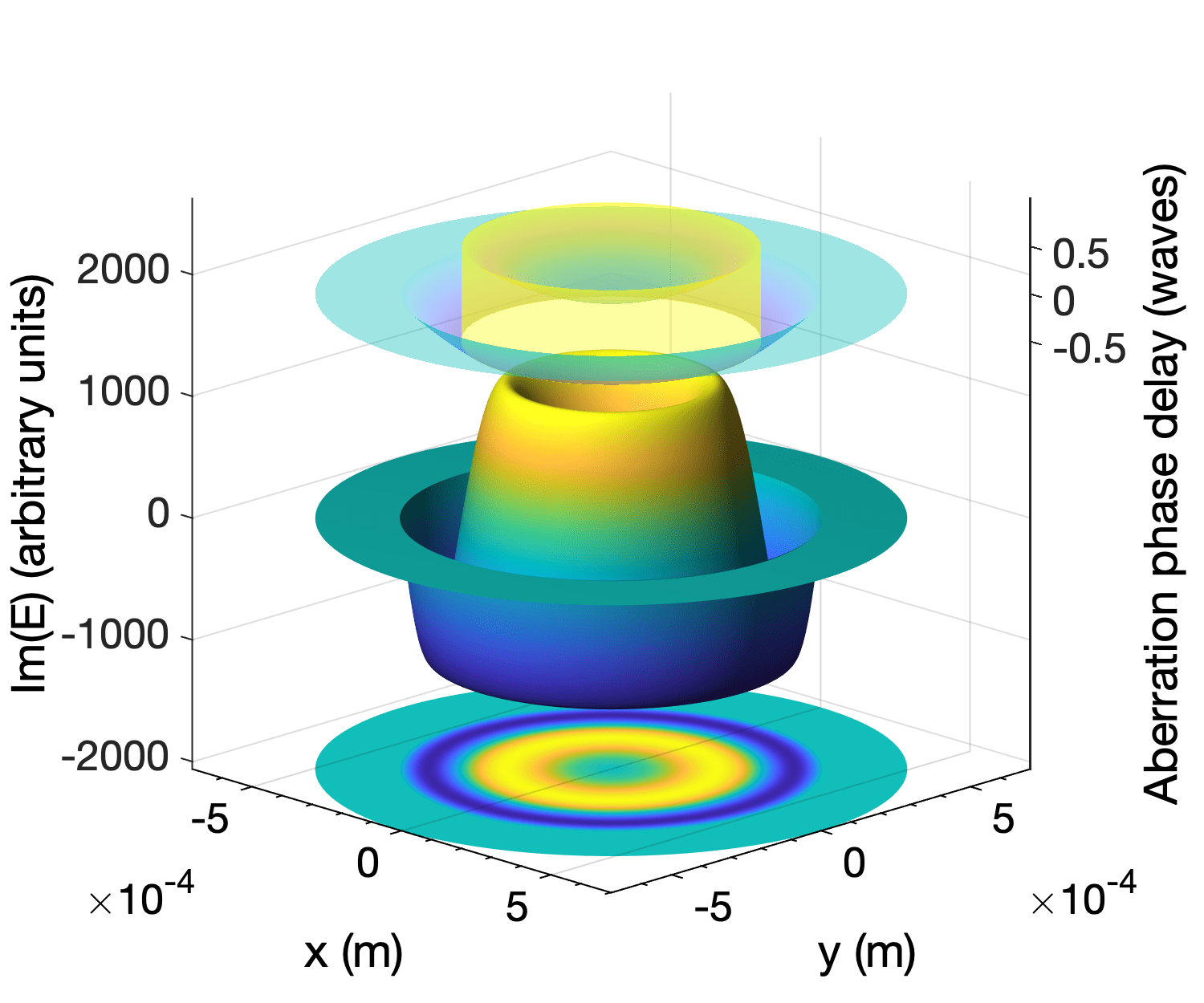}
		\caption{1 wave}
		\label{fig:Fieldcurvature_planewave_fields_1}
	\end{subfigure} \\
	\par\bigskip %
	\begin{subfigure}[c]{0.5\textwidth}
		\centering
		\includegraphics[width=6.5cm]{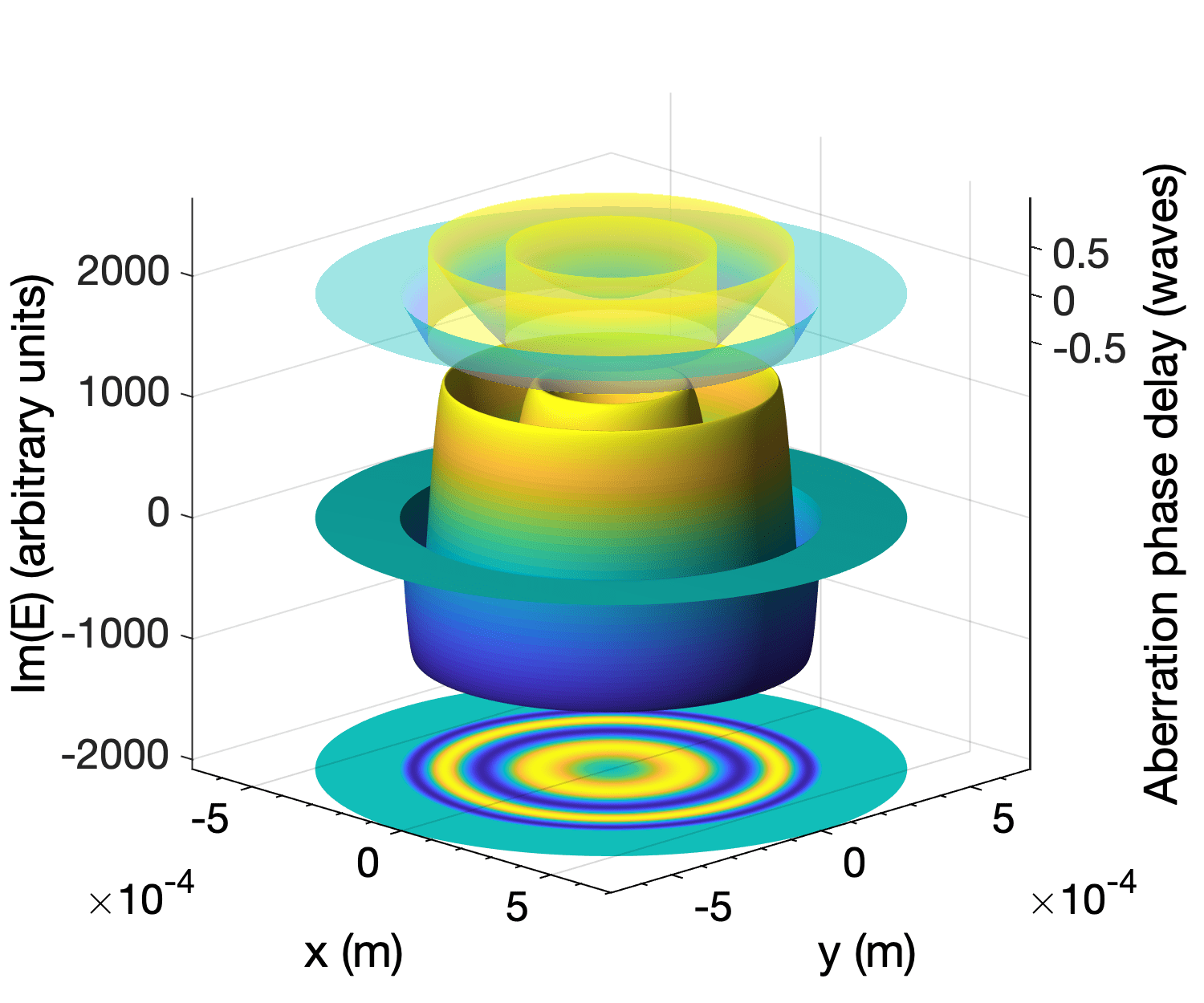}
		\caption{2 waves}
		\label{fig:Fieldcurvature_planewave_fields_2}
	\end{subfigure} %
	\begin{subfigure}[c]{0.5\textwidth}
		\centering
		\includegraphics[width=6.5cm]{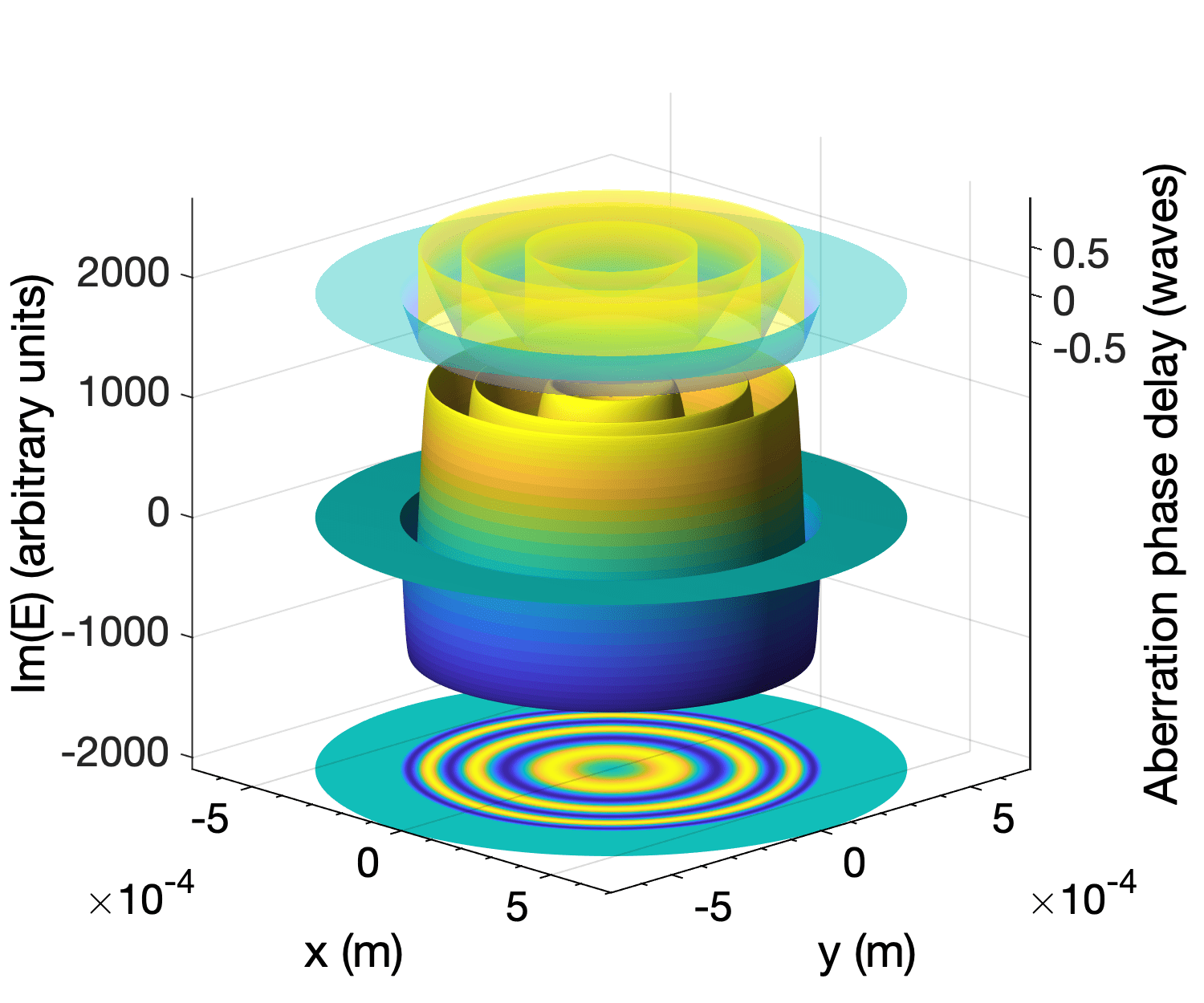}
		\caption{3 waves}
		\label{fig:Fieldcurvature_planewave_fields_3}
	\end{subfigure} %
\caption{The counterparts of the plots in Fig.~\ref{fig:Distortion_planewave_fields} with field curvature applied instead of distortion.}
\label{fig:Fieldcurvature_planewave_fields}
\vspace{-2mm}
\end{figure}

\FloatBarrier %

The cavity mode decompositions for the aberrated plane waves of Fig.~\ref{fig:Distortion_planewave_fields}~and~\ref{fig:Fieldcurvature_planewave_fields} are shown in Fig.~\ref{fig:Distortion_planewave_decomps}~and~\ref{fig:Fieldcurvature_planewave_decomps} below. Mode activity for distortion spreads quickly, mostly along the $\alpha$ index, as the aberration strength increases; while the pattern does not travel upward in $n$, as it did with a Laguerre-Gaussian beam, it is notably still symmetric about $\alpha=0$. Field curvature displays no inclination to excite modes other than those with $\alpha=0$, consistent with its depiction in Fig.~\ref{fig:Fieldcurvature_LGbasecase_decomps}, which is logical given the circular symmetry of the aberration. However, the single highest-power mode is no longer guaranteed to have $n=0$, a departure from its Laguerre-Gaussian behavior.

\begin{figure}[!ht]
	\begin{subfigure}[c]{0.5\textwidth}
		\centering
		\includegraphics[width=6.5cm]{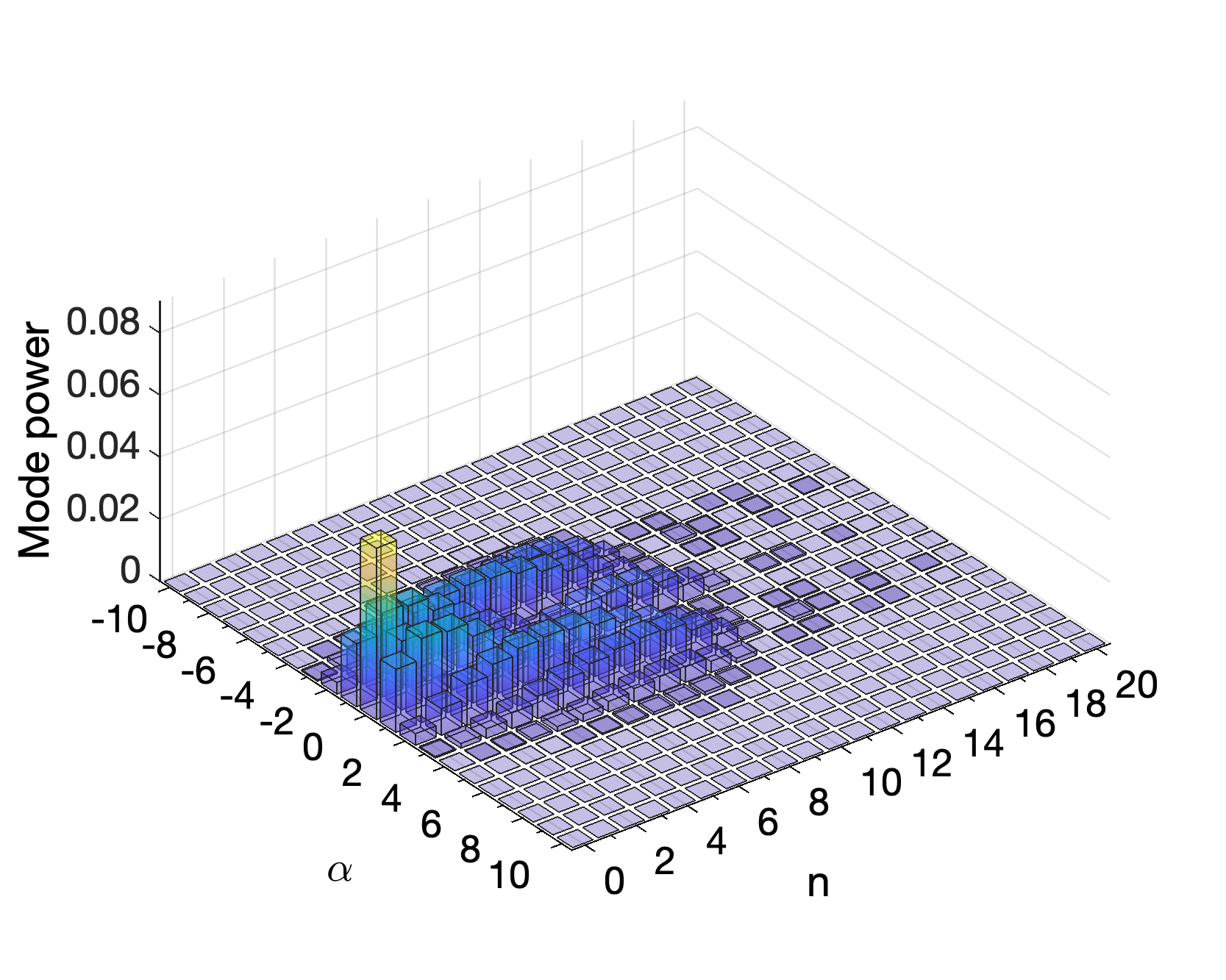}
		\caption{0.5 waves}
		\label{fig:Distortion_planewave_decomp_05}
	\end{subfigure} %
	\begin{subfigure}[c]{0.5\textwidth}
		\centering
		\includegraphics[width=6.5cm]{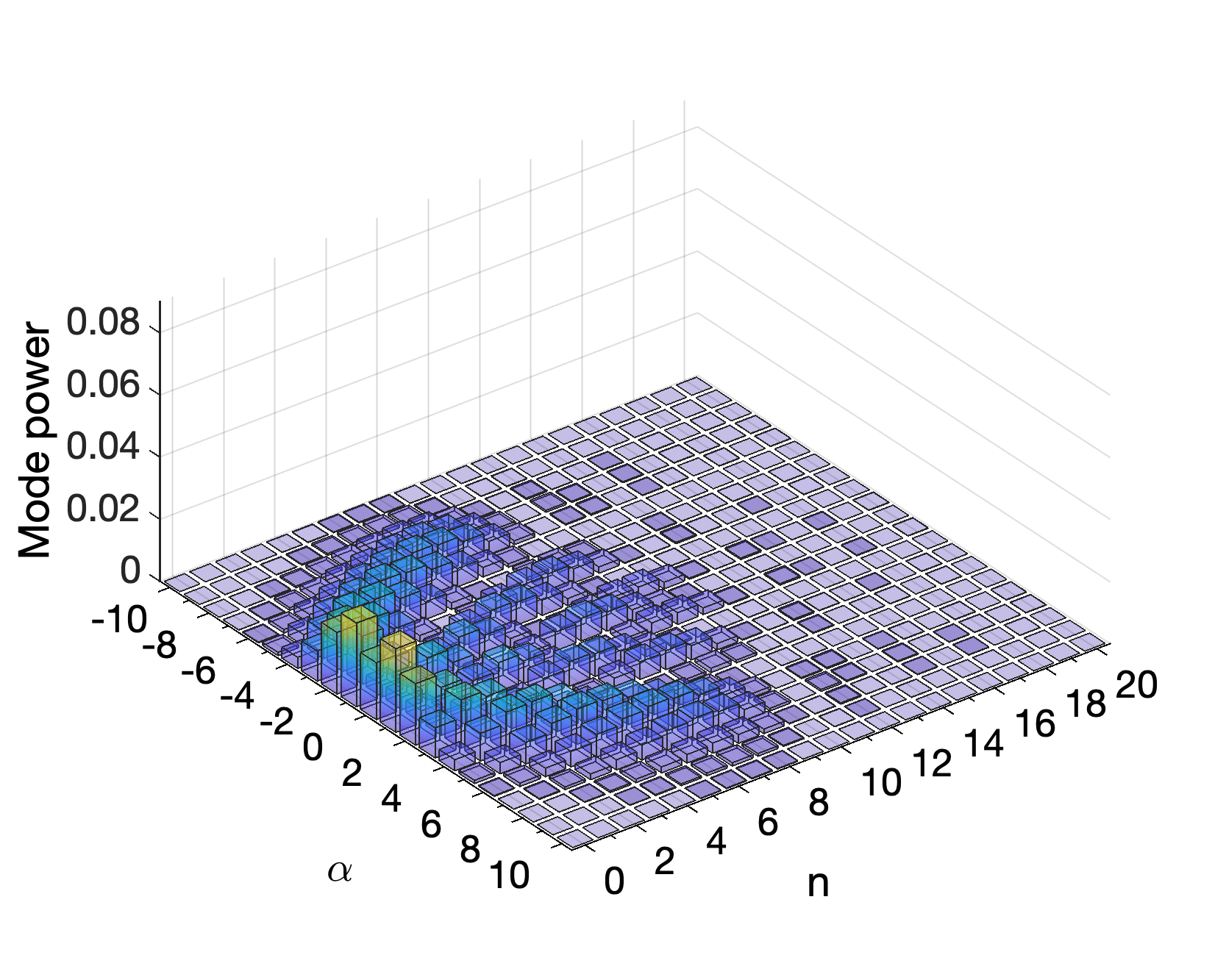}
		\caption{1 wave}
		\label{fig:Distortion_planewave_decomp_1}
	\end{subfigure} \\
	\par\bigskip %
	\begin{subfigure}[c]{0.5\textwidth}
		\centering
		\includegraphics[width=6.5cm]{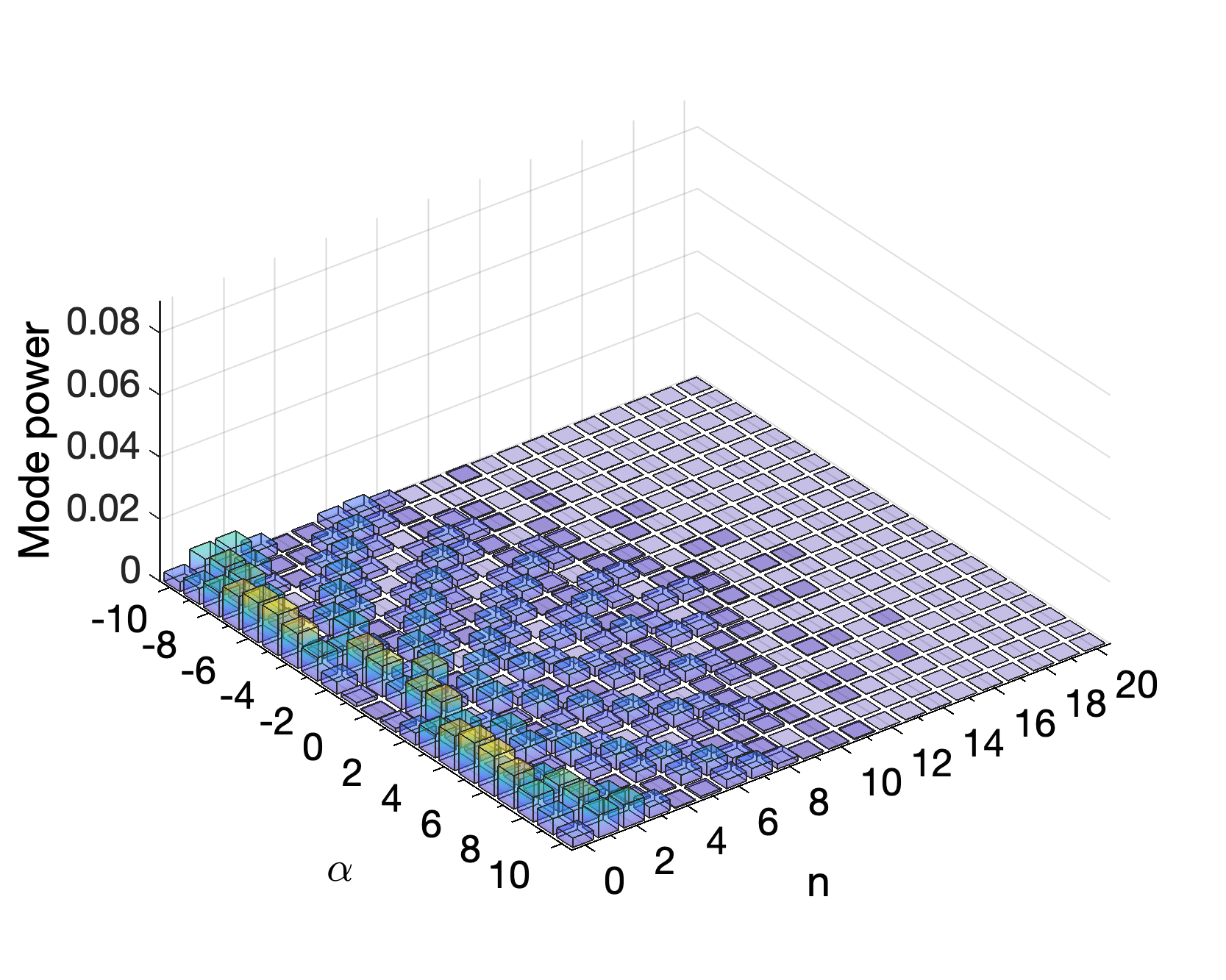}
		\caption{2 waves}
		\label{fig:Distortion_planewave_decomp_2}
	\end{subfigure} %
	\begin{subfigure}[c]{0.5\textwidth}
		\centering
		\includegraphics[width=6.5cm]{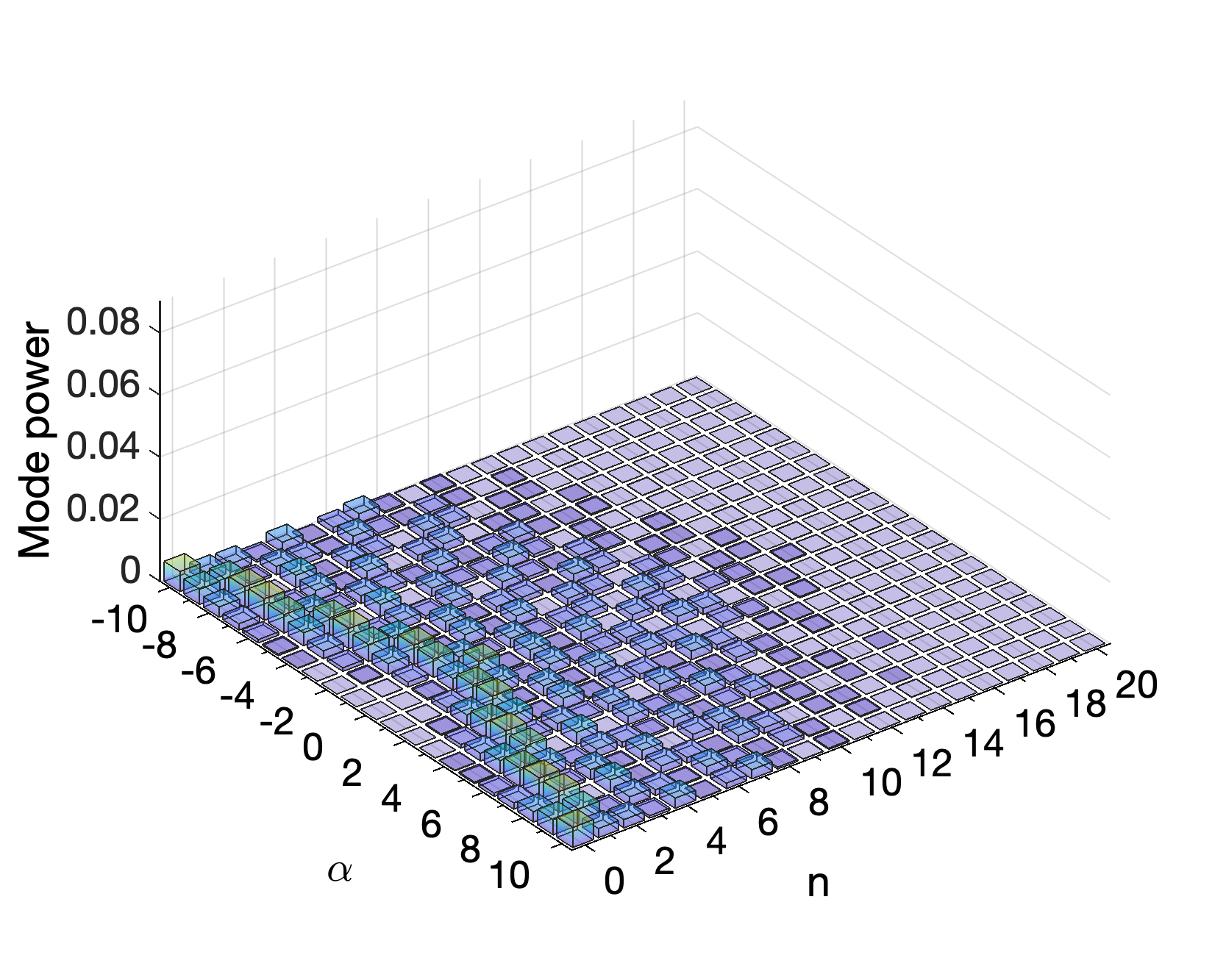}
		\caption{3 waves}
		\label{fig:Distortion_planewave_decomp_3}
	\end{subfigure} %
\caption{The power distribution amongst the transverse cavity modes as the strength of distortion increases from one to five waves of maximum phase delay. }
\label{fig:Distortion_planewave_decomps}
\end{figure}

\begin{figure}[!ht]
	\begin{subfigure}[c]{0.5\textwidth}
		\centering
		\includegraphics[width=6.5cm]{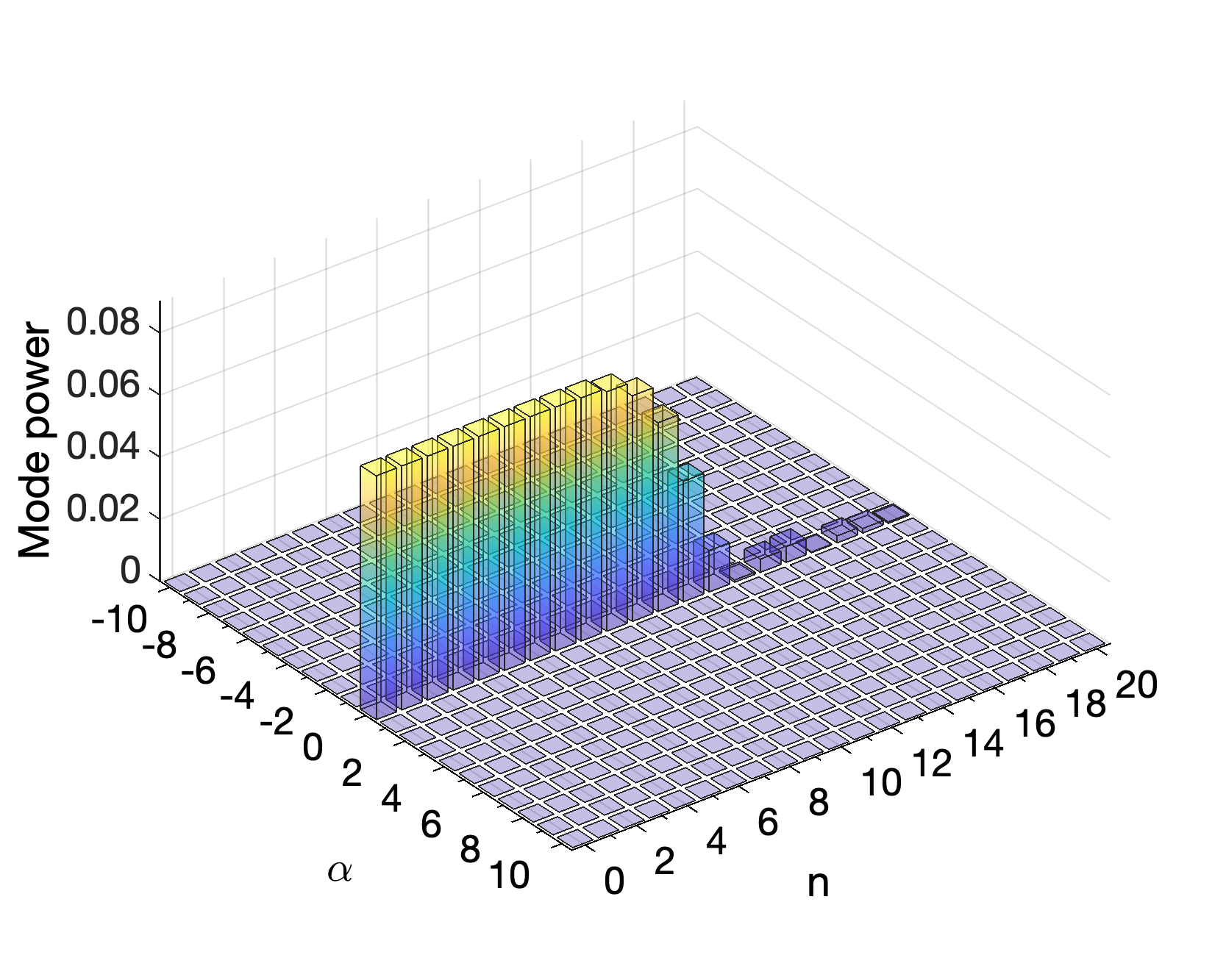}
		\caption{0.5 waves}
		\label{fig:Fieldcurvature_planewave_decomp_05}
	\end{subfigure} %
	\begin{subfigure}[c]{0.5\textwidth}
		\centering
		\includegraphics[width=6.5cm]{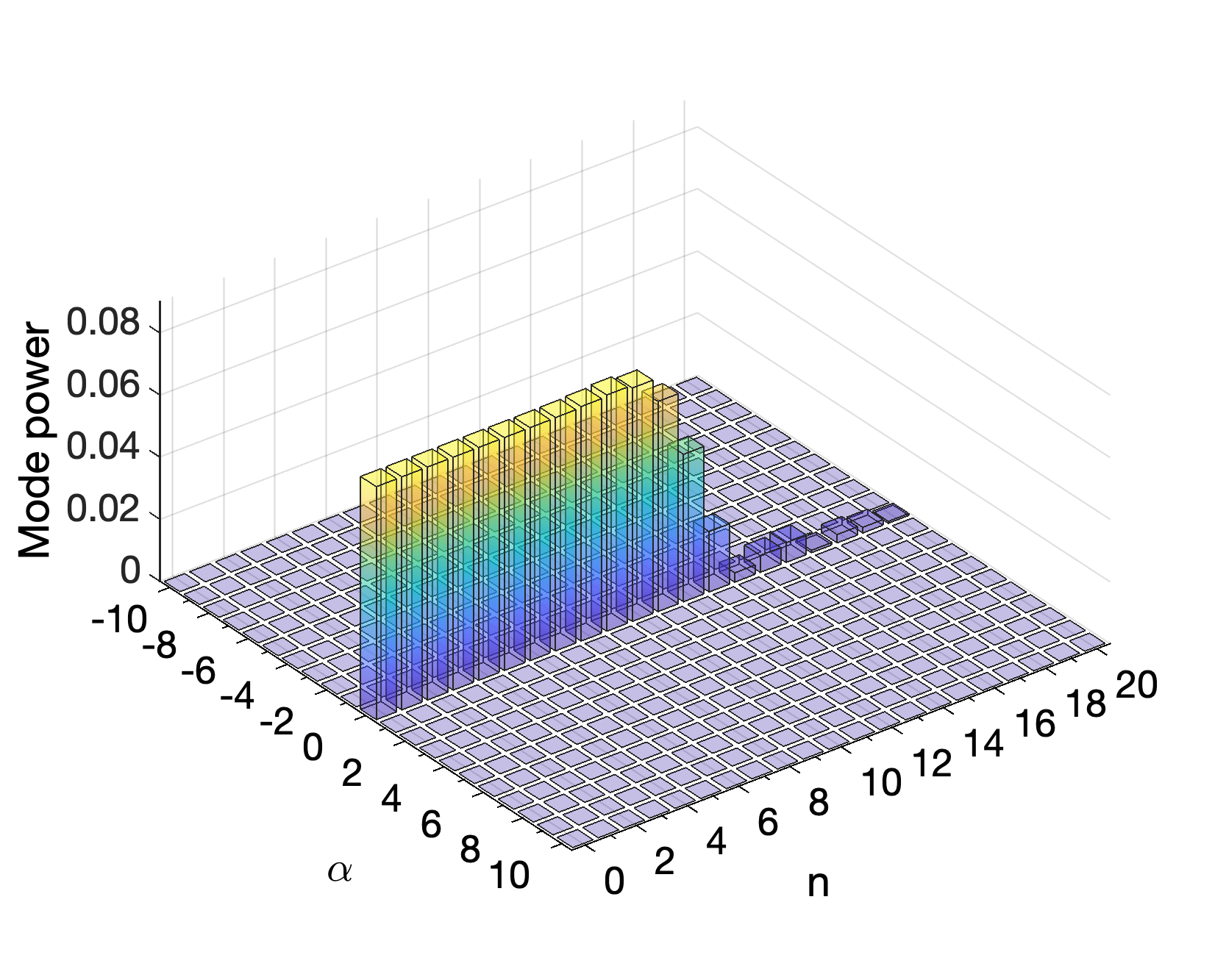}
		\caption{1 wave}
		\label{fig:Fieldcurvature_planewave_decomp_1}
	\end{subfigure} \\
	\par\bigskip %
	\begin{subfigure}[c]{0.5\textwidth}
		\centering
		\includegraphics[width=6.5cm]{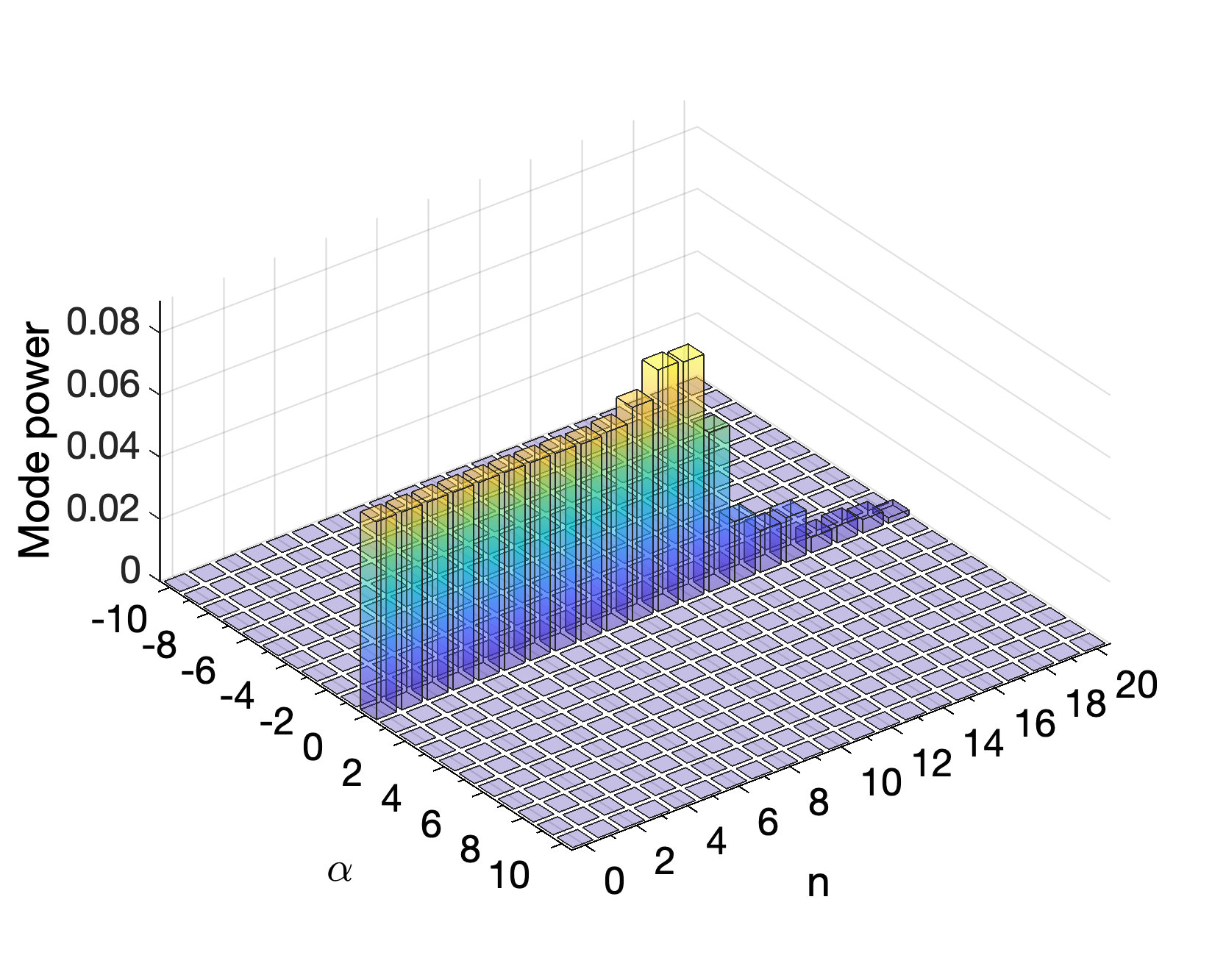}
		\caption{2 waves}
		\label{fig:Fieldcurvature_planewave_decomp_2}
	\end{subfigure} %
	\begin{subfigure}[c]{0.5\textwidth}
		\centering
		\includegraphics[width=6.5cm]{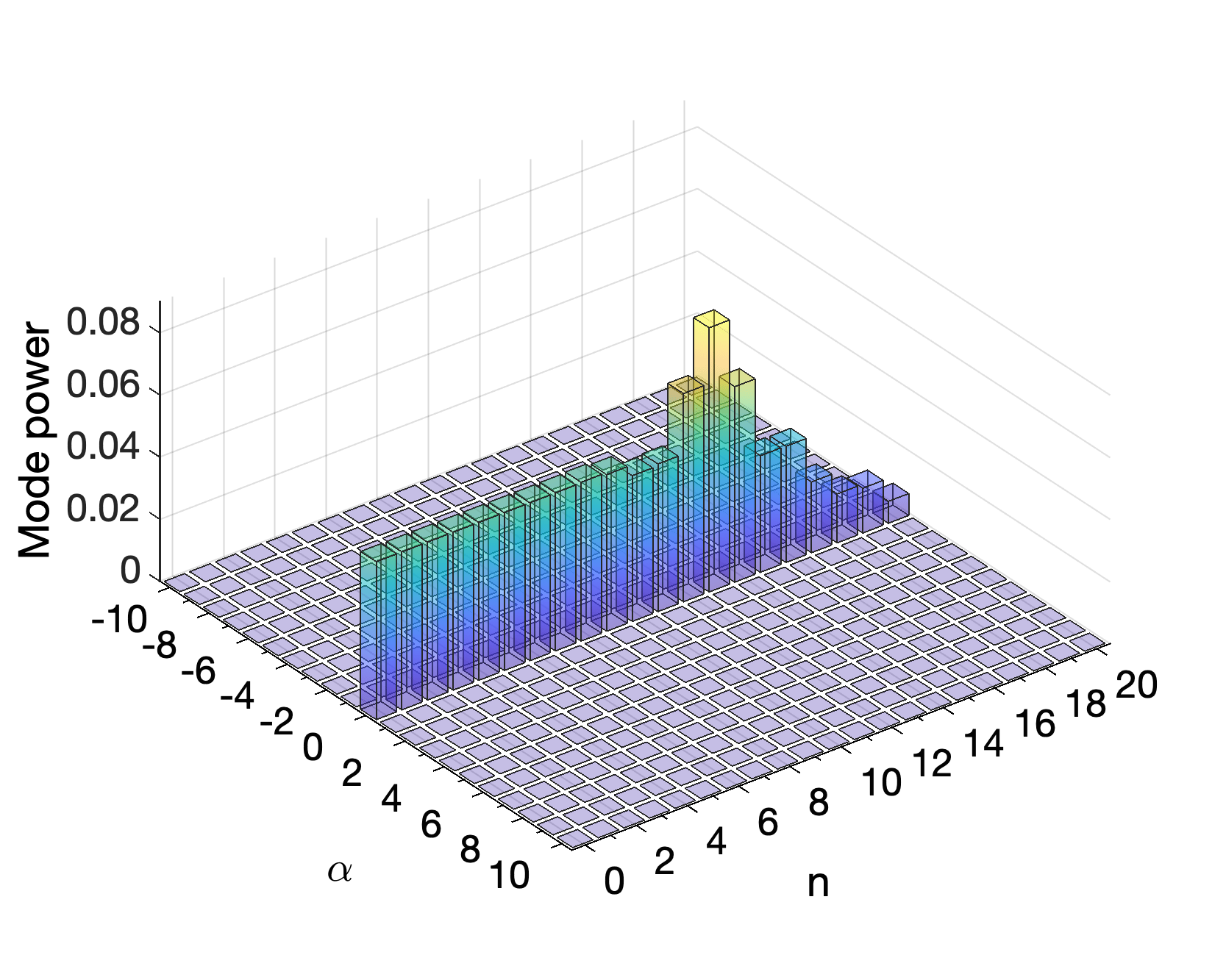}
		\caption{3 waves}
		\label{fig:Fieldcurvature_planewave_decomp_3}
	\end{subfigure} %
\caption{The transverse cavity mode power distribution caused by half a wave to three waves of maximum phase delay, shaped as field curvature.}
\vspace{-2mm}
\label{fig:Fieldcurvature_planewave_decomps}
\end{figure}

\FloatBarrier %

As before, we may translate the cavity mode decomposition into the observable spectral transmission. Despite the differences in mode activity behavior, Fig.~\ref{fig:Distortion_planewave_spectra} shows recognizable similarities with the Laguerre-Gaussian-borne spectrum: power is still concentrated into a packet of limited width which detaches completely from the fundamental to drift upward in frequency. Field curvature's spectrum in Fig.~\ref{fig:Fieldcurvature_planewave_spectra} also exhibits a local maximum of power traveling upwards in frequency, but it is easily discerned from distortion by the significant and evenly-distributed power remaining in the modes down to the fundamental. 

\begin{figure}[h!]
	\begin{subfigure}[c]{0.5\textwidth}
		\centering
		\includegraphics[width=7cm]{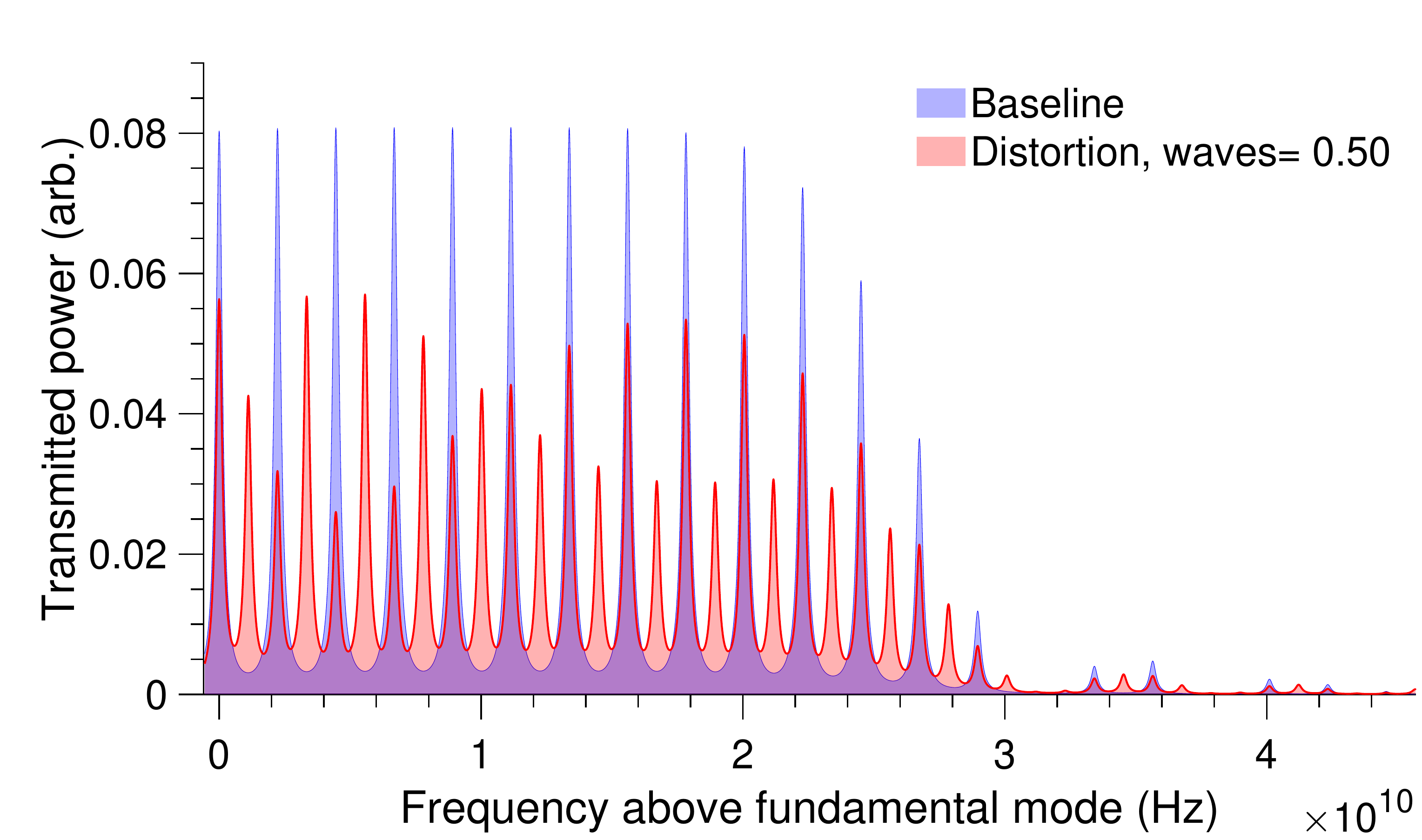}
		\caption{0.5 waves}
		\label{fig:Distortion_planewave_spectrum_05}
	\end{subfigure} %
	\begin{subfigure}[c]{0.5\textwidth}
		\centering
		\includegraphics[width=7cm]{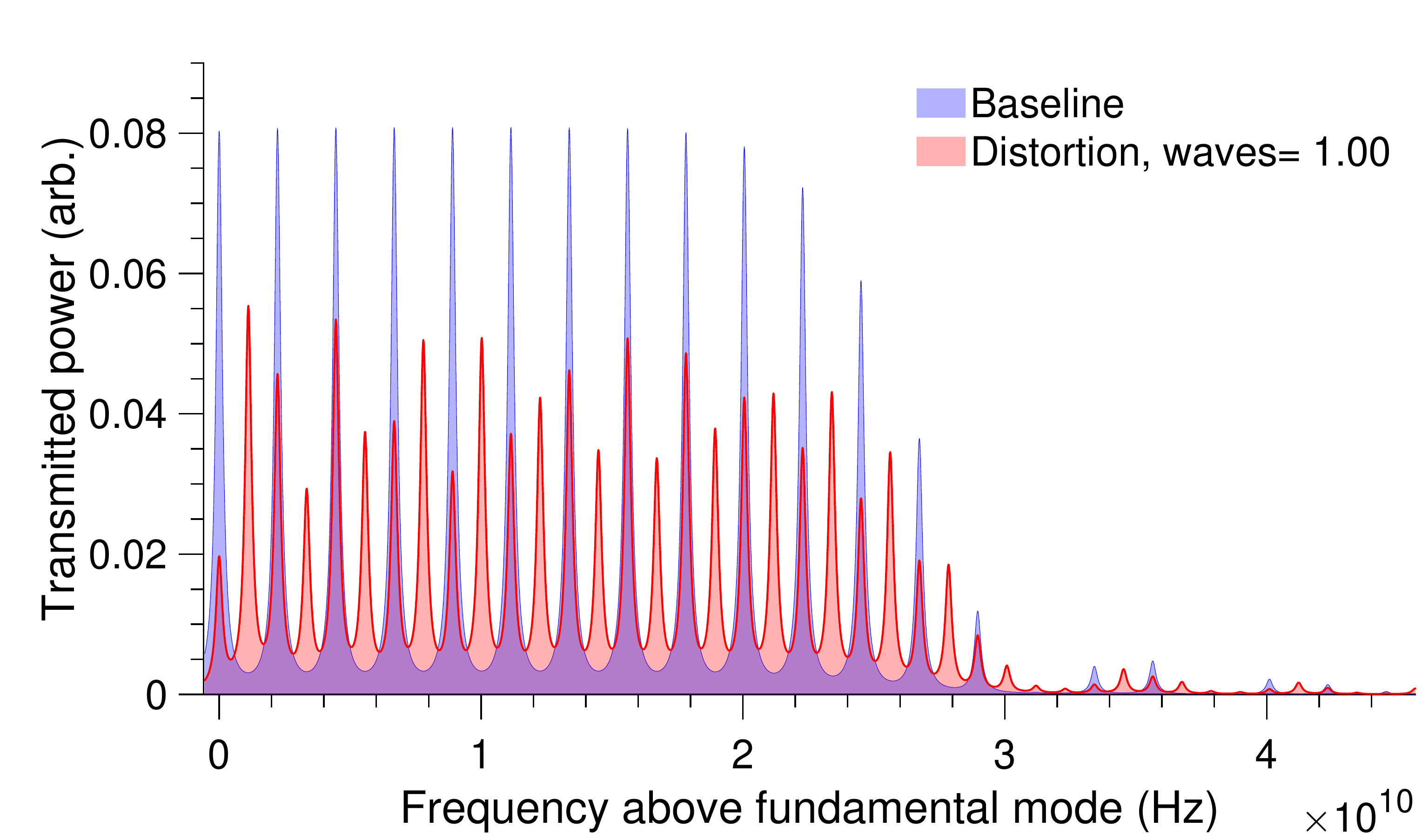}
		\caption{1 wave}
		\label{fig:Distortion_planewave_spectrum_1}
	\end{subfigure} \\
	\par\bigskip %
	\begin{subfigure}[c]{0.5\textwidth}
		\centering
		\includegraphics[width=7cm]{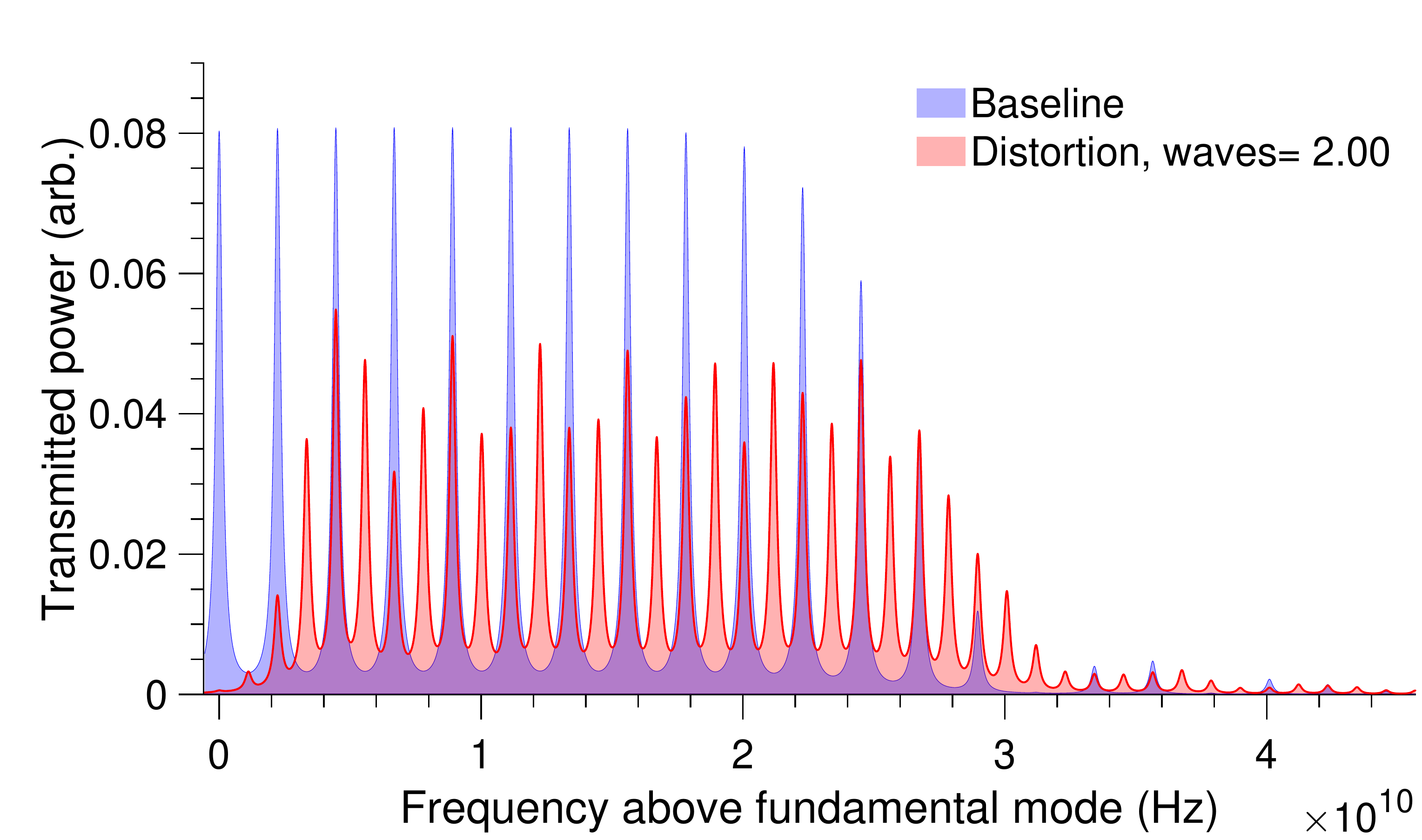}
		\caption{2 waves}
		\label{fig:Distortion_planewave_spectrum_2}
	\end{subfigure} %
	\begin{subfigure}[c]{0.5\textwidth}
		\centering
		\includegraphics[width=7cm]{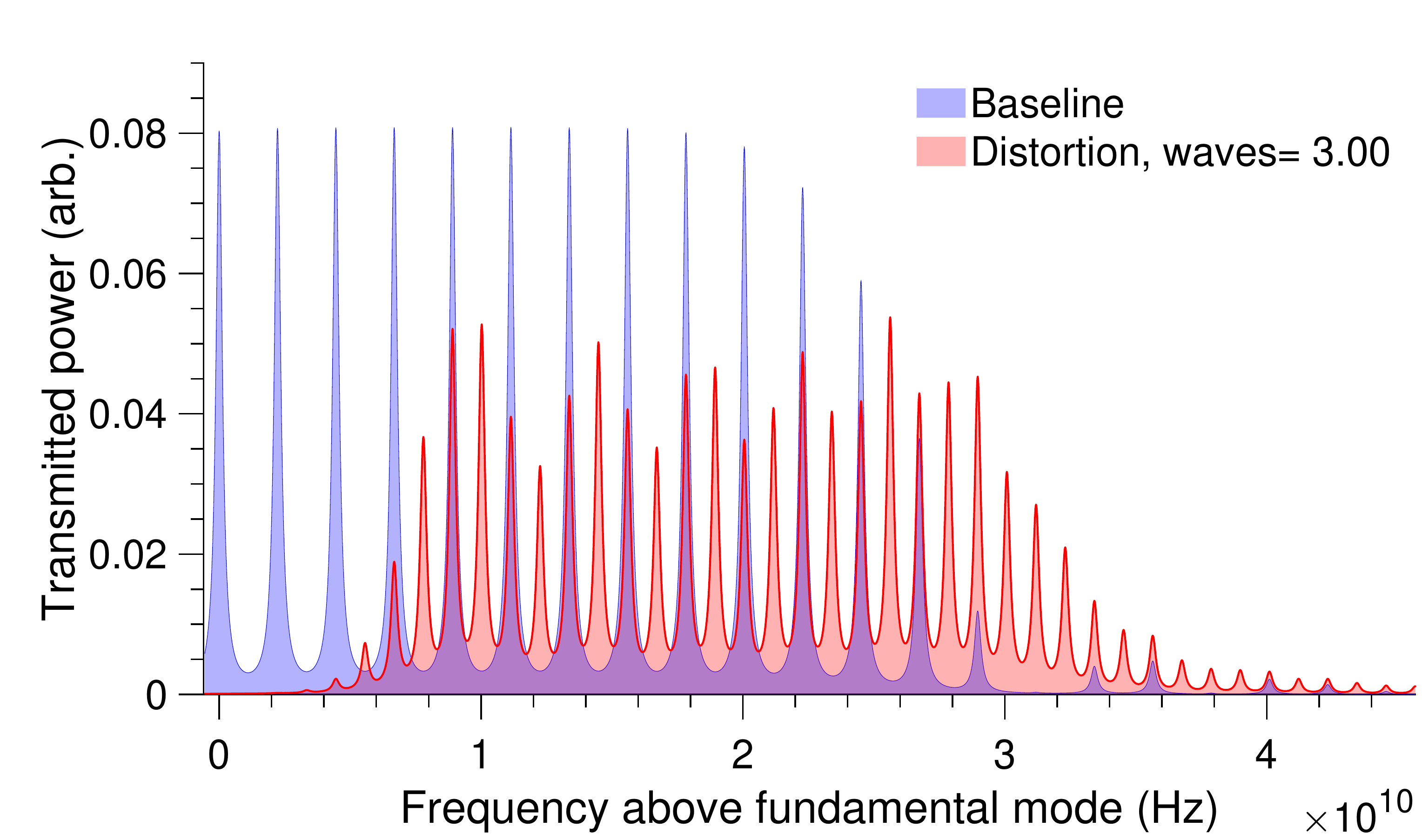}
		\caption{3 waves}
		\label{fig:Distortion_planewave_spectrum_3}
	\end{subfigure} %
\caption{The intensity spectrum transmitted by the cavity as a result of transverse mode excitation by a pupiled plane wave, with and without varying amounts of distortion. It should be noted that the spectral range shown accounts for all the modes contributing degenerately to these frequencies, which is a larger range of mode indices than shown in the preceding Figs.~\ref{fig:Distortion_planewave_decomps}~and~Fig.~\ref{fig:Fieldcurvature_planewave_decomps}.}
\label{fig:Distortion_planewave_spectra}
\end{figure}

\begin{figure}[h!]
	\begin{subfigure}[c]{0.5\textwidth}
		\centering
		\includegraphics[width=7cm]{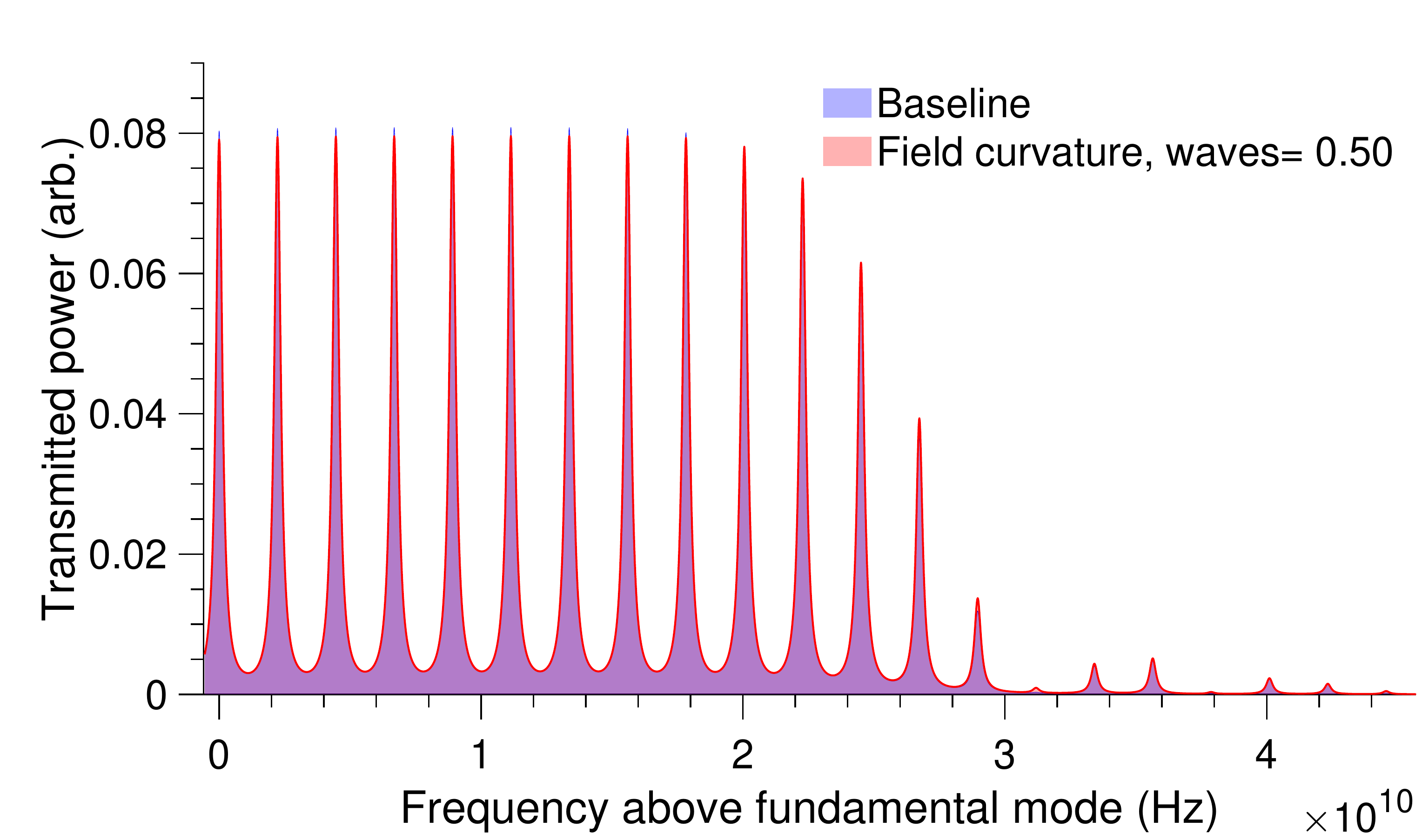}
		\caption{0.5 waves}
		\label{fig:Fieldcurvature_planewave_spectrum_05}
	\end{subfigure} %
	\begin{subfigure}[c]{0.5\textwidth}
		\centering
		\includegraphics[width=7cm]{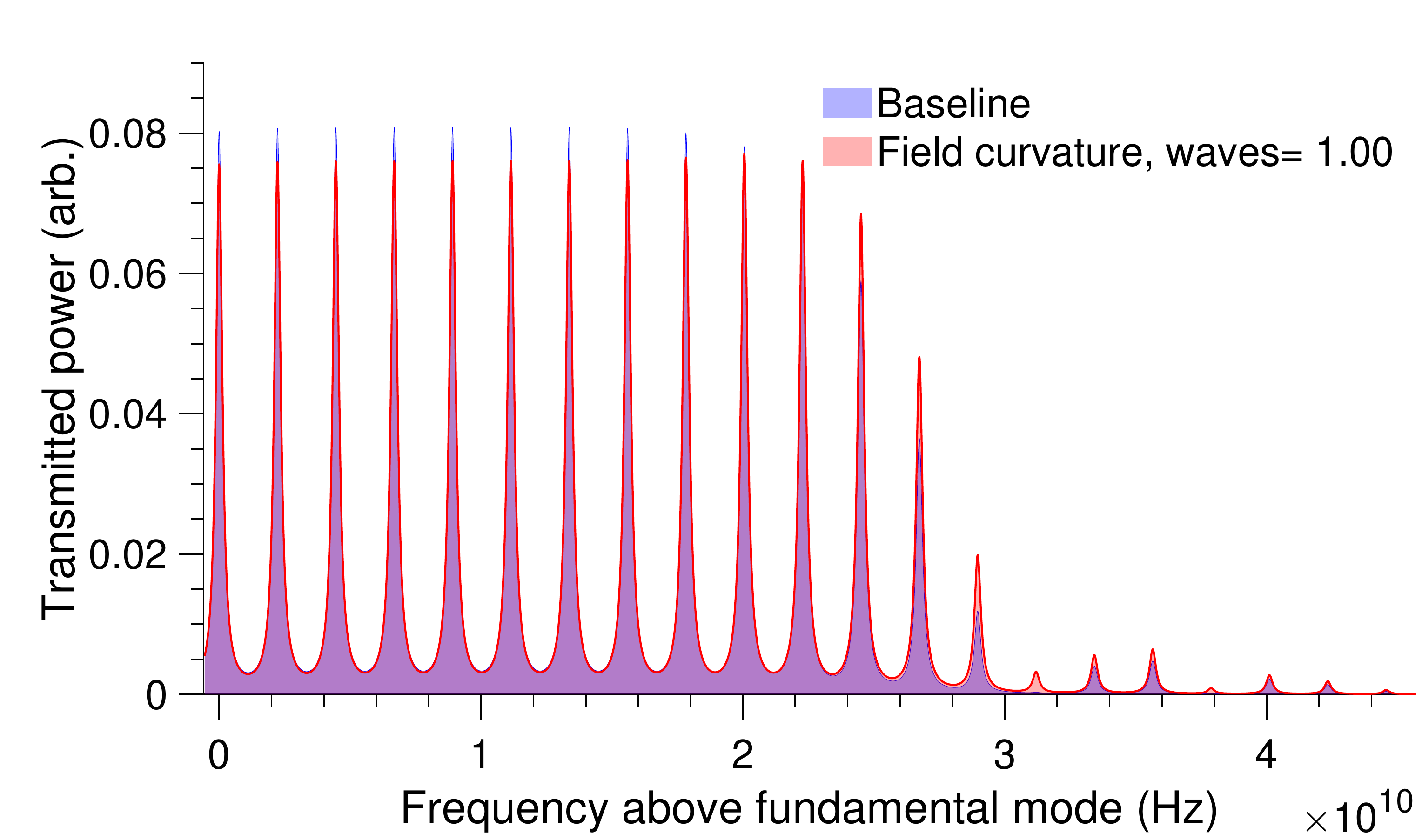}
		\caption{1 waves}
		\label{fig:Fieldcurvature_planewave_spectrum_1}
	\end{subfigure} \\
	\par\bigskip %
	\begin{subfigure}[c]{0.5\textwidth}
		\centering
		\includegraphics[width=7cm]{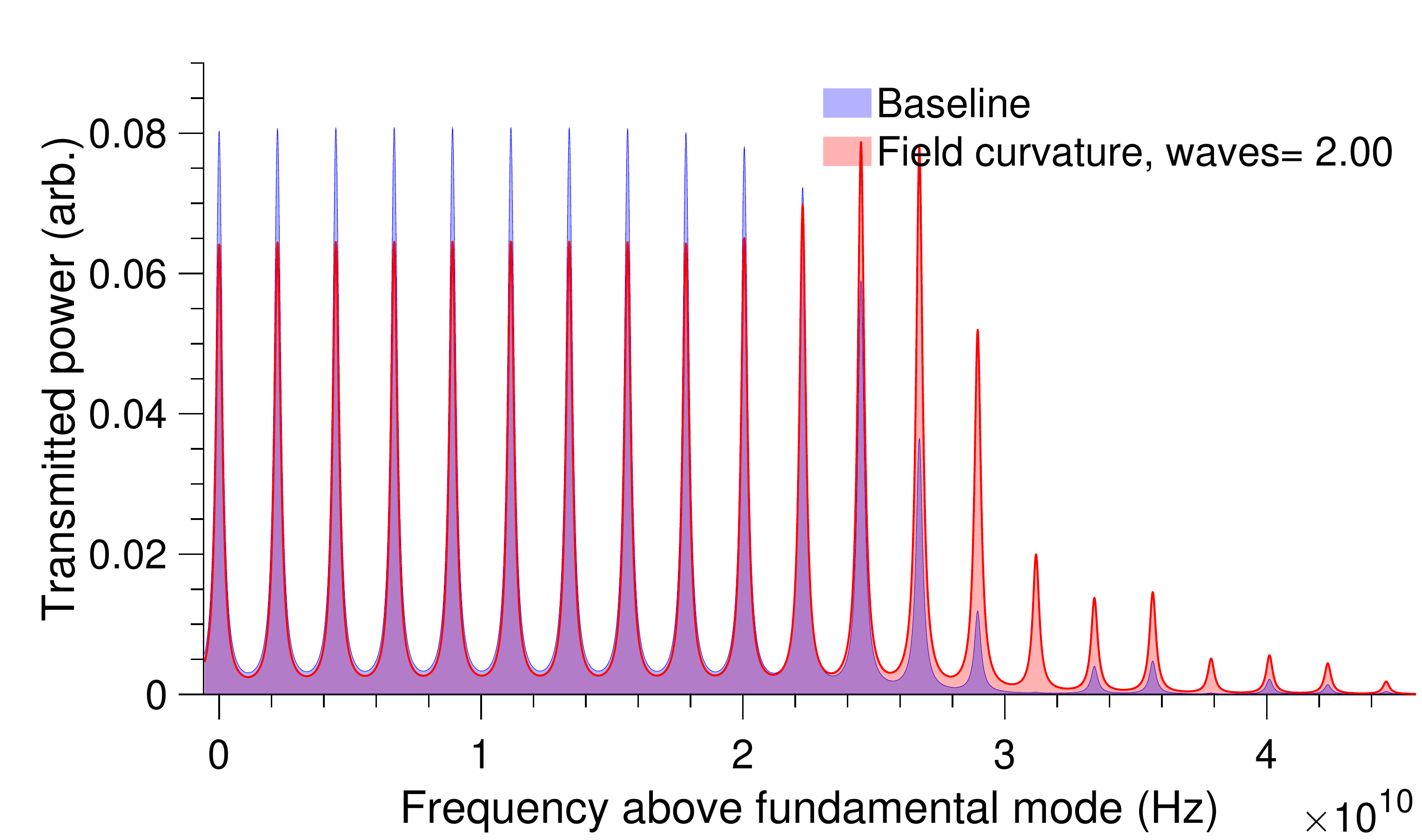}
		\caption{2 waves}
		\label{fig:Fieldcurvature_planewave_spectrum_2}
	\end{subfigure} %
	\begin{subfigure}[c]{0.5\textwidth}
		\centering
		\includegraphics[width=7cm]{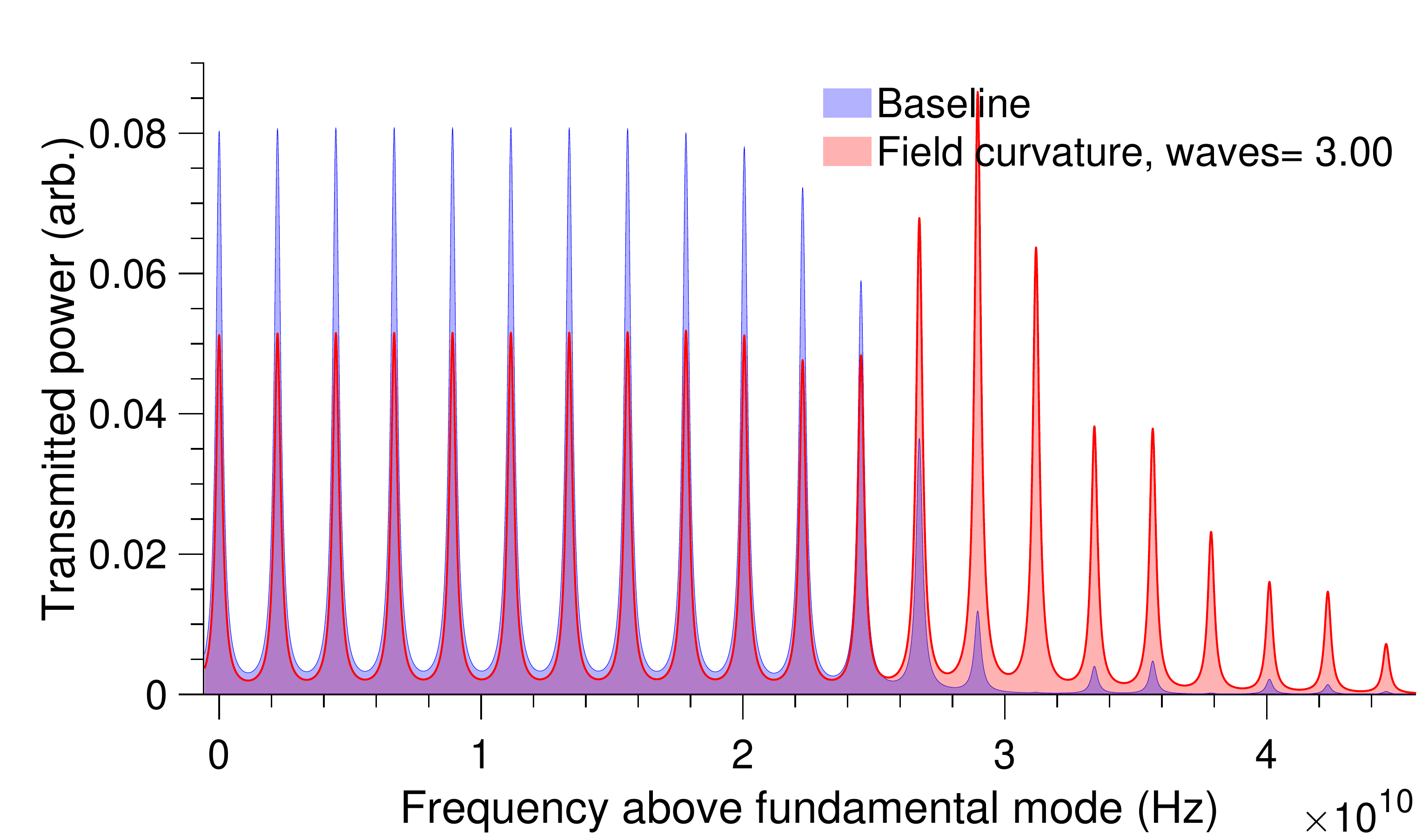}
		\caption{3 waves}
		\label{fig:Fieldcurvature_planewave_spectrum_3}
	\end{subfigure} %
\caption{The intensity spectrum transmitted by the cavity as a result of transverse mode excitation by a pupiled plane wave, with and without varying amounts of field curvature.}
\label{fig:Fieldcurvature_planewave_spectra}
\end{figure}

\FloatBarrier %

\begin{figure}[h!]
	\begin{subfigure}[c]{0.5\textwidth}
		\centering
		\includegraphics[width=7cm]{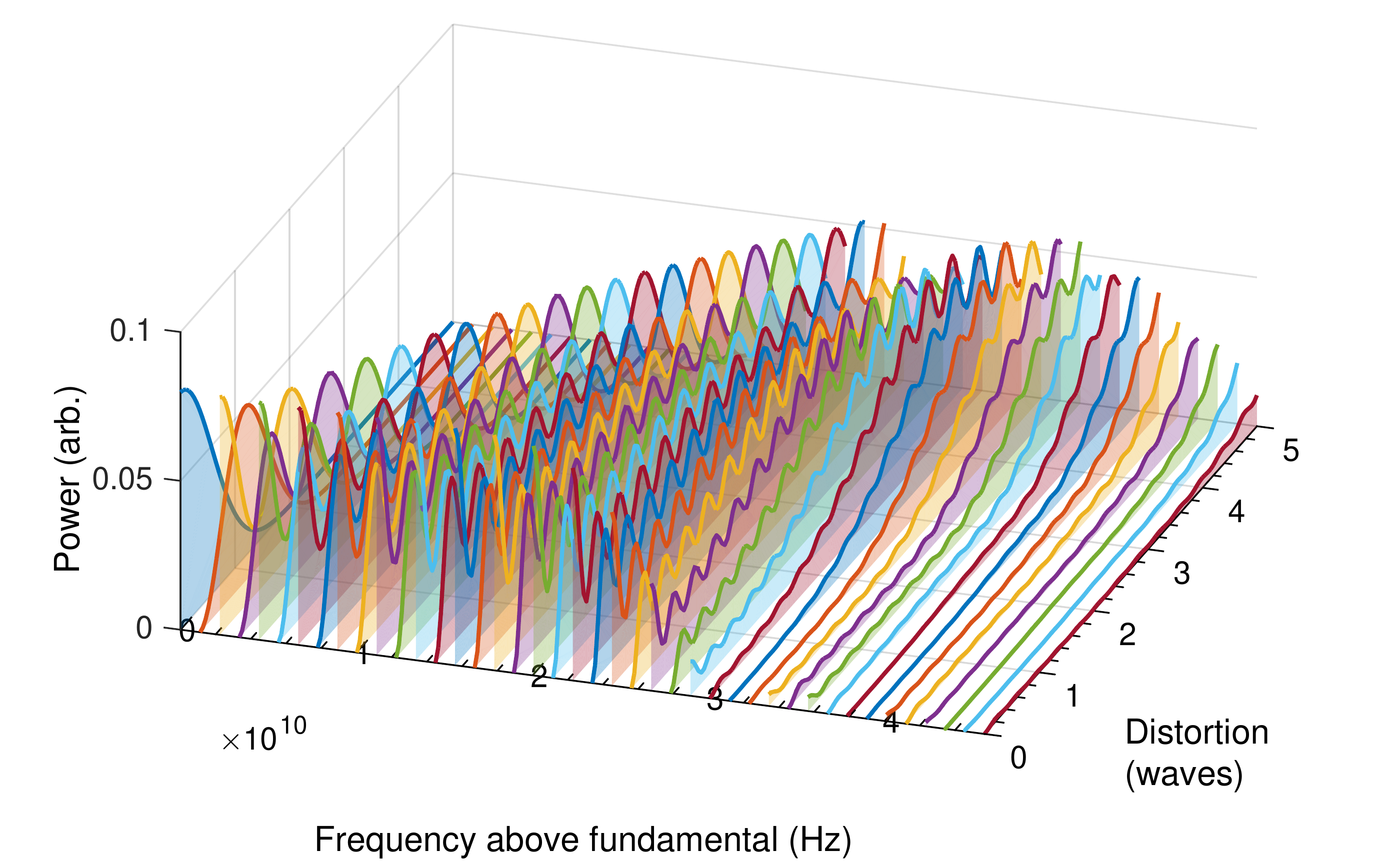}
		\caption{Distortion}
		\label{fig:Distortion_planewave_spectrum_history}
	\end{subfigure} %
	\begin{subfigure}[c]{0.5\textwidth}
		\centering
		\includegraphics[width=7cm]{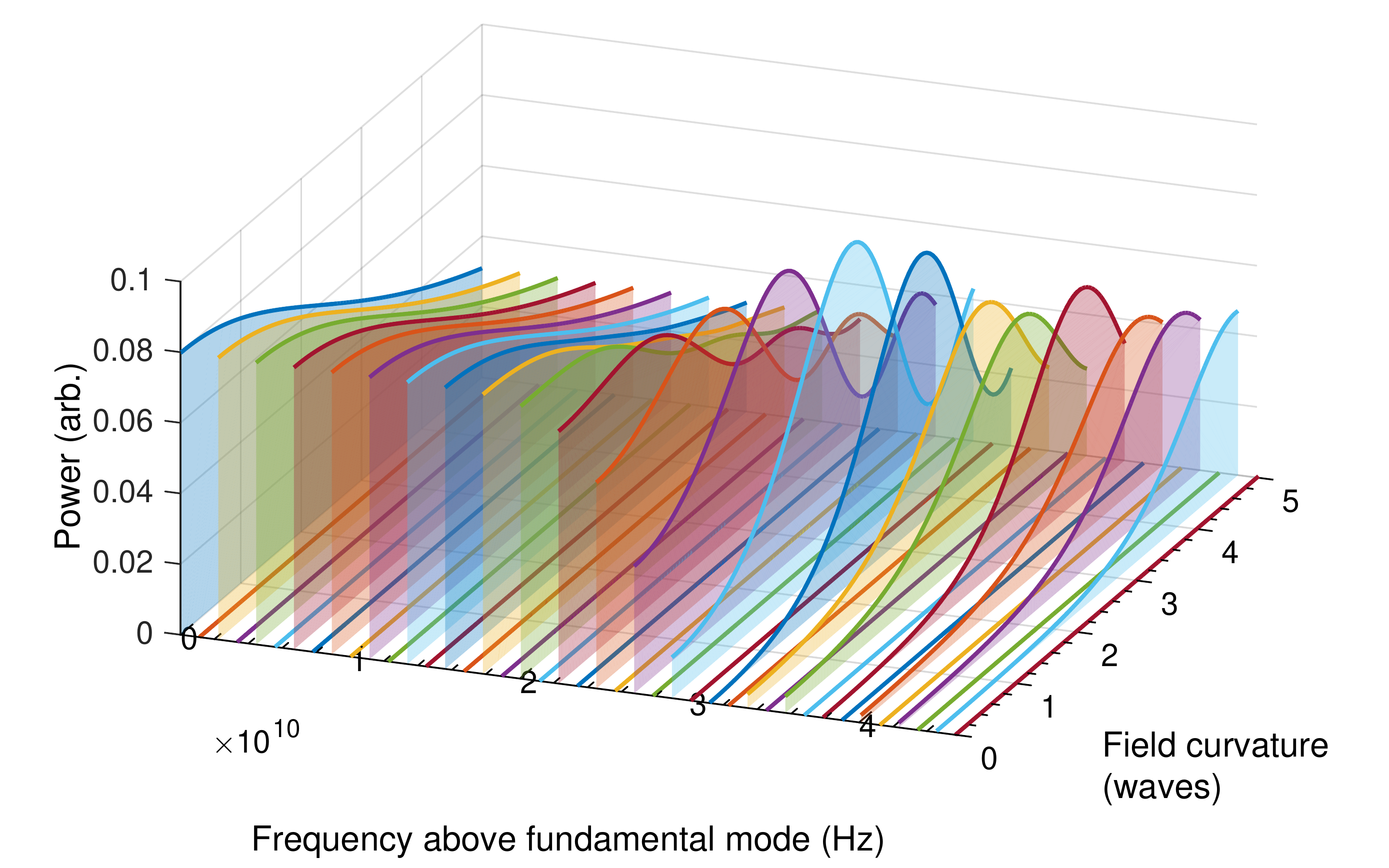}
		\caption{Field curvature}
		\label{fig:Fieldcurvature_planewave_spectrum_history}
	\end{subfigure} %
\caption{The spectral intensities of optical cavity transmission versus the varying amounts of distortion (left) and field curvature (right) applied to a pupiled plane wave. In the latter case, the mode activity being limited to $\alpha=0$ makes the spectrum much simpler than in the former, where activity is spread across both mode indices.}
\label{fig:Planewave_spectrum_histories}
\end{figure}

\FloatBarrier %

The differences in the spectral evolution of distortion and field curvature, summarized in Fig.~\ref{fig:Distortion_planewave_spectrum_history}~and~\ref{fig:Fieldcurvature_planewave_spectrum_history} respectively, suggest that a single metric can again serve as a good indicator of aberration strength and perhaps enabling limited ability to distinguish dominant aberration characteristics. This suspicion is supported in  Figs.~\ref{fig:Planewave_spectral_singlemetrics} below. The spectral power distribution of distortion now departs the vertical axis---i.e., contains no power in the lowest-frequency modes---starting with much lower aberration strengths, exaggerating its differentiation from field curvature; aberration strengths greater than half a wave may be estimated by summing the power within a fixed frequency of the fundamental. Field curvature, in contrast, never reaches zero on the vertical axis, instead offering to be identified by its differing rates of power collection per bandwidth. 

\begin{figure}[h!]
	\begin{subfigure}[c]{0.5\textwidth}
		\centering
		\includegraphics[width=7cm]{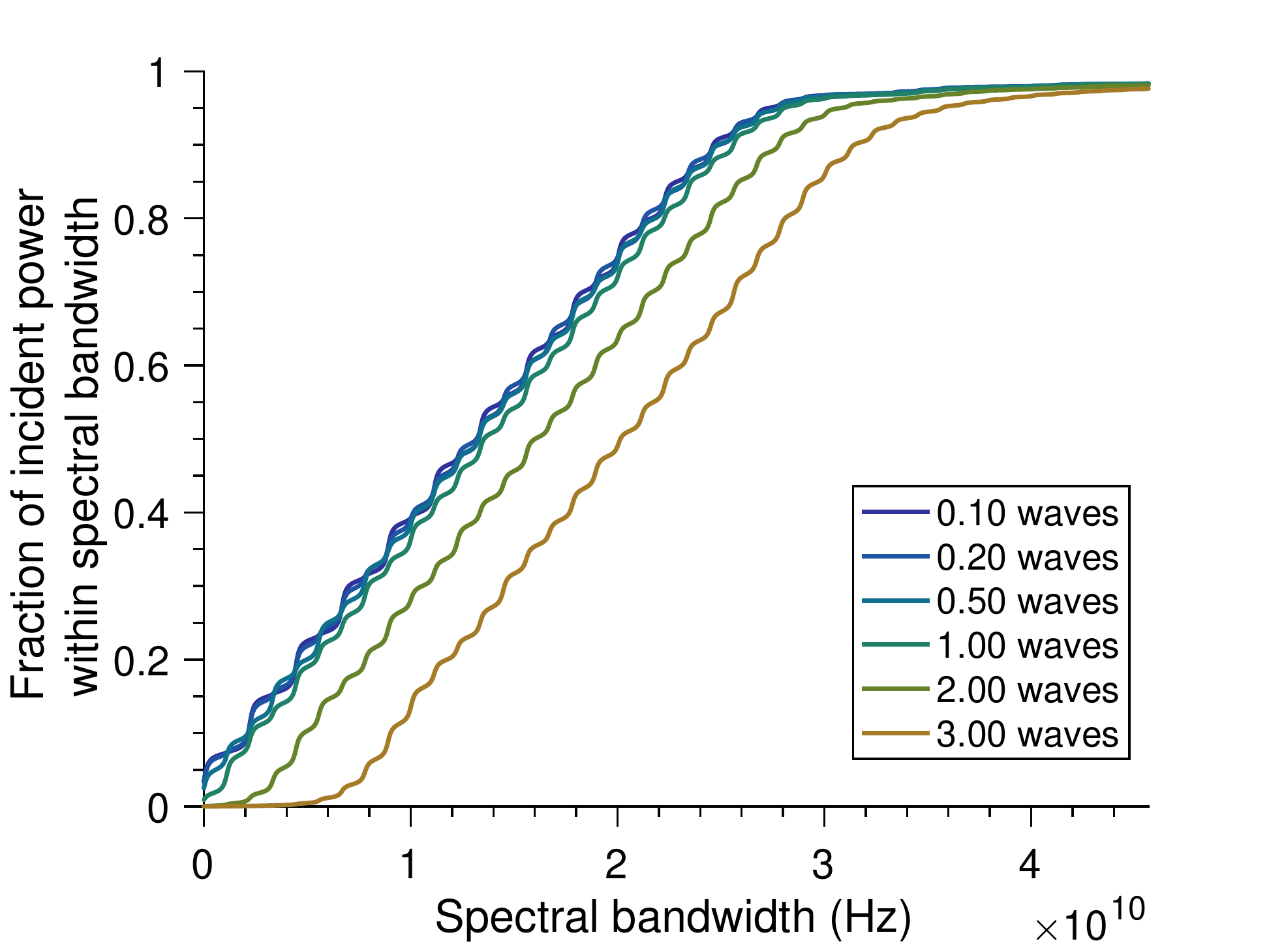}
		\caption{Distortion}
		\label{fig:Distortion_planewave_spectral_power_distr}
	\end{subfigure} %
	\begin{subfigure}[c]{0.5\textwidth}
		\centering
		\includegraphics[width=7cm]{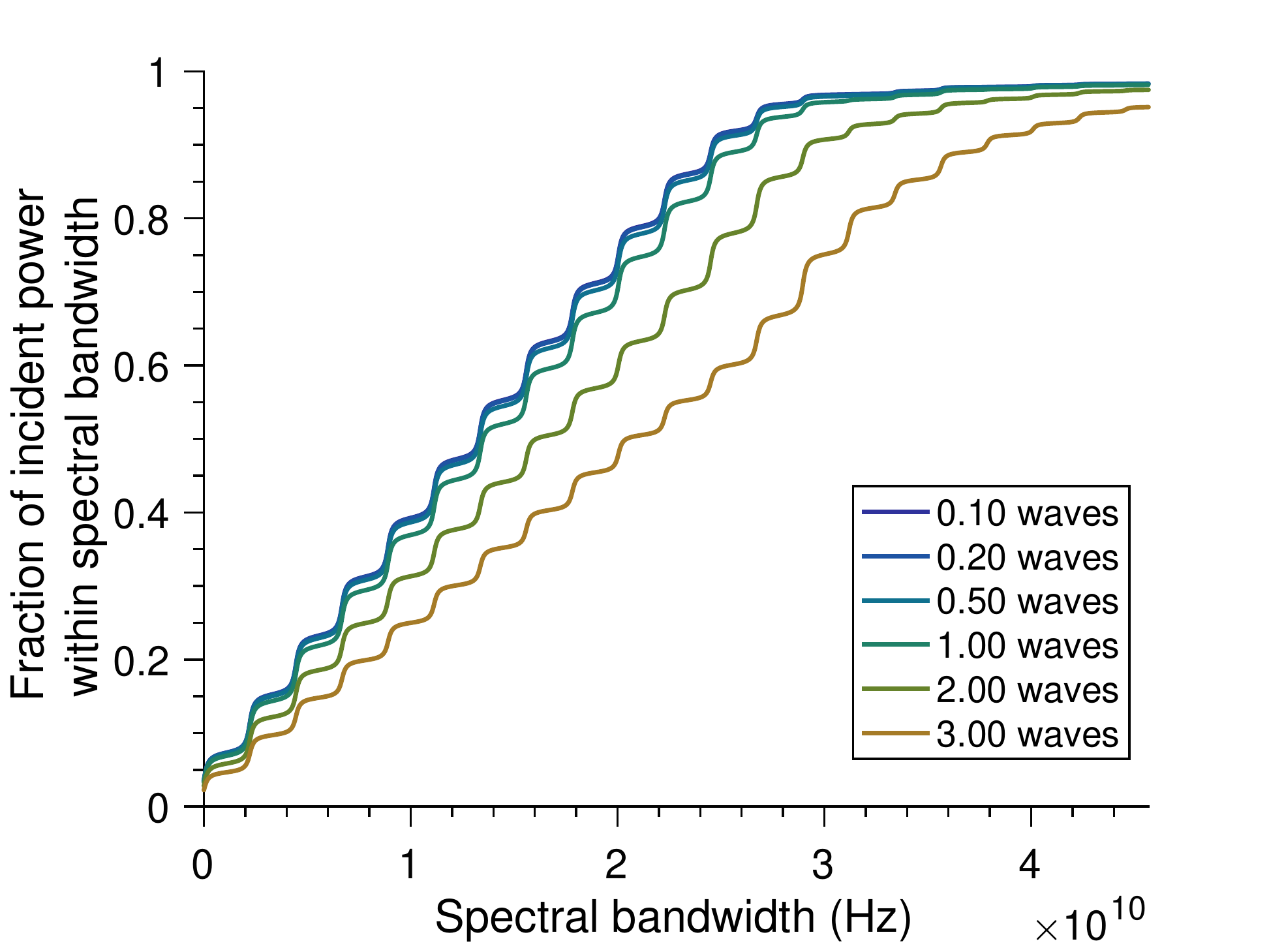}
		\caption{Field curvature}
		\label{fig:Fieldcurvature_planewave_spectral_power_distr}
	\end{subfigure} %
	\caption{The spectral power distribution plots introduced in Fig.~\ref{fig:LGbasecase_spectral_singlemetrics}, this time with a plane wave baseline beam instead of a cavity-matched Laguerre-Gaussian.}
	\label{fig:Planewave_spectral_singlemetrics}
\end{figure}

\FloatBarrier %

\section{Concept Demonstration - Many Aberrations}

The single-aberration scenarios examined above were chosen to have simple and clear geometries, but real-world systems are unlikely to produce wavefronts with such characteristics. For example, wavefront aberrations could be created with thermal distortion in a complex optical system or be created over long distances of horizontal propagation through atmosphere, where many aberrations would exist simultaneously. 

These more elborate phasefronts---including those described by Kolmogorov theory \cite{Noll_Zernikepolynomialsatmospheric_1976}---can be modeled using the Zernike polynomials, a family of surface functions which can be superimposed to describe any smooth, well-behaved surface on the unit circle. The Zernikes are often given in modified form for various purposes; however, the changes often reduce to differences in normalization and index definition. \cite{Noll_Zernikepolynomialsatmospheric_1976} \cite{Wang_WavefrontinterpretationZernike_1980} \cite{Boreman_ZernikeexpansionsnonKolmogorov_1996} For our purposes, the Zernike term $Z_{n}^{m}$ is given for $m \geq 0$ by \cite{Lakshminarayanan_Zernikepolynomialsguide_2011a}
\begin{align}
Z_{n}^{m}(\rho, \theta) = \left( \frac{2(n+1)}{1+\delta_{m,0}} \right)^{1/2} R_m^n ( \rho ) \cos m \theta
\label{eqn:Zernike_posm_def}
\end{align}

\noindent and for $m < 0$ by 
\begin{align}
Z_{n}^{-m}(\rho, \theta) = \left( \frac{2(n+1)}{1+\delta_{m,0}} \right)^{1/2} R_m^n ( \rho ) \sin m \theta
\label{eqn:Zernike_negm_def}
\end{align}

\noindent where $\delta_{m,0}$ is the Kronecker delta ($=1$ if $m=0$, and $=0$ otherwise) and $R_m^n$ is the radial polynomial given by 
\begin{align}
R_{n}^{m}(\rho) = \sum_{s=0}^{(n-m)/2} \frac{(-1)^n (n-s)!}{s! (n+m)/2 - s)! ((n-m)/2 - s)!} \rho^{(n-2s)}
\label{eqn:Zernike_Rterm_def}
\end{align}

The Seidel aberrations can each be expressed as single, low-order Zernike terms; for instance, distortion is equivalent to $Z_{1}^{1}$, and field curvature is assigned $Z_{2}^{0}$. Conversely, a more complicated surface can be created by summing a number of Zernike polynomials with randomly-chosen weightings. We can use this surface as a phase delay function, with a constant strength coefficient prepended, in a way exactly analogous to the preceding Seidel examples:
\begin{align}
E_\text{aberrated} (r, \phi) = & E_\text{source} (r, \phi, z = z_\text{cavity})  \cdot \exp \left(i W \cdot \sum_{n=0}^{n=n_\text{max}} \, \sum_{a=0}^{a=n} Z_n^{m=-n+2a} \right)
\label{eqn:Pollock_Zernike_def}
\end{align}

An arbitrary set of Zernike term weightings shown in Fig.~\ref{fig:10_Zernikes_weightings} results in the surface depicted by Fig.~\ref{fig:10_Zernikes_surface}. Using the same initial distant-source plane wave and parameters as before, we apply eqn.~\ref{eqn:Pollock_Zernike_def} to obtain the aberrated fields in Fig.~\ref{fig:Zernikes_planewave_fields}. We note that it would be very useful to analyze multiple sets of Zernikes produced by common types of aberrations, say those produced by thermal expansion effects in camera lens trains or by Kolmogorov turbulence in the atmosphere. However, the treatment of any one of these would add many pages to this work and take us far afield from the basic cavity vs aberration concept. These additions will be taken up in a subsequent work.

\begin{figure}[h!]
	\begin{subfigure}[c]{0.5\textwidth}
		\centering
		\includegraphics[width=7cm]{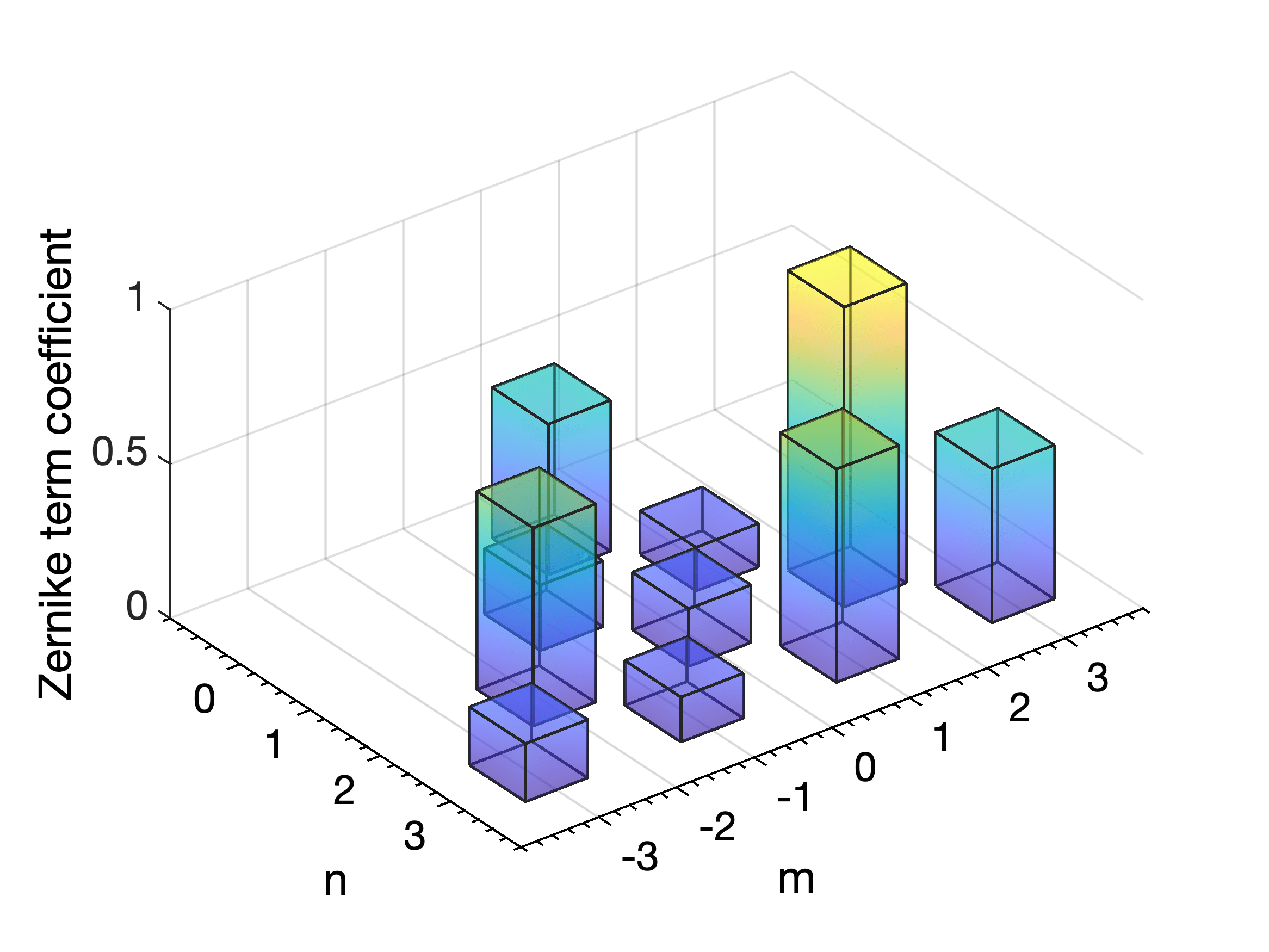}
		\caption{Spectral power distribution}
		\label{fig:10_Zernikes_weightings}
	\end{subfigure} %
	\begin{subfigure}[c]{0.5\textwidth}
		\centering
		\includegraphics[width=7cm]{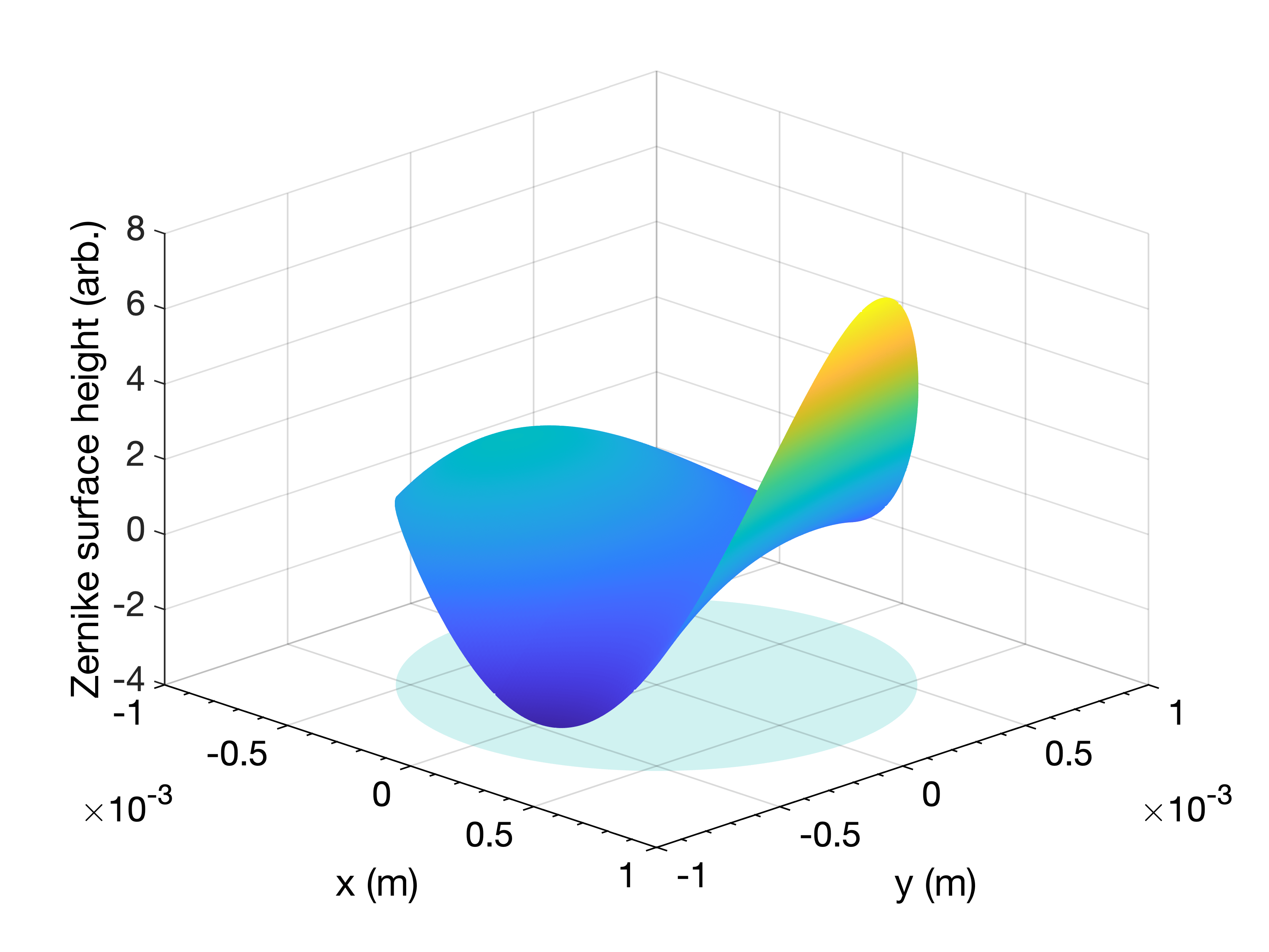}
		\caption{Spectral width}
		\label{fig:10_Zernikes_surface}
	\end{subfigure} %
	\caption{At left, the randomly chosen coefficients we will use for the first ten Zernike terms. Note the triangular shape of the valid Zernike indices. At right, the resulting surface plotted on a unit disk rescaled to our spatially-dimensioned sampling area. }
	\label{fig:10_Zernikes}
\end{figure}

\begin{figure}[h!]
	\begin{subfigure}[c]{0.5\textwidth}
		\centering
		\includegraphics[width=6.5cm]{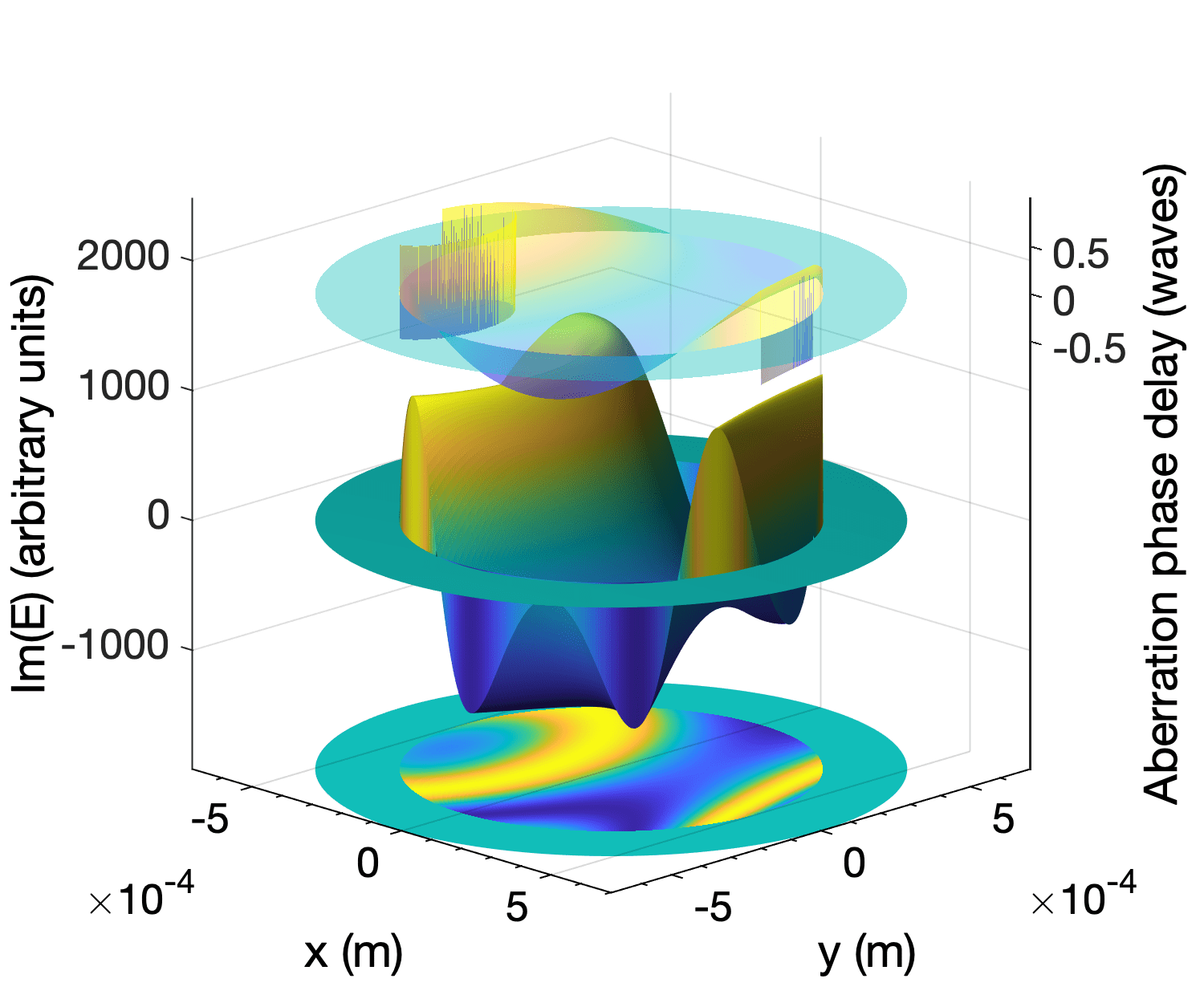}
		\caption{0.5 waves}
		\label{fig:Zernikes_planewave_fields_05}
	\end{subfigure} %
	\begin{subfigure}[c]{0.5\textwidth}
		\centering
		\includegraphics[width=6.5cm]{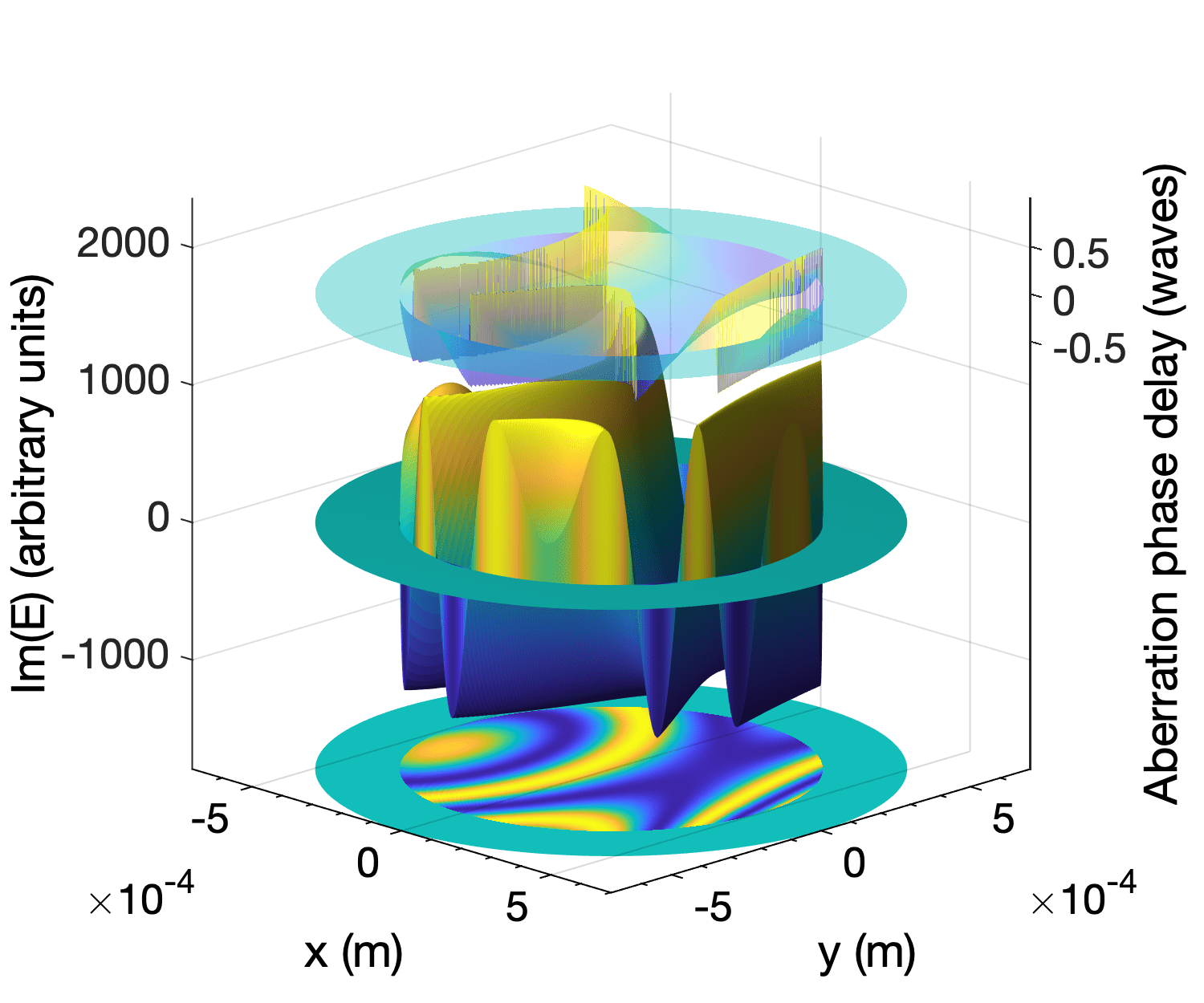}
		\caption{1 wave}
		\label{fig:Zernikes_planewave_fields_1}
	\end{subfigure} \\
	\par\bigskip %
	\begin{subfigure}[c]{0.5\textwidth}
		\centering
		\includegraphics[width=6.5cm]{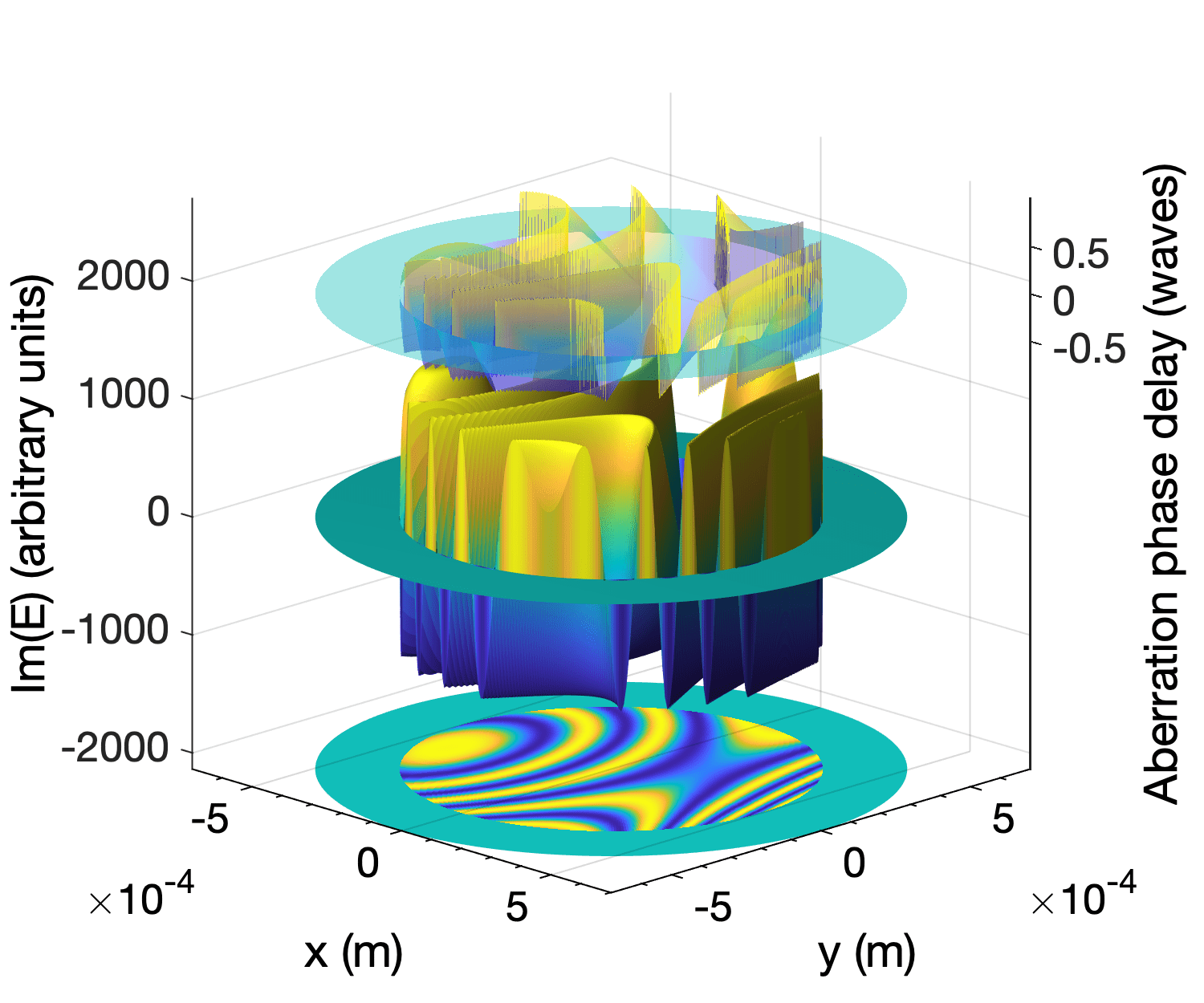}
		\caption{2 waves}
		\label{fig:Zernikes_planewave_fields_2}
	\end{subfigure} %
	\begin{subfigure}[c]{0.5\textwidth}
		\centering
		\includegraphics[width=6.5cm]{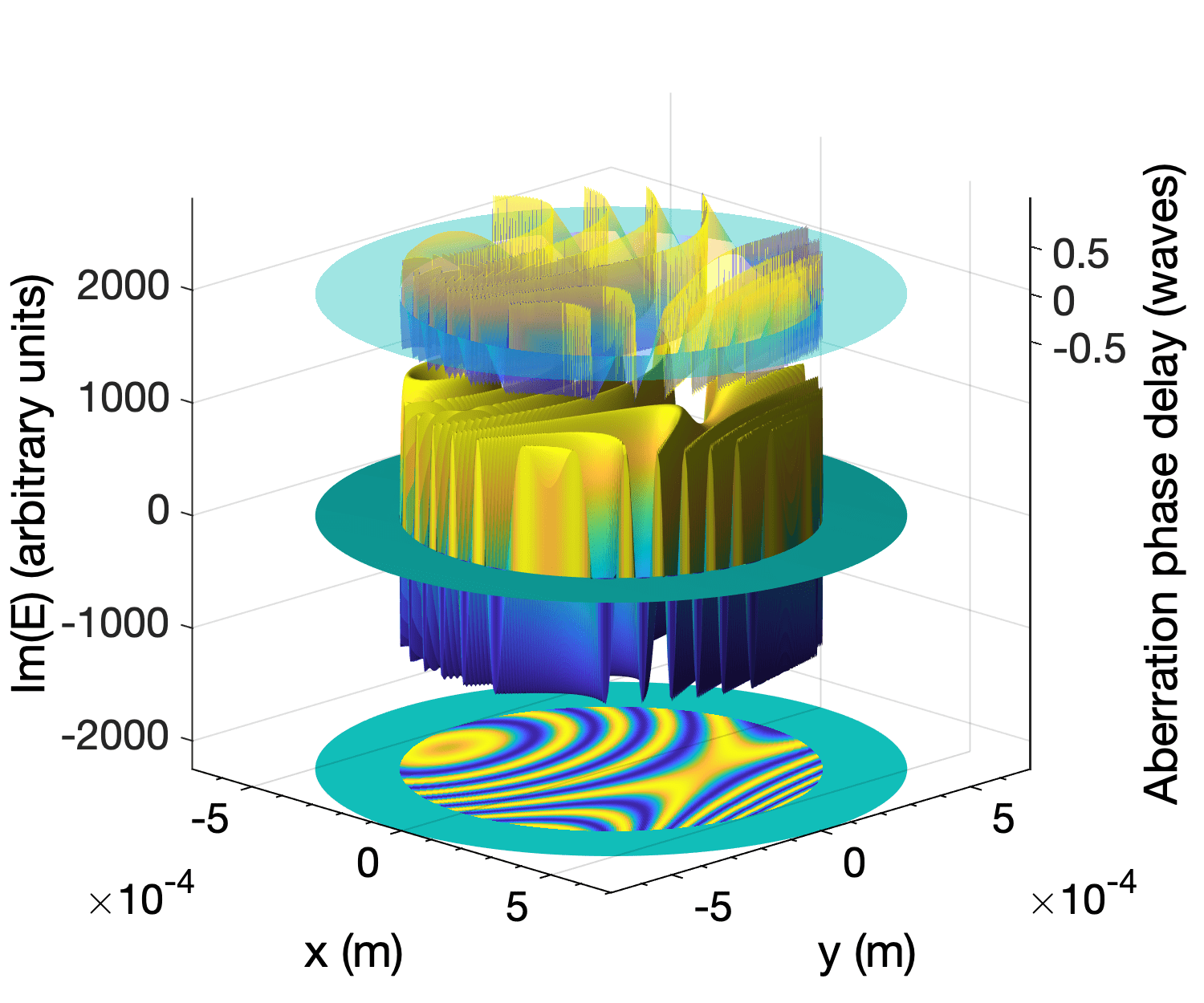}
		\caption{3 waves}
		\label{fig:Zernikes_planewave_fields_3}
	\end{subfigure} %
\caption{The shape and effects on imaginary field component of a phase delay, defined by randomly-weighting the first ten Zernike terms and scaling their sum to achieve a given maximum number of waves, applied to a plane wave bounded by a circular pupil function with negligible frequency component loss from diffraction.}
\label{fig:Zernikes_planewave_fields}
\end{figure}

These fields are brought to our optical cavity, as before, and the resulting mode excitations are shown in Fig.~\ref{fig:Zernikes_planewave_decomps}. Unlike our geometrically-simple Seidels, these mode patterns display no discernible symmetry along either index. Some similarity to the progression of distortion is suggested by the partial "ridge" of mode power which seems to spread downwards in $n$ and outwards in $\alpha$. 

\begin{figure}[h!]
	\begin{subfigure}[c]{0.5\textwidth}
		\centering
		\includegraphics[width=6.5cm]{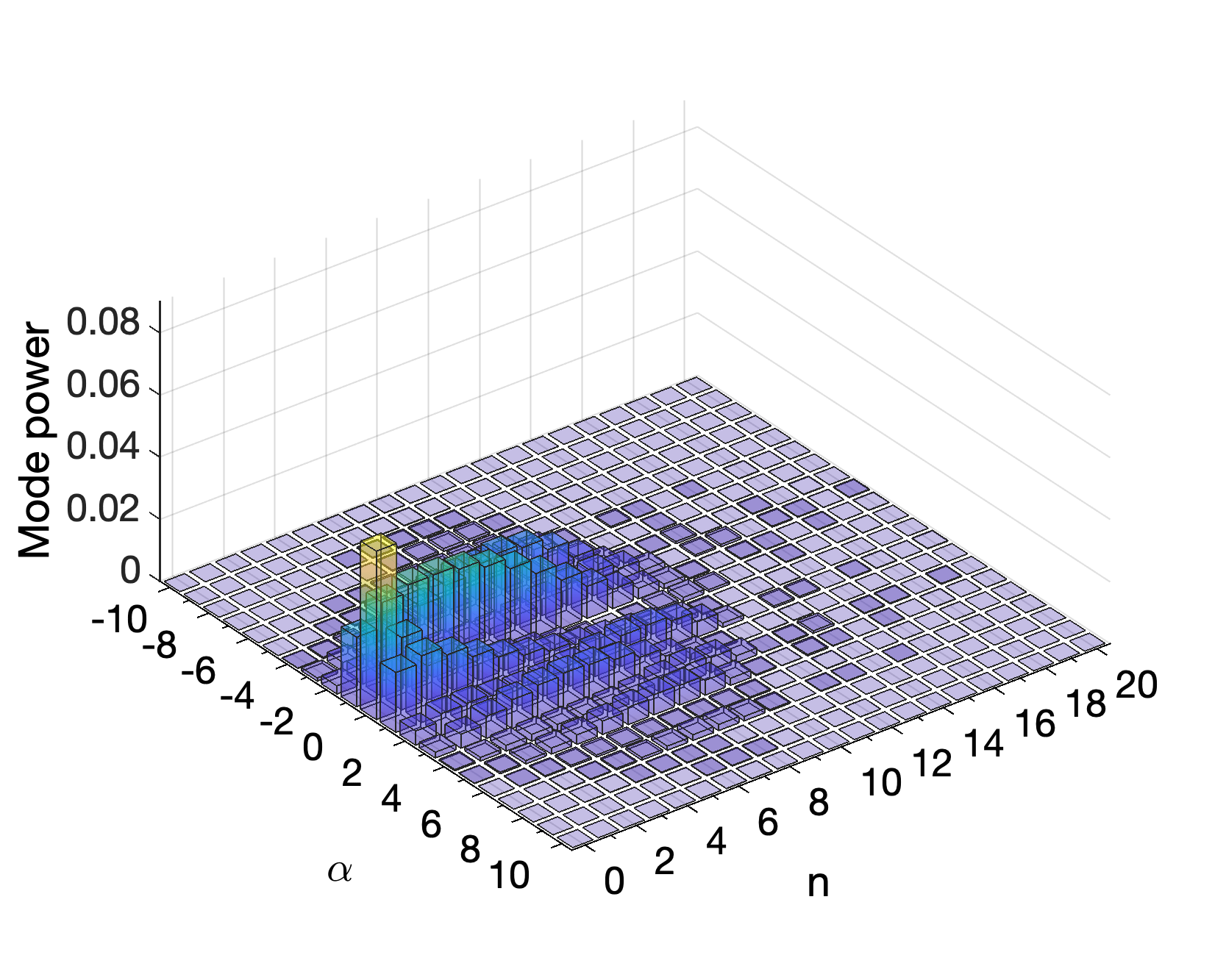}
		\caption{0.5 waves}
		\label{fig:Zernikes_planewave_decomp_05}
	\end{subfigure} %
	\begin{subfigure}[c]{0.5\textwidth}
		\centering	
		\includegraphics[width=6.5cm]{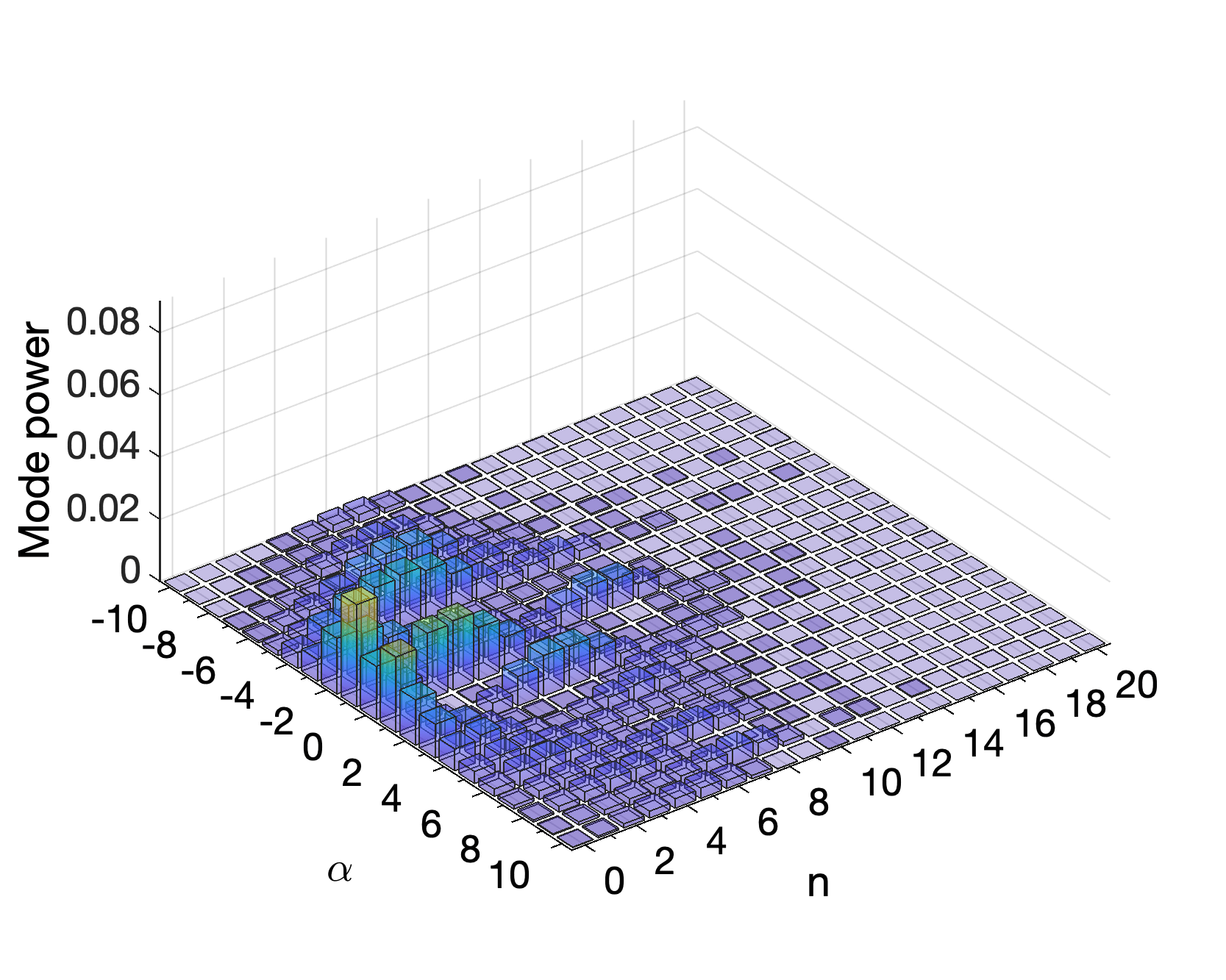}
		\caption{1 wave}
		\label{fig:Zernikes_planewave_decomp_1}
	\end{subfigure} \\
	\par\bigskip %
	\begin{subfigure}[c]{0.5\textwidth}
		\centering
		\includegraphics[width=6.5cm]{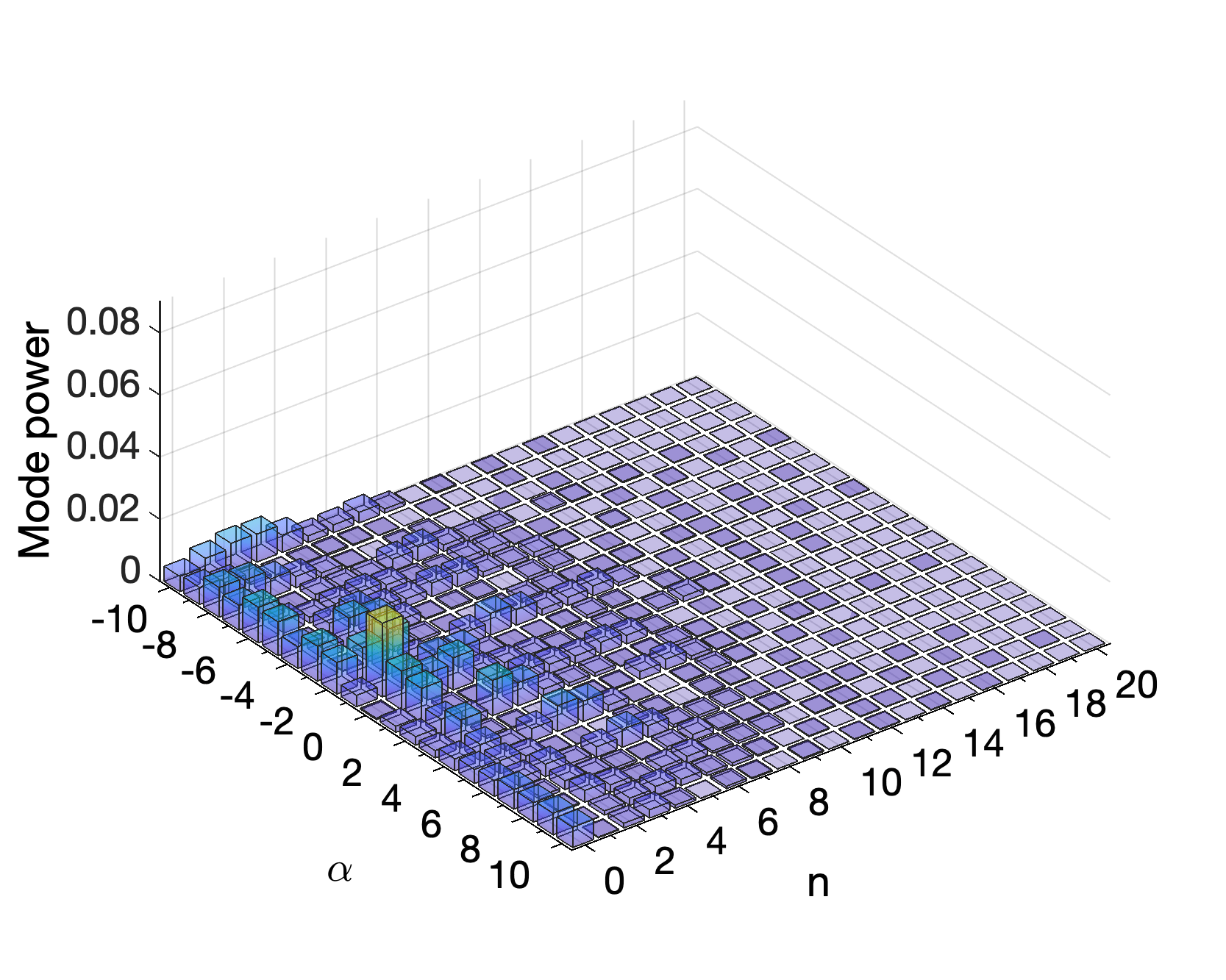}
		\caption{2 waves}
		\label{fig:Zernikes_planewave_decomp_2}
	\end{subfigure} %
	\begin{subfigure}[c]{0.5\textwidth}
		\centering
		\includegraphics[width=6.5cm]{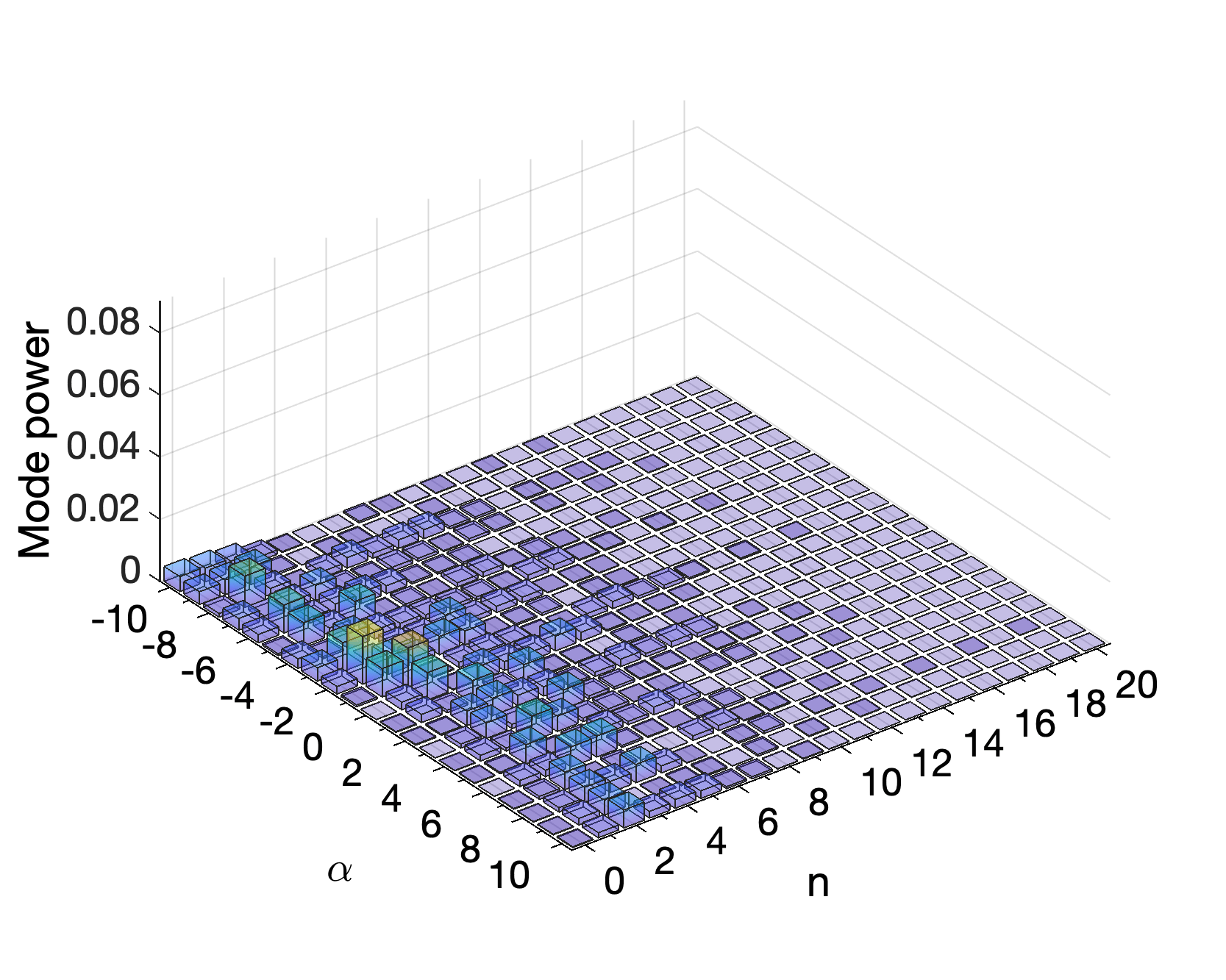}
		\caption{3 waves}
		\label{fig:Zernikes_planewave_decomp_3}
	\end{subfigure} %
\caption{Amplitudes of transverse cavity modes excitation resulting from a fundamental Gaussian input beam with varying strengths of phase delay in the shape of 10~randomly-weighted Zernike terms.}
\label{fig:Zernikes_planewave_decomps}
\end{figure}

\FloatBarrier %

Finally, Fig.~\ref{fig:Zernikes_planewave_spectra} examines the intensity spectrum transmitted. The spectrum develops as a slow, rather uneven combination of a "traveling wave packet" and an equilibrating spread toward higher order modes, which were respectively the qualitative patterns of distortion and field curvature. Recalling that the 10~Zernike terms of our phase delay include rather weighty distortion $Z_{1}^{1}$ and field curvature $Z_{2}^{0}$ terms, and that the remaining three Seidels can be seen as products of those two, this hybrid behavior makes intuitive sense. 

\begin{figure}[h!]
	\begin{subfigure}[c]{0.5\textwidth}
		\centering
		\includegraphics[width=7cm]{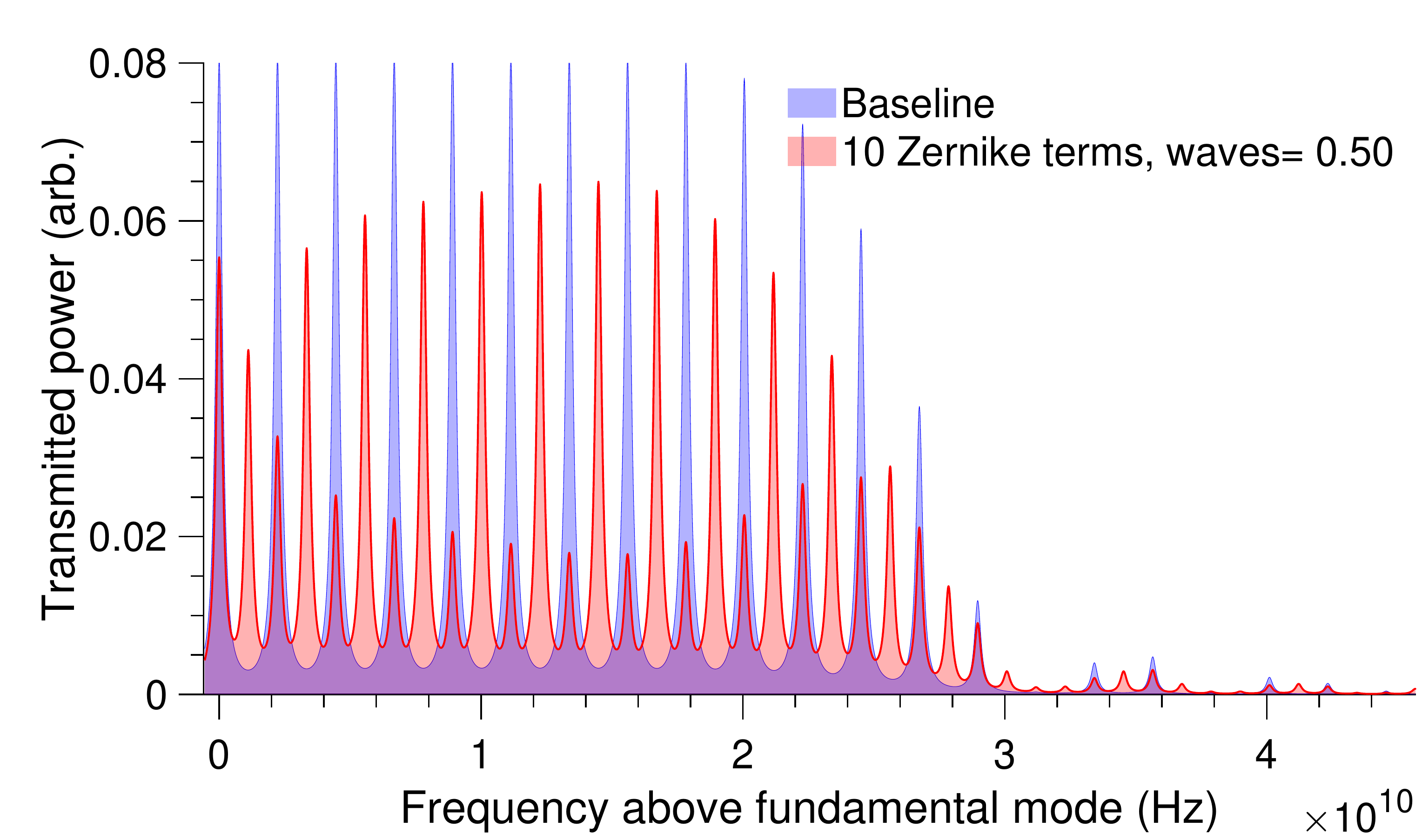}
		\caption{0.5 waves}
		\label{fig:Zernikes_planewave_spectra_05}
	\end{subfigure} %
	\begin{subfigure}[c]{0.5\textwidth}
		\centering
		\includegraphics[width=7cm]{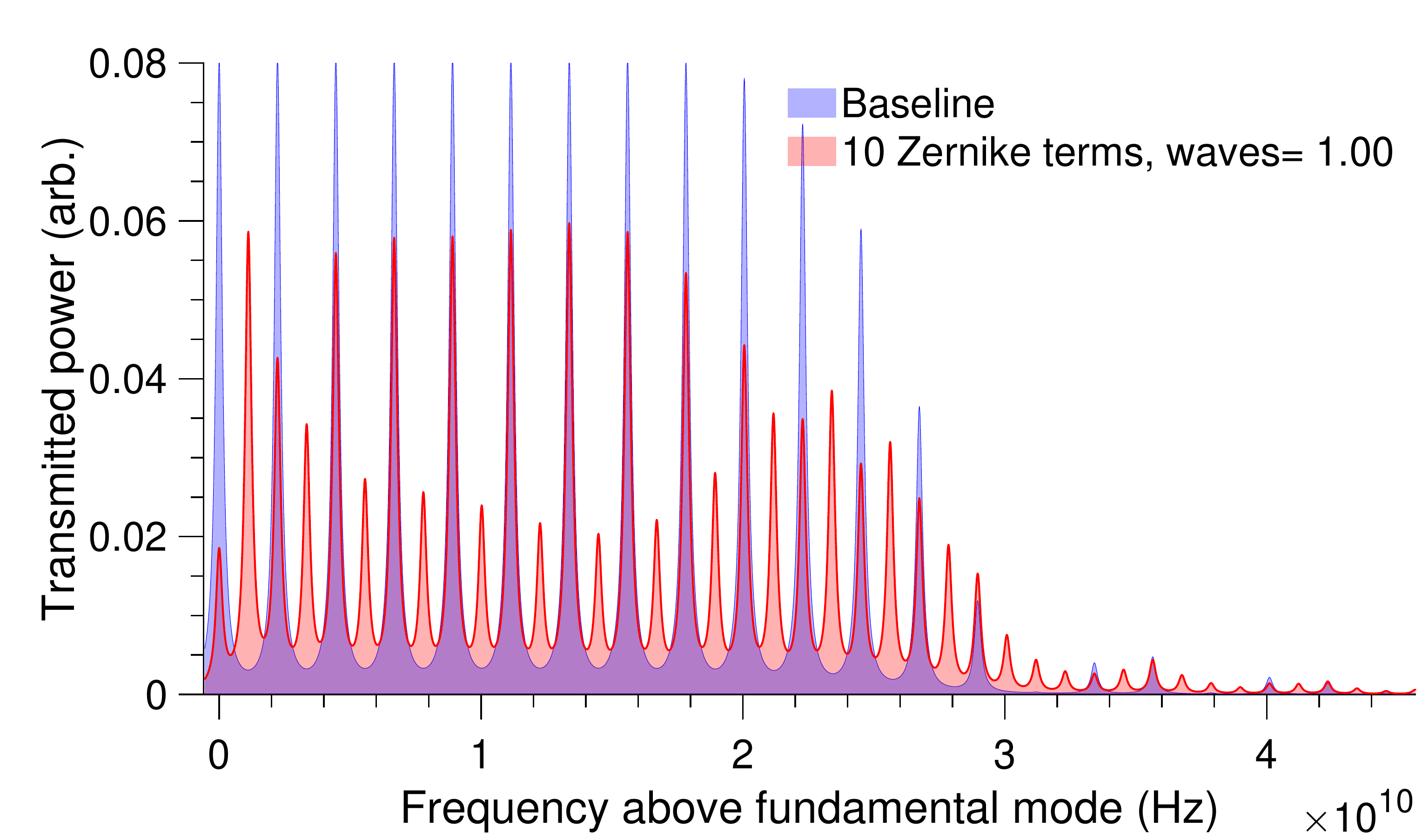}
		\caption{1 wave}
		\label{fig:Zernikes_planewave_spectra_1}
	\end{subfigure} \\
	\par\bigskip %
	\begin{subfigure}[c]{0.5\textwidth}
		\centering
		\includegraphics[width=7cm]{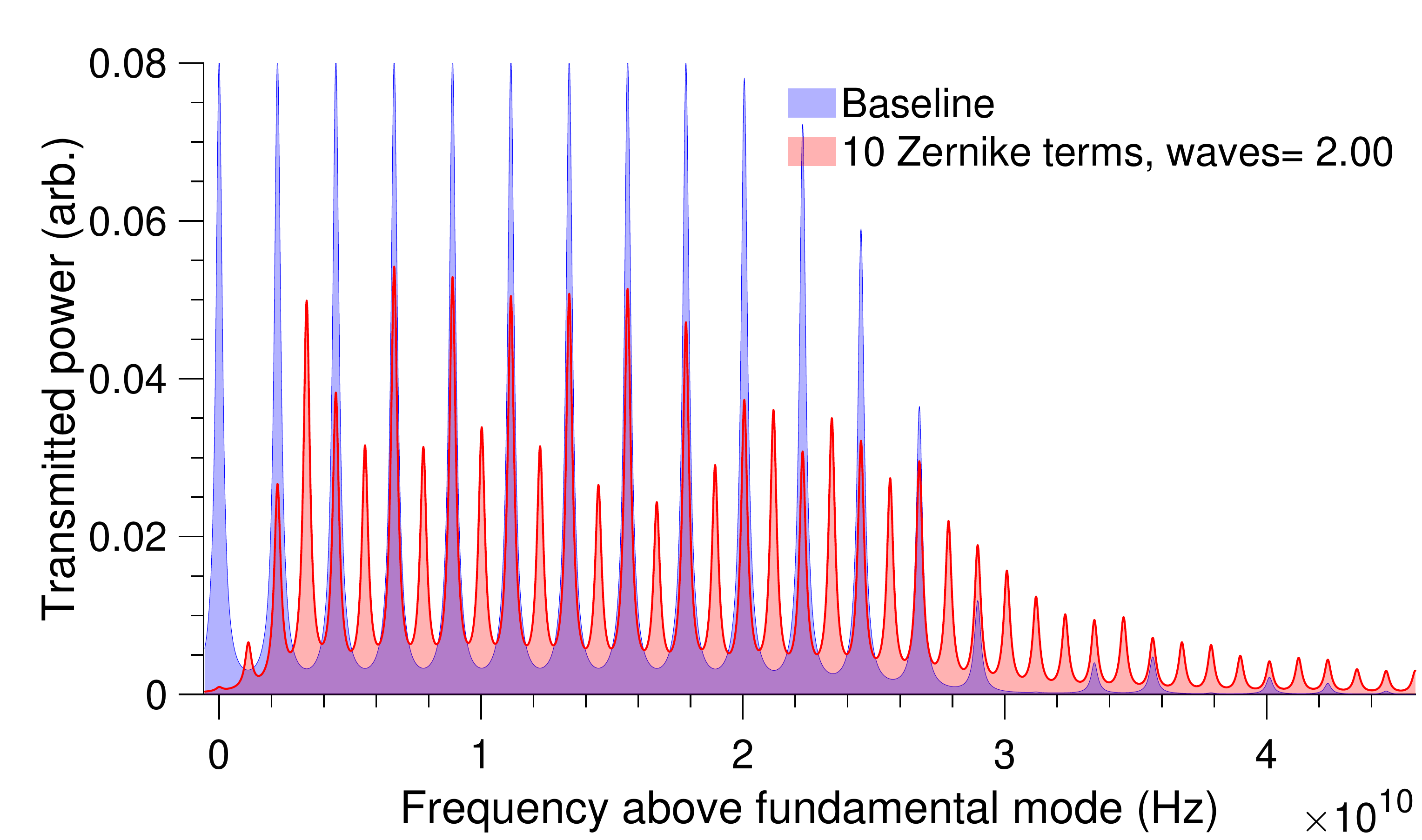}
		\caption{2 waves}
		\label{fig:Zernikes_planewave_spectra_2}
	\end{subfigure} %
	\begin{subfigure}[c]{0.5\textwidth}
		\centering
		\includegraphics[width=7cm]{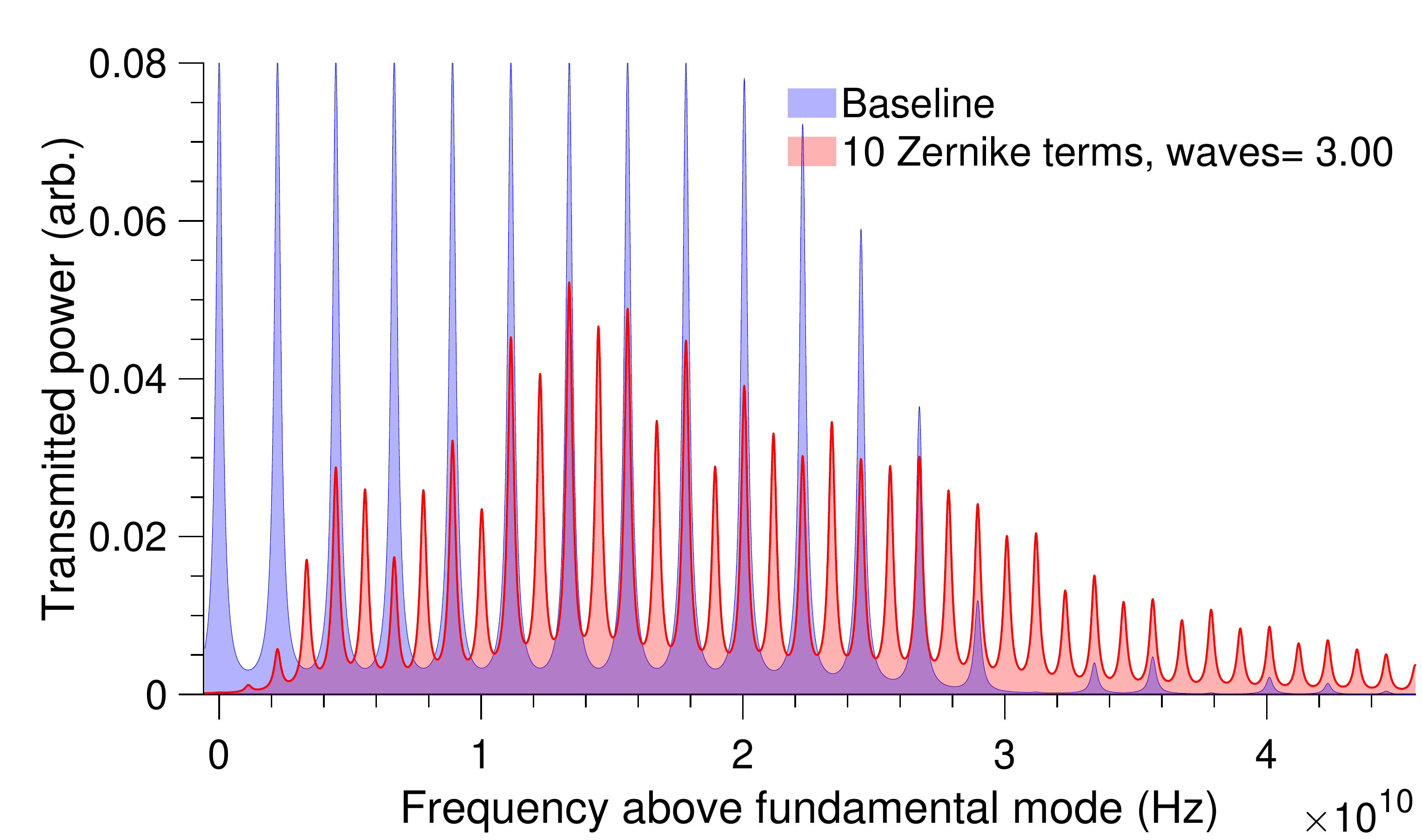}
		\caption{3 waves}
		\label{fig:Zernikes_planewave_spectra_3}
	\end{subfigure} %
\caption{The intensity spectrum transmitted by the cavity for a plane wave with a phasefront derived from ten randomly-weighted Zernike terms.}
\label{fig:Zernikes_planewave_spectra}
\end{figure}

Plotting the spectrum's progression versus aberration strength in Fig.~\ref{fig:Zernikes_planewave_spectrum_history}, we see an echo of distortion as the share of power in the lowest frequencies drops to zero with increasing aberration strength. Beyond that, a casual observer could conclude with confidence only that the spectra evolves in a generally noisier fashion, suggesting that a single spectral metric might be especially appreciated. 

\begin{figure}[h!]
\savebox{\largestimage}{\includegraphics[width=7cm]{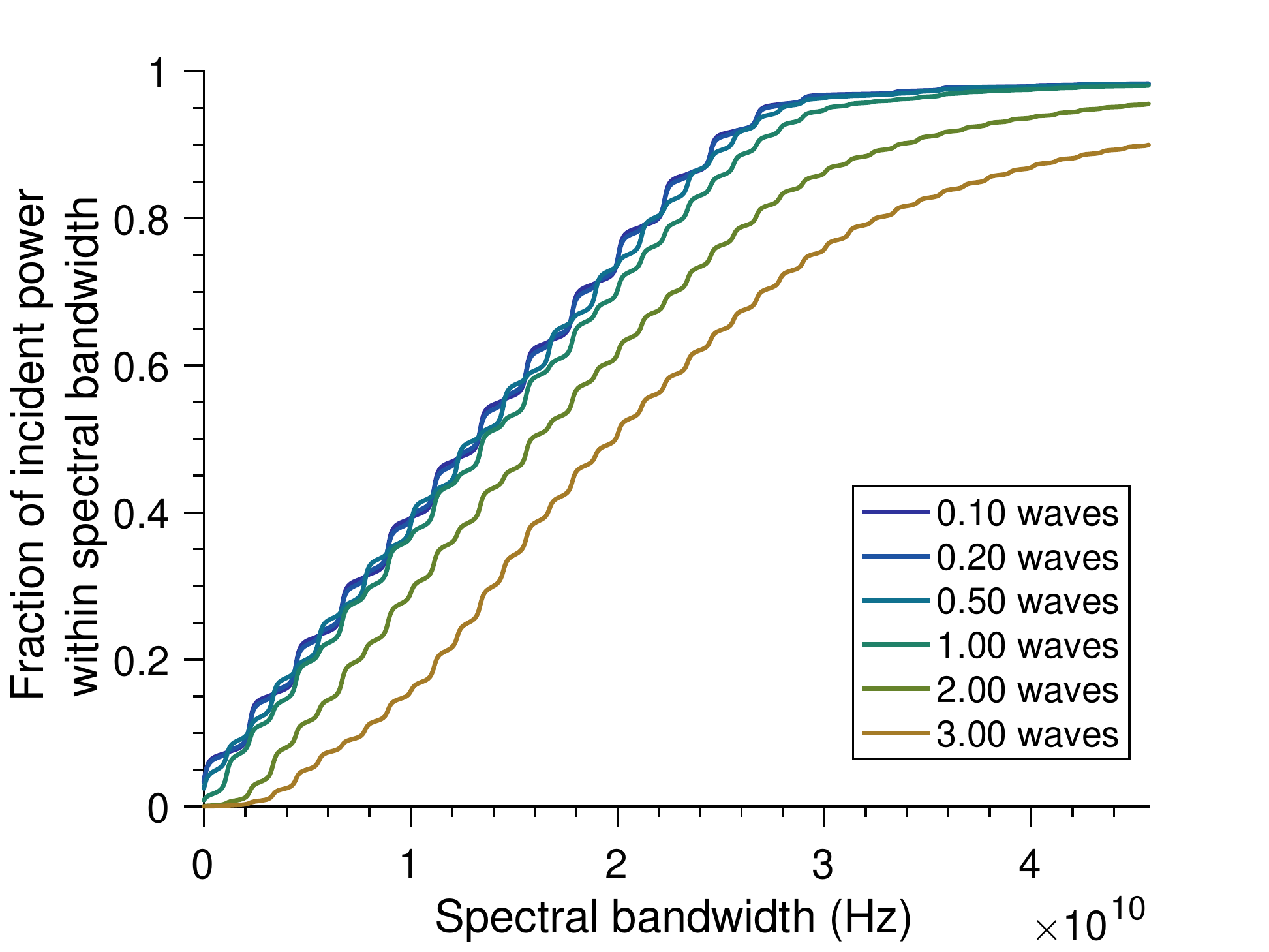}} %
	\begin{subfigure}[b]{0.5\textwidth}
		\centering
		\raisebox{\dimexpr.5\ht\largestimage-.5\height}{\includegraphics[width=7cm]{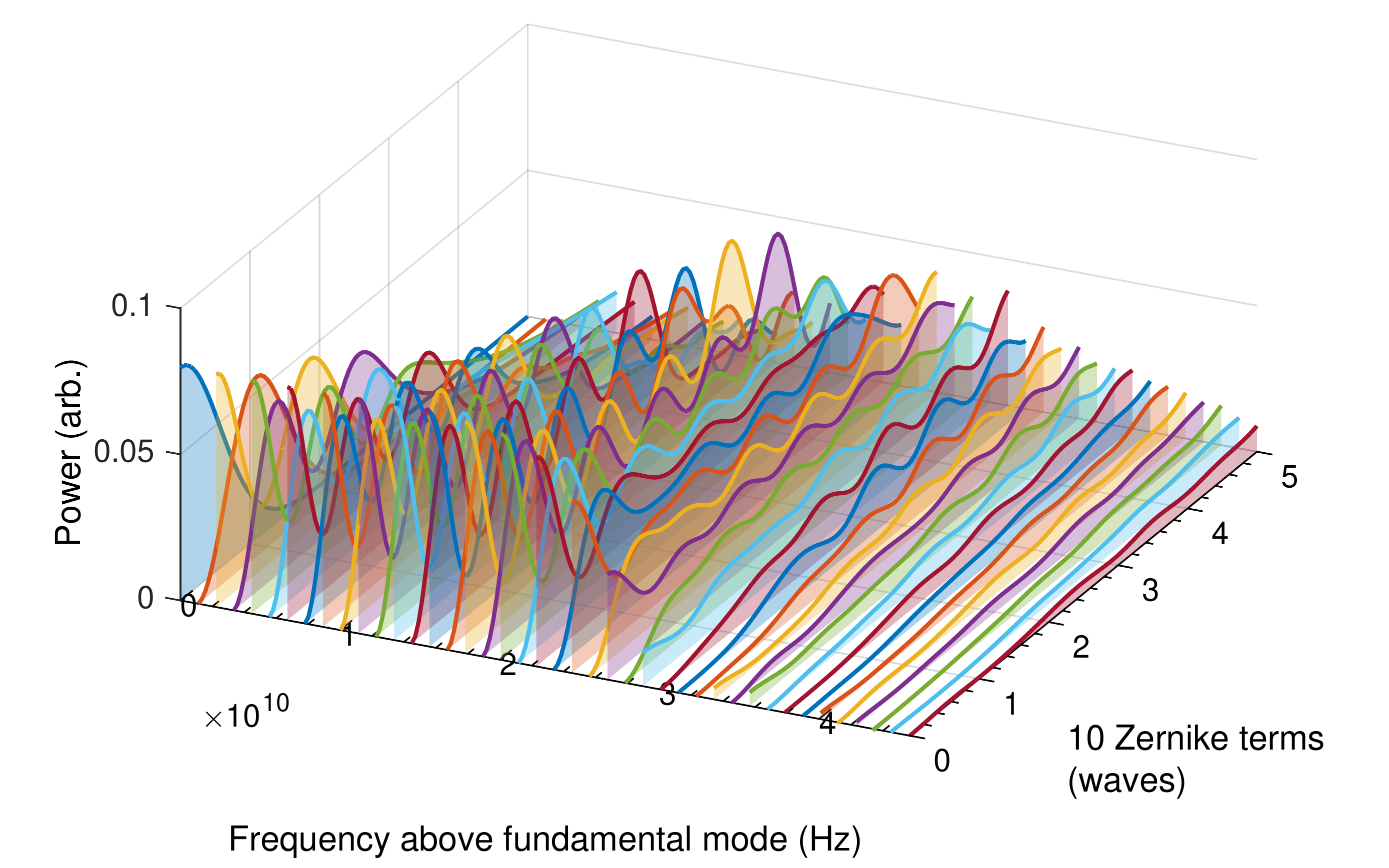}}
		\caption{Spectrum vs phase delay}
		\label{fig:Zernikes_planewave_spectrum_history}
	\end{subfigure} %
	\begin{subfigure}[b]{0.5\textwidth}
		\centering
		\usebox{\largestimage}
		\label{fig:Zernikes_planewave_spectral_power_distr}
		\caption{Spectral power distribution}
		\label{fig:Zernikes_planewave_spectral_singlemetrics}
	\end{subfigure} %
\vspace{-2mm}
\label{fig:Zernikes_planewave_spectrum_history_plus_singlemetrics}
\caption{The spectral intensities (left) and power distribution (right) of optical cavity output when stimulated by a pupiled plane wave with varying amounts of phase delay, shaped by arbitrarily weighting 10 Zernike polynomials.}
\end{figure}

The spectral power distribution, shown in Fig.~\ref{fig:Zernikes_planewave_spectral_singlemetrics}, does indeed prove easier to interpret. At lower values of $W$, the power distributions seen are difficult to distinguish from those of a slightly higher strength of distortion; as the phase delay becomes more significant, departures such as a non-uniform slope to the spectral power line---which we previously witnessed in Fig.~\ref{fig:Fieldcurvature_planewave_spectral_power_distr}, as the result of field curvature---begin to appear. Judging by these qualitative interpretations, single spectral metrics do indeed appear able to serve as a good indicator of aberration strength. 

\FloatBarrier %

\section{Conclusions and future work}

The results presented above show that aberrations induced in a beam will excite high-order modes when that beam is coupled into an optical cavity. On the one hand, this is a perfectly natural result of how aberrations are merely spatial variations in an otherwise uniform phase (and intensity) front. The more variation there is (in the form of spatial frequency content and amplitude) across the phasefront, the more spatial modes will be excited in the cavity, and the higher their amplitude coefficients will be. On the other hand, the optical cavity performs the valuable and somewhat surprising task of converting the spatial variations across the phase/intensity front into a distribution of spectral frequencies. Let us be very clear that no new frequencies are created that were not in the initial bandwidth of the light source; typical high power lasers have bandwidths of many tens of GHz, so it is easy to design a cavity that has many spatial modes within this bandwidth yet avoids overlapping into the next longitudinal mode. The simulations presented here on such a cavity suggest that input aberration strength, and perhaps form, can be well estimated from simple spectral measurements of its transmitted light. 

In our efforts to introduce the concept of aberration sensing with optical cavity modes, we have assumed that an imaging optical system is used to transfer the incident wavefront onto the front cavity mirror. This will work well, but it requires imaging optics, which are at least modestly complex. The simplest aberration sensors based on cavity spatial mode spectra may have optical systems limited to a single focusing lens. The cavity would then analyze the Fourier transform of the incident coherent field, which would convert the base plane wave to a delta function and the added aberrations would have somewhat different spectral features. The mathematics of this process are very similar to what we have performed here, except for the Fourier transform. 

The next steps in this work are to connect the theory introduced in this work to the practical aberrations seen in real systems. The daunting part of this effort is the sheer number of types of systems and therefore immediate efforts will concentrate on mathematically well-characterized sets of aberration-inducing phenomena, such as Kolmogorov turbulence.

\section{Acknowledgements}

The authors thank Joseph Pe\~{n}ano, Steve Hammel, Al Ogloza, Brad Tiffany, and Mint Kunkel for enlightening conversations. This work was supported by the Joint Directed Energy Transition Office (DE-JTO) and the Office of Naval Research under grant N00014-17-1-2438.

\printbibliography

\end{document}